\documentclass[twocolumn]{aastex7}

\usepackage{physics}
\usepackage{bm}
\usepackage[dvipsnames]{xcolor}
\usepackage{mdframed}
\usepackage{multirow}
\usepackage{amssymb}
\usepackage{subcaption}
\usepackage{siunitx}
\usepackage{cleveref}

\newcommand{\Mpch}{\ensuremath{h^{-1}\,\mathrm{Mpc}}}
\newcommand{\cMpch}{\ensuremath{h^{-1}\,\mathrm{cMpc}}}

\newcommand{\kpch}{\ensuremath{h^{-1}\,\mathrm{kpc}}}
\newcommand{\ckpch}{\ensuremath{h^{-1}\,\mathrm{ckpc}}}

\newcommand{\kmsecMpc}{\ensuremath{\mathrm{km}\,\mathrm{s}^{-1}\,\mathrm{Mpc}^{-1}}}
\newcommand{\Msunh}{\ensuremath{h^{-1}\,M_\odot}}

\newcommand{\Om}{\Omega_{\rm m}}
\newcommand{\Dvir}{\Delta_{\rm vir}}
\newcommand{\Rs}{R_{\rm s}}
\newcommand{\Mvir}{M_{\rm vir}}
\newcommand{\spin}{\lambda_{\rm Bullock}}
\newcommand{\Vmax}{V_{\rm max}}
\newcommand{\Vrms}{V_{\rm rms}}

\newcommand{\Lo}{\bm{\Lambda}}
\newcommand{\Lt}{\bm{\Lambda}^\prime}
\newcommand{\zo}{\hat{z}}
\newcommand{\zt}{\hat{z}^\prime}
\newcommand{\SC}{\texttt{SC-SAM}}
\newcommand{\CS}{\texttt{CAMELS-SAM}}
\newcommand{\Bacco}{\texttt{BACCO}}
\newcommand{\LCDM}{$\Lambda$CDM}

\begin{document}

\title{Learning the Universe with cosmological rescaling of merger trees and semi-analytic galaxy formation models}

\author[orcid=0000-0002-0986-314X, gname=Richard, sname=Stiskalek]{Richard Stiskalek}
\affiliation{Astrophysics, University of Oxford, Denys Wilkinson Building, Keble Road, Oxford, OX1 3RH, UK}
\affiliation{Center for Computational Astrophysics, Flatiron Institute, 162 5th Avenue, New York, NY 10010, USA}
\email[show]{\href{mailto:richard.stiskalek@physics.ox.ac.uk}{richard.stiskalek@physics.ox.ac.uk}}
\author[0000-0002-8449-1956]{Lucia A. Perez}
\affiliation{Center for Computational Astrophysics, Flatiron Institute, 162 5th Avenue, New York, NY 10010, USA}
\affiliation{Department of Astrophysical Sciences, Princeton University, 4 Ivy Lane, Princeton, NJ 08544, USA}
\email[]{}
\author[0000-0002-3185-1540]{Shy Genel}
\affiliation{Center for Computational Astrophysics, Flatiron Institute, 162 5th Avenue, New York, NY 10010, USA}
\affiliation{Columbia Astrophysics Laboratory, Columbia University, 550 West 120th Street, New York, NY 10027, USA}
\email[]{sgenel@flatironinstitute.org}
\author[0000-0002-6748-6821]{Rachel S. Somerville}
\affiliation{Center for Computational Astrophysics, Flatiron Institute, 162 5th Avenue, New York, NY 10010, USA}
\email[]{}
\author[0000-0003-2953-3970]{Raul E. Angulo}
\affiliation{Donostia International Physics Center (DIPC), Paseo Manuel de Lardizabal 4, 20018 Donostia-San Sebasti\'{a}n, Spain}
\affiliation{IKERBASQUE, Basque Foundation for Science, E-48013 Bilbao, Spain}
\email[]{}
\author[0000-0001-7511-7025]{Sergio Contreras}
\affiliation{Facultad de F\'isica, Universidad de Sevilla, Campus de Reina Mercedes, Av. Reina Mercedes s/n 41012 Seville, Spain}
\email[]{}

\begin{abstract}
    Learning cosmology from galaxy surveys requires large suites of simulations spanning the cosmological and astrophysical parameter space, yet hydrodynamical simulations of galaxy formation remain prohibitively expensive.
    Semi-analytic models offer an inexpensive, physically grounded alternative, but still require halo merger trees from $N$-body simulations, and densely sampling cosmological parameters in sufficient volume remains expensive.
    We address this by extending cosmological rescaling to operate directly on merger trees and applying it in the $\Om$--$\sigma_8$ plane, running the Santa Cruz semi-analytic model for galaxy formation on the rescaled trees to produce galaxy populations across new cosmological and astrophysical parameters at negligible additional cost.
    A novel halo-profile-based correction, controlled by a single free parameter, suppresses systematic bias in rescaled halo masses to below the per cent level.
    We apply the method to parameter estimation of $\Om$ and $\sigma_8$ given either the stellar mass function or the two-point correlation function, finding that as few as 64, and potentially fewer, base $N$-body simulations, rescaled to ${\sim}1000$ training samples, match the accuracy of 750 dedicated $N$-body simulations; rescaling to 3200 realisations improves the prediction of $\Om$ by ${\sim}25\%$.
    Rescaling all merger trees from a single \texttt{CAMELS-SAM} $N$-body simulation costs ${\sim}0.1$~CPUh, compared to several thousand~CPUh to run the simulation itself.
    We demonstrate a practical route to obtaining predictions of galaxy summary statistics across cosmological and astrophysical parameters, even with a relatively small number of base $N$-body simulations.
\end{abstract}



\section{Introduction}\label{sec:introduction}

Modern cosmology relies on precision measurements from large-scale structure surveys, necessitating accurate theoretical predictions across a wide range of cosmological and astrophysical parameters.
Cosmological hydrodynamical simulations such as \texttt{Horizon-AGN}~\citep{Dubois_2014}, \texttt{EAGLE}~\citep{Schaye_2015}, \texttt{IllustrisTNG}~\citep{Pillepich_2018,Springel_2018}, \texttt{SIMBA}~\citep{Dave_2019}, \texttt{FLAMINGO}~\citep{Schaye_2023}, and \texttt{MillenniumTNG}~\citep{Pakmor_2023} remain the gold standard for modelling galaxy formation, capturing complex baryonic physics including gas cooling, star formation, and feedback from supernovae and active galactic nuclei.
However, such simulations are prohibitively expensive for dense sampling of the cosmological and astrophysical parameter space: a single high-resolution run can require tens to hundreds of millions of CPUh, making comprehensive exploration of parameter space impractical.
The demand for large simulation suites has surged with the advent of machine learning, particularly simulation-based inference~\citep[SBI;][]{Papamakarios_2019,Greenberg_2019,Alsing_2018,Alsing_2019,Cranmer_2020}, which bypasses explicit likelihood evaluation but typically requires thousands of forward-model realisations for training~\citep{Villaescusa-Navarro_2020,Bairagi_2025,Massara_2025}.
Transfer learning from lower-fidelity simulations can reduce this budget~\citep{Saoulis_2024}, yet even such approaches benefit from maximising the effective number of independent realisations.

The \texttt{CAMELS} project~\citep{CAMELS_2021} represents a leading effort to address this challenge, providing thousands of hydrodynamical simulations that systematically sample two cosmological parameters ($\Om$, $\sigma_8$) and four astrophysical feedback parameters across multiple galaxy formation models, though the \texttt{CAMELS} suite has since expanded to additional parameters~\citep{Ni_2023}.
However, even this ambitious programme remains limited in cosmological volume, parameter coverage and resolution.
The \CS\ suite~\citep{Perez_2023} adopts an alternative strategy: coupling $N$-body simulations with semi-analytic galaxy formation models~\citep[SAMs;][]{White_1991,Kauffmann_1993,Somerville_1999,Cole_2000,Croton_2006,Bower_2006,DeLucia_2007,Guo_2011,Benson_2012,Henriques_2015,Lacey_2016,Croton_2016,Hirschmann_2016,Lagos_2018,Pandya_2026}, which predict galaxy properties---e.g., stellar masses, star formation rates, gas content, and morphologies---by solving coupled differential equations along halo merger trees rather than evolving baryons directly through a hydrodynamical solver.
This approach reduces the computational cost per realisation by orders of magnitude while retaining physical prescriptions for gas cooling, star formation, and feedback, and thus enables denser sampling of the parameter space or larger volumes.
When calibrated against cosmological hydrodynamical simulations, SAMs reproduce the predictions of the latter to comparable accuracy~\citep{Pandya_2020,Hadzhiyska_2021,Gabrielpillai_2022,Oren_2026}.
Nevertheless, each realisation still requires an underlying $N$-body simulation, and running the hundreds to thousands needed for robust cosmological inference at the resolution required to sufficiently resolve merger trees and in sufficient volume remains computationally prohibitive.

Cosmological rescaling offers a promising alternative to running expensive simulations from scratch.
This approach transforms the outputs of an $N$-body simulation run under an original cosmology to approximate structure formation in a target cosmology, without rerunning the simulation~\citep{Angulo_2010}.
Computationally expensive simulations can thus be reused across a range of nearby cosmological parameters.
The core idea is to find a length scale such that the variance of the linear matter field of the target cosmology is matched by that of the original at a suitably chosen redshift.

Rescaling has proven effective for constructing mock catalogues and training emulators, with validation studies demonstrating per-cent-level agreement in halo mass functions and clustering statistics for modest parameter shifts~\citep{Angulo_2010,Ruiz_2011,Mead_2014,Contreras_2020,Contreras_2023,Mokeddem_2025}.
Nevertheless, rescaling accuracy degrades as the target cosmology deviates further from the original~\citep{Angulo_2010,Zennaro_2019}: differences in growth history alter halo formation times and internal structure in ways that a simple coordinate transformation cannot fully capture.

A prominent application of the rescaling is the \Bacco\ project, which uses rescaling for emulator construction~\citep{Angulo_2021}.
Starting from three high-resolution $N$-body simulations ($1440~\Mpch$ boxes with $4320^3$ particles), \Bacco\ applies rescaling to span an eight-dimensional cosmological parameter space and build emulators for the non-linear matter power spectrum, halo mass function, halo clustering, galaxy clustering and galaxy-galaxy lensing~\citep{Contreras_2020,Arico_2021,Zennaro_2023,Contreras_2023,Ortega_2026,Mahony_2026}.
The resulting emulators achieve $2$--$3\%$ accuracy for $\Lambda$CDM and extended models including dynamical dark energy and massive neutrinos~\citep{Zennaro_2019}.

Most prior work, including \Bacco, applies rescaling to halo catalogues or particle snapshots.
Here, we instead directly scale the dark matter halo merger trees, allowing rapid re-evaluation of galaxy formation models under new cosmologies without rerunning expensive $N$-body or hydrodynamical simulations.
Operating at the tree level also bypasses the halo-finding and merger-tree construction stages that particle-level rescaling must repeat, which constitute the dominant post-processing cost.
Because merger trees encode the full assembly history of halos, rescaling must not only preserve instantaneous halo properties but also merger timing and mass ratios, both of which influence SAM predictions~\citep{Benson_2016,Chandro-Gomez_2025}.

In this work, we develop and apply a modified rescaling algorithm that operates directly on halo merger trees from the \CS\ suite.
We re-run SAMs on the rescaled trees to generate galaxy populations and assess the fidelity of rescaled simulations when processed to produce galaxy properties.
Our main contributions are: (i) extending the rescaling formalism to merger trees, including a novel NFW-based correction that suppresses systematic bias in rescaled halo mass; (ii) validating rescaled galaxy populations against SAMs run on direct $N$-body simulations across a range of $\Om$ and $\sigma_8$ values; and (iii) demonstrating that a small number of base $N$-body simulations, in combination with the rescaling method, suffices to train accurate neural network estimators of cosmological parameters from galaxy statistics.

This work is part of the Learning the Universe collaboration\footnote{\url{https://learning-the-universe.org/}}, which seeks to infer the cosmological parameters and initial conditions of the Universe by combining state-of-the-art simulations, machine-learning emulation, and Bayesian forward modelling.
The merger-tree rescaling method developed here contributes to this programme by reducing the computational cost of generating training samples of galaxy observables across cosmological and astrophysical parameters for the machine-learning models used in cosmological inference.

The paper is organised as follows.
Section~\ref{sec:data} describes the \CS\ simulations and SAM outputs used in this study.
Section~\ref{sec:rescaling} presents the rescaling algorithm and our modifications for merger trees.
Section~\ref{sec:method} details the validation methodology, including summary statistics and the neural network parameter estimator.
Section~\ref{sec:results} validates the rescaling algorithm on individual objects and population statistics.
Section~\ref{sec:parameter_estimation} applies the rescaling to parameter estimation from galaxy summary statistics.
Section~\ref{sec:discussion} discusses implications and limitations, and Section~\ref{sec:conclusion} concludes.
We define $h \equiv H_0 / \left(100~\kmsecMpc\right)$.


\section{Simulation data}\label{sec:data}

We use the \CS\ suite~\citep{Perez_2023}, which consists of \num{1000} $N$-body simulations of volume $(100~\cMpch)^3$, each evolved with $640^3$ dark matter particles.
The simulations quasi-uniformly sample across a Latin hypercube\footnote{Latin hypercube sampling is a stratified sampling method that divides the range of each parameter into $N$ equally probable intervals and uniformly samples once from each interval. The samples from different parameters are then randomly paired, ensuring more uniform coverage of the parameter space than purely random sampling.} the cosmological parameter space of matter density $\Omega_{\rm m} \in [0.1, 0.5]$ and amplitude of matter fluctuations $\sigma_8 \in [0.6, 1.0]$, while fixing the background to a flat \LCDM\ cosmology with baryonic density $\Omega_{\rm b} = 0.049$, Hubble constant $H_0 = 67.11~\kmsecMpc$, scalar spectral index $n_{\rm s} = 0.9624$, sum of the neutrino masses $m_{\nu} = 0$, and dark energy equation-of-state parameter $w = -1$.

The \CS\ $N$-body simulations were run with \texttt{Arepo}~\citep{Springel_2010,AREPO} using only cold dark matter particles, and the initial conditions were generated with second-order Lagrangian perturbation theory (2LPT) at $z = 127$ using power spectra from \texttt{CAMB}~\citep{CAMB}.
The gravitational softening length is fixed to $4~\ckpch$ until $z = 1$, and held constant at $2~\kpch$ thereafter.
The resulting dark matter particle mass is $3.18 \times (\Om / 0.3) \times 10^{8}~\Msunh$.
Each simulation stores 100 snapshots between $z = 27$ and $z = 0$.
Halos and subhalos are identified using \texttt{Rockstar}~\citep{Behroozi_2013}, and merger trees are constructed with \texttt{ConsistentTrees}~\citep{Behroozi_2013_cs}.

Within \CS, galaxy populations are generated using the Santa Cruz SAM (\SC;~\citealt{Somerville_1999,Somerville_2008,Somerville_2015,Somerville_2021}). \SC\ evolves baryons along dark matter merger trees by solving coupled differential equations governing gas accretion, cooling, star formation, stellar and active galactic nucleus feedback.
The \SC\ implementation in \CS\ follows~\citet{Gabrielpillai_2022}, excluding all satellite halos from the \texttt{ConsistentTrees} merger trees and tracking only segments of the main branches that contain more than 100 dark matter particles.
Following accretion into the main halo, satellite radial positions evolve under a semi-analytic ``orphan model''.
This model, detailed in \citet{Somerville_2008}, employs a modified Chandrasekhar formula~\citep{Boylan_Kolchin_2008} for the orbital angular momentum loss rate due to dynamical friction, which depends on the satellite's initial orbital energy and angular momentum.
While the orbital energy and angular momentum could in principle be extracted from the $N$-body subhalo histories, the current \SC\ instead draws the orbital circularity from the distribution calibrated to subhalo populations in cosmological $N$-body simulations by \citet{Zentner_2005}, and the orbital energy from a uniform distribution in the circular-orbit radius $r_{\rm circ}(E)/R_{\rm vir} \in [0.6, 1.0]$.
Galaxy merger rates and derived properties such as stellar mass and star formation rate therefore depend on the random seed of the orbit draws.
For the $N$-body resolution of \CS, the \SC\ model is numerically converged for halos with mass greater than $10^{11}~\Msunh$ (Appendix~A of \citealt{Perez_2023}), corresponding to a stellar mass limit of approximately $10^8~\Msunh$ for the fiducial \CS\ model.
In \CS, three astrophysical parameters are varied to probe the impact of feedback on galaxy formation.
Two control the amplitude and slope of the stellar-driven wind mass outflow rate, while the third sets the amplitude of thermal heating in active galactic nuclei radio-mode feedback.
These parameters are sampled over broad priors using a Latin hypercube design; see~\citet{Perez_2023} for how the Latin hypercube sampling was set up in \CS.

In \CS, each simulation uses unique phases of the initial density field (so-called initial conditions).
However, to benchmark the rescaling algorithm, we require simulations with identical initial phases but different cosmological parameters, enabling bijective halo matching between runs.
We run paired $N$-body simulations using the same configuration as \texttt{CAMELS-SAM}: each pair shares the same initial phases but differs in cosmological parameters, with one simulation serving as the original ($\Lo$) and the other as the target ($\Lt$).
We consider three types of cosmological steps for initial testing: (i) a step in $\Om$ from $0.25$ to $0.3$ at fixed $\sigma_8 = 0.8$, (ii) steps in $\sigma_8$ at fixed $\Om = 0.3$, from $\sigma_8 = 0.8$ to target values of up to $1.0$, and (iii) a joint step in both parameters.
Because the cosmological rescaling algorithm scales the simulation box size by a factor $s$ (described in~\autoref{sec:rescaling}), the target simulation is run with box size $(s\times100~\cMpch)^3$, scaling the force resolution with the box size, while keeping all other settings fixed.

Additionally, for the subsequent parameter estimation (\autoref{sec:method_parameter_estimation}), we run 15 simulations with random initial conditions, all with box size $(100~\cMpch)^3$.
These simulations sample $\Om$ from 0.15 to 0.5 in steps of 0.025, all at fixed $\sigma_8 = 1.0$.
We place these simulations at $\sigma_8 = 1.0$ because rescaling to lower $\sigma_8$ values requires selecting an earlier snapshot, since structure formation proceeds more rapidly in higher $\sigma_8$ cosmologies; this avoids the need to evolve the simulations into the future.
For the phase-mixing experiment of~\autoref{sec:results_phase_mixing}, we additionally run a set of simulations matching the \CS\ setup but sharing a single initial-condition phase across all of them.


\section{Rescaling algorithm}\label{sec:rescaling}

Here, we give an overview of the rescaling algorithm.
In \autoref{sec:rescaling_particle}, we first describe the cosmological rescaling algorithm applied to particle snapshots from $N$-body simulations (though in this work we apply it directly to halos, not particles).
Then, in \autoref{sec:rescaling_halo_catalogues}, we describe its application directly to halo catalogues.
Lastly, in \autoref{sec:rescaling_trees}, we introduce the application of the cosmological rescaling algorithm to merger trees.

\subsection{Rescaling of particle snapshots}\label{sec:rescaling_particle}

The purpose of the cosmological rescaling algorithm is to map a particle snapshot from an $N$-body simulation run with an original cosmology to a (nearby) target cosmology, without rerunning the simulation~\citep{Angulo_2010}.

We denote the original cosmology by $\Lo$ and the target cosmology, to which the original snapshots are to be rescaled, by $\Lt$.
We denote redshifts in the original and target cosmologies by $\zo$ and $\zt$, respectively\footnote{Hats denote the specific snapshot redshifts of the rescaling pair, in contrast to $z$ used as a generic redshift variable elsewhere.}.
Throughout, ``unprimed'' quantities refer to the original cosmology and ``primed'' quantities to the target cosmology.

For a linear matter power spectrum $P(k)$ at $z = 0$, the redshift-dependent linear power spectrum is
\begin{equation}
    P(k, z) = D^2(z) P(k),
\end{equation}
where $D(z)$ is the linear growth factor, normalised such that $D(0) = 1$.
The variance of the linear density field smoothed over a comoving scale $R$ is
\begin{equation}
    \sigma^2(R, z) \equiv \frac{D^2(z)}{4\pi} \int_0^{\infty} \dd k\ k^2 P(k) W^2(kR),
\end{equation}
where $W(kR)$ is the Fourier transform of the real-space top-hat filter,
\begin{equation}
    W(kR) = 3\left[\frac{\sin(kR) - kR \cos(kR)}{(kR)^3}\right].
\end{equation}

Given a target cosmology $\Lt$, for which we want a particle snapshot at redshift $\zt$, we seek to find a redshift $\zo$ in the original cosmology $\Lo$ such that upon rescaling lengths by a factor of $s$, the linear variances of the rescaled and target fields agree over a range of scales.
This is achieved by minimising the following residual variance with respect to both $s$ and $\zo$,
\begin{equation}\label{eq:delta_rms}
    \delta^2_{\mathrm{rms}}(s, \zo)
    =
    \frac{1}{\ln(R_2 / R_1)} \int_{R_1}^{R_2} \frac{\mathrm{d}R}{R}
    \left( 1 - \frac{\sigma(s^{-1} R, \zo)}{\sigma^\prime(R, \zt)} \right)^2,
\end{equation}
where $R_1$ and $R_2$ define the scale range over which the linear variances are matched in the target cosmology.
Under the approximation of a universal mass function, as in Press--Schechter theory, the halo mass function is determined by $\sigma(R,\,z)$, so \autoref{eq:delta_rms} approximately matches the HMFs over the corresponding mass range~\citep{Angulo_2010}.
At this linear-theory level the growth factor $D(z)$ is not an independent factor, since it enters only through the variance $\sigma(R,\,z)$ that \autoref{eq:delta_rms} matches at the target redshift; the mean matter density $\bar{\rho}_{\rm m}\propto\Om$, which sets the scale-to-mass conversion $M = (4\pi/3)\,\bar{\rho}_{\rm m}\,R^3$, is absorbed by the mass rescaling of \autoref{eq:scaling_m}; and the collapse threshold $\delta_{\rm c}$ carries the residual cosmology dependence.
The equivalence is therefore only as accurate as the assumption of a universal mass function.
We use the \Bacco\ code~\citep{Angulo_2021} to minimise \autoref{eq:delta_rms}. We implement the remainder of the rescaling algorithm in \texttt{JAX}\footnote{\url{https://github.com/jax-ml/jax}}.

Given the original and target cosmologies, there are two rescaling directions. In the forward direction, one fixes the original redshift $\zo$ (i.e.\ selects a snapshot of the original simulation) and solves for $s$ and $\zt$; this determines the target redshift of the rescaled simulation. In the backward direction, one fixes the desired target redshift $\zt$ and solves for $s$ and $\zo$; this identifies which snapshot of the original simulation to rescale from.

To construct the rescaled snapshot, we take the original snapshot at redshift $\zo$, reassign it to redshift $\zt$, and scale the simulation box size as
\begin{equation}\label{eq:length_scaling}
    L^\prime = s L.
\end{equation}
Next, we correspondingly scale particle positions $\bm{x}$, velocities $\bm{v}$, and masses $m$.
In the reduced version of the rescaling algorithm~\citep{Ruiz_2011}, which omits large-scale corrections, these are rescaled as
\begin{align}
    \bm{x}^\prime &= s \bm{x}, \\
    \bm{v}^\prime &= \left( \frac{\Omega'_{\mathrm{m}} {L'}^2 (1 + \zt)}{\Omega_{\mathrm{m}} L^2 (1 + \zo)} \right)^{1/2} \bm{v},\\
    m^\prime &= s^3 \frac{\Om^\prime}{\Om} m.
    \label{eq:scaling_m}
\end{align}
The mass rescaling ensures that $\Om^\prime$ satisfies its definition as the ratio of matter to critical density, given the unchanged particle number.
By reassigning the redshift and box size and rescaling particle properties, we obtain a snapshot matching the target cosmology.
This outlines a reduced version of the rescaling algorithm, whereas the original formulation of~\citet{Angulo_2010} includes additional large-scale corrections to the particle rescaling to match the power spectra in the linear regime.
The reduced version is appropriate for simulation boxes with $L \lesssim 100~\Mpch$, which do not sample the relevant large-scale modes, making the large-scale correction unnecessary~\citep{Ruiz_2011}.

\subsection{Rescaling of halo catalogues}\label{sec:rescaling_halo_catalogues}

Rescaling particle snapshots directly would require rerunning the halo finder, and because the rescaling uses linear theory, it would not guarantee accurate results within halos, where the density field is non-linear.
To address both issues, direct rescaling of the halo catalogues has been proposed (e.g.~\citealt{Ruiz_2011,Mead_2014}).
In this approach, halo position and velocity are rescaled as in the original formalism (optionally including large-scale corrections).
On the other hand, the halo mass no longer follows the rescaling of \autoref{eq:scaling_m}, and additional prescriptions are derived for the rescaling of internal halo properties, typically assuming the Navarro--Frenk--White (NFW;~\citealt{Navarro_1997}) density profile.
The rescaling of halo position (even with the large-scale corrections) is expected to break down on small scales, particularly for satellite halos.
However, recall that the \SC\ implementation we use does not track satellite halo orbits through the merger trees, but instead via the orphan model described in \autoref{sec:data}.

\vspace{3mm}
We now describe the halo rescaling method employed in this work, introducing a novel correction to the density profile rescaling to reduce bias in the rescaled halo mass when varying $\Om$ and $\sigma_8$.
To rescale the halo virial mass $\Mvir$, we assume that halos are described by the NFW profile (though we do not enforce this explicitly),
\begin{equation}
    \rho(r) = \frac{\rho_0}{\frac{r}{\Rs}(1 + \frac{r}{\Rs})^2},
\end{equation}
where $\rho_0$ is a characteristic NFW density and $\Rs$ is the scale radius. The virial mass $\Mvir$ follows the spherical overdensity mass definition
\begin{equation}\label{eq:virial_mass}
    \Mvir \equiv \frac{4\pi R_{\rm vir}^3}{3} \Delta_{\rm vir}(z) \rho_{\rm crit}(z),
\end{equation}
where $R_{\rm vir}$ is the virial radius, $\Delta_{\rm vir}(z)$ is the virial overdensity, and $\rho_{\rm crit}(z)$ is the critical density. The virial radius is the radius of a sphere enclosing an average density of $\Delta_{\rm vir}(z) \times \rho_{\rm crit}(z)$.
We use the virial overdensity threshold of~\citet{Bryan_1998},
\begin{equation}\label{eq:delta_vir}
    \Delta_{\rm vir}(z) = 18\pi^2 + 82x - 39x^2,
\end{equation}
where $x = \Omega_{\rm m}(z) - 1$.
Lastly, the critical density is given by
\begin{equation}
    \rho_{\rm crit}(z) = \frac{3H^2(z)}{8\pi G},
\end{equation}
where $H(z)$ is the Hubble parameter at redshift $z$ and $G$ is the Newtonian constant of gravitation.
A halo described by the NFW profile can be characterised either by $\rho_0$ and $\Rs$, or, equivalently, by its mass $\Mvir$ and concentration, defined as
\begin{equation}
    c \equiv R_{\rm vir} / \Rs.
\end{equation}

For a halo from an $N$-body simulation (identified by \texttt{Rockstar}), the NFW profile can be fitted to obtain its best-fit parameters.
We introduce a rescaling of the NFW parameters,
\begin{align}
    \rho_0^\prime &= \rho_0 \left(\frac{\Om^\prime}{\Om} \mathcal{A}_{\rm scale}\right),\\
    \Rs^\prime &= \Rs \left(s \mathcal{A}_{\rm scale}^{-1/3}\right),
\end{align}
where $\mathcal{A}_{\rm scale}$ is a correction factor defined as
\begin{equation}\label{eq:A_param}
    \mathcal{A}_{\rm scale} \equiv \left[\frac{\Dvir(z)}{\Dvir^\prime(z^\prime)}\right]^\alpha,
\end{equation}
with $\alpha$ being a free parameter.
The parametrisation of $\mathcal{A}_{\rm scale}$ is chosen empirically to minimise the bias in the rescaled halo mass (described in \autoref{sec:NFW_scaling_correction_test}).

We denote by $M_{\rm vir,NFW}$ the virial mass computed by integrating the NFW profile defined by $\rho_0$ and $\Rs$ using \autoref{eq:virial_mass}.
Analogously, $M_{\rm vir,NFW}^\prime$ is computed in the target cosmology from the rescaled NFW parameters $\rho_0^\prime$ and $\Rs^\prime$.
We then rescale the virial masses reported by \texttt{Rockstar} as
\begin{equation}\label{eq:Mvir_scaling}
    \Mvir^\prime = \Mvir \left(\frac{M_{\rm vir,NFW}^\prime}{M_{\rm vir,NFW}}\right).
\end{equation}
Fixed-threshold mass definitions such as $M_{\rm 200c}$ and $M_{\rm 200b}$ are obtained from the same rescaled density profile: since $\mathcal{A}_{\rm scale}$ is calibrated on $\Mvir$, the rescaled NFW parameters are fully determined, and the remaining masses follow from the analogous ratio to \autoref{eq:Mvir_scaling}, evaluated at the radius enclosing the respective fixed overdensity.
Given $\Mvir^\prime$ and $\Rs^\prime$, we could compute the rescaled concentration $c^\prime$.
However, because concentration encodes the formation history of an object and the rescaling as described so far is calibrated only on the halo mass function (power spectrum), this would yield a halo population with a biased mass-concentration relation~\citep{Contreras_2020}.
Thus, following~\citet{Angulo_2021}, we correct the average concentration at fixed halo mass using the mass-concentration relation of~\citet{Ludlow_2016},
\begin{equation}
    c^\prime
    =
    c\left(\frac{c^\prime_{\rm Ludlow}(M_{\rm 200c}^\prime)}{c_{\rm Ludlow}(M_{\rm 200c})}\right),
\end{equation}
where $c_{\rm Ludlow}(M_{\rm 200c})$ denotes concentration as a function of $M_{\rm 200c}$ following the mass-concentration relation of~\citeauthor{Ludlow_2016}.
We then update $\rho_0^\prime$ and $\Rs^\prime$ to be consistent with this $c^\prime$.

For an NFW profile, the maximum circular velocity is $\Vmax \approx 1.64 \Rs \sqrt{G \rho_0}$.
Using $\Rs^\prime$ and $\rho_0^\prime$, we assume $\Vmax$ scales as
\begin{equation}
    \Vmax^\prime = \Vmax \left(\frac{\Rs^\prime \sqrt{\rho_0^\prime}}{\Rs \sqrt{\rho_0}}\right).
\end{equation}
For the velocity dispersion $\Vrms$, a similar closed-form solution for an NFW profile exists.
Under the assumption of spherical symmetry and isotropic orbits, the velocity dispersion of an NFW halo can be approximated to scale as
\begin{equation}
    \Vrms^\prime = \Vrms \left( \frac{M_{\rm 200c}^\prime H^\prime(z^\prime)}{M_{\rm 200c} H(z)} \right)^{1/3}.
\end{equation}
We do not scale the halo spin parameter $\lambda$, whose distribution in \LCDM\ is approximately log-normal with only a weak dependence on mass and redshift~\citep{Bett_2007}, although $\lambda$ correlates with merger history~\citep{Vitvitska_2002,DOnghia_2007}.
We do not consider the rescaling of additional halo properties, as they are not used in the version of \SC\ employed in this paper.

\subsection{Rescaling of merger trees}\label{sec:rescaling_trees}

\begin{figure*}
    \centering
    \includegraphics[width=\textwidth]{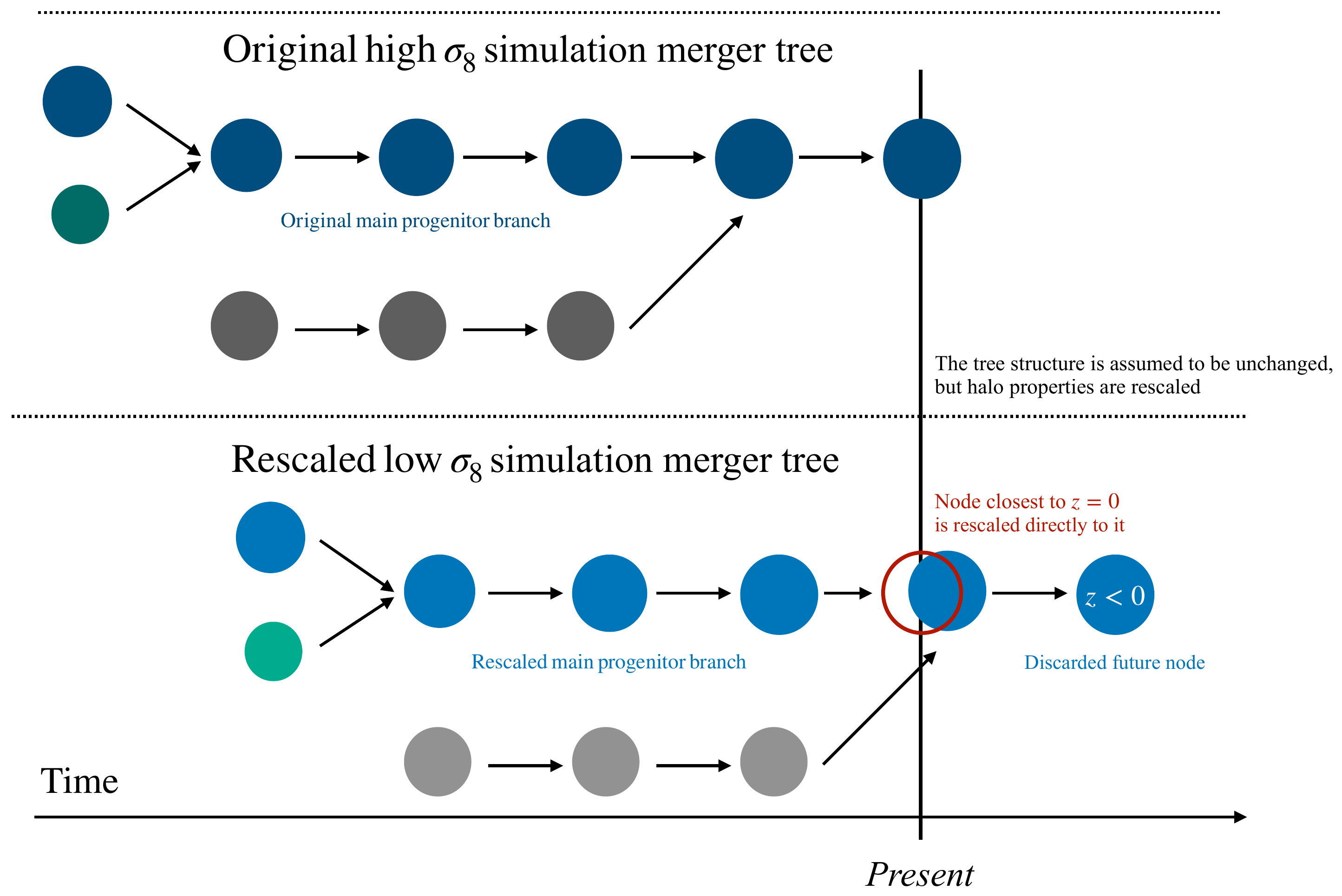}
    \caption{Schematic of the merger tree rescaling from a high $\sigma_8$ simulation (top) to a lower $\sigma_8$ target cosmology (bottom).
    Circles represent halos at discrete snapshots, with arrows indicating the progenitor--descendant linkages.
    The main progenitor branch is highlighted in blue, while secondary branches (grey and green) merge into it at earlier times.
    Rescaling to lower $\sigma_8$ (a cosmology with slower growth of structure) shifts all snapshots forward in time, rescaling halo properties while keeping the merger tree morphology unchanged.
    The node closest to $z = 0$ (red circle) is interpolated using the two bracketing snapshots to obtain halo properties at $z = 0$.
    All other future nodes are retained in the tree but are discarded when processing it with the semi-analytic model (\SC).
    }
    \label{fig:rescaling_schematic}
\end{figure*}

We now extend the halo catalogue rescaling to halo merger trees.
We assume that the merger tree morphology (the linkage between progenitors and descendants) remains unchanged, but that the halo properties (nodes of the tree) are rescaled as described above.
Thus, rescaling a merger tree can be treated as an iterative process over all snapshots.
The $n$\textsuperscript{th} snapshot at redshift $\zo_n$ is mapped to a redshift $\zt_n$, while the rescaling parameter $s$ remains the same for all $n$ (as it is redshift-independent).
Depending on the rescaling direction in the $\Om$--$\sigma_8$ plane, all snapshots are mapped either to the past or to the future.
In \CS, particle snapshots at future times ($z < 0$) are not available.
Therefore, to obtain a snapshot at $z = 0$ after rescaling, the mapping must be forward in time.
Consequently, a given simulation can only be rescaled toward smaller $\Om$ and/or $\sigma_8$, where the growth of structure proceeds more slowly.
For example, a snapshot at $z > 0$ from a simulation with larger $\sigma_8$ resembles the $z = 0$ snapshot of a simulation with smaller $\sigma_8$.
In the \Bacco\ suite, some simulations were run into the future to mitigate this limitation~\citep{Angulo_2021}.

After rescaling a merger tree forward in time, some nodes will lie in the future, resulting in one rescaled snapshot just below and one just above $z = 0$; \autoref{fig:rescaling_schematic} illustrates this procedure.
To enable a comparison at the present time, we linearly interpolate between these two bracketing snapshots to exactly $z = 0$ as follows.
If a halo maps to a unique descendant in the next snapshot, its properties are interpolated as a function of the scale factor.
If multiple halos share the same descendant, only the most massive progenitor is interpolated, while the others have their redshift snapped to $z = 0$ but retain their original properties.
Halos at the $z > 0$ snapshot with no descendants are snapped directly to $z = 0$.
In practice, ${\sim}99.5\%$ of halos are interpolated---either as single progenitors (${\sim}99\%$) or as the most massive progenitor of a shared descendant (${\sim}0.5\%$)---while only ${\sim}0.5\%$ are secondary progenitors whose properties are snapped without interpolation; the impact of these approximations is therefore negligible.
In effect, the snapshot just above $z = 0$ is adjusted to lie exactly at $z = 0$,
while the remainder of the tree---including nodes in the future---is left intact.
The full tree is then passed to \SC, and any output corresponding to $z < 0$ is discarded.
Since \SC\ solves its set of differential equations forward in time, the future tree nodes (which may not be handled correctly) have no impact on the $z = 0$ output.

\subsection{Suites of rescaled simulations}\label{sec:rescaled_suites}

So far, we have outlined how to apply the rescaling algorithm to the merger trees of an $N$-body simulation to map it to a nearby cosmology.
We now turn to cosmological parameter estimation and repurpose the rescaling algorithm as a training-set generator, enabling large suites of rescaled simulations to be produced from only a few $N$-body simulations.
Similar ideas underpin the \texttt{BACCO} framework, which uses rescaled simulations to build emulators of the non-linear matter power spectrum~\citep{Angulo_2021}, including baryonic effects~\citep{Arico_2021} and models with massive neutrinos or dynamical dark energy~\citep{Zennaro_2023}.
Here, however, we aim to construct a bank of rescaled merger trees that are then processed with the \SC\ model to generate diverse galaxy populations across a wide range of cosmological and astrophysical parameters at little additional computational cost.

Our starting point is the suite of \num{900} $N$-body simulations from \CS\footnote{The full \CS\ suite contains \num{1000} simulations, but we exclude 100 with $\Om < 0.15$ due to a minor numerical issue in the \texttt{BACCO} rescaling at high baryon fractions.}, though we also consider smaller subsets to mimic situations where only a few simulations are computationally feasible and to reserve a fraction for testing.
These simulations span a Latin hypercube in the $\Om$--$\sigma_8$ plane, and each has a unique set of initial conditions.
We require the suite of rescaled simulations to follow the same Latin hypercube design to ensure quasi-uniform coverage of parameter space and avoid biases.
Accordingly, we draw the cosmological and astrophysical parameters of each rescaled realisation from a new Latin hypercube realisation following the same design as \CS, so that the parameter values are distinct from those of the original suite.
For the parameter estimation, we use the backward rescaling direction: we fix $\zt = 0$ and solve for $s$ and $\zo$, so that the rescaled simulation always produces a snapshot at $z = 0$.
For each base simulation, this fixes the original snapshot to rescale from.
Anchoring at $\zt = 0$ places the rescaling in the most dark-energy-dominated regime, where it is least accurate; at higher redshifts, where the Universe is matter-dominated, the rescaling becomes more accurate (see \autoref{sec:results_individual_objects}).

For each target cosmology, we assign a base simulation that approximately minimises
the expected rescaling error while avoiding reusing the same base simulations.
A base simulation is admissible if: (i) the scale factor to $\zt = 0$ lies within $[0.7,\,1.0]$, (ii) the length rescaling lies within $[0.8,\,1.25]$, and (iii) the predicted power spectrum error at $k = 1~h\,\mathrm{Mpc}^{-1}$~\citep{Contreras_2020} is within ten times the minimum error across all candidate base simulations for that target.
From admissible candidates, we select randomly with weights inversely proportional to prior usage; if fewer than five candidates exist, we iteratively relax the power spectrum error tolerance.

Thus, this procedure produces a Latin hypercube suite of rescaled simulations spanning new cosmological and astrophysical parameters.
For each target, we identify a base simulation from the original \CS\ suite together with the corresponding rescaling parameters, apply the rescaling algorithm to its merger trees, and then process the rescaled trees with the \SC\ model and the associated astrophysical parameters to generate galaxy populations.
In addition to the \num{900} \CS\ simulations, we also use 15 supplementary $N$-body simulations at fixed $\sigma_8 = 1.0$ spanning $\Om \in [0.15, 0.5]$, which serve as base simulations for rescaling to lower $\sigma_8$ values.


\section{Validation methodology}\label{sec:method}

Here, we describe our procedure for assessing the reliability of the rescaling algorithm.
First, in \autoref{sec:method_summary_statistics}, we describe the summary statistics used to test the population properties.
In \autoref{sec:method_matching}, we describe the matching of halos between simulations with identical initial conditions, evolved under different cosmologies.
Finally, in \autoref{sec:method_parameter_estimation}, we describe the machine learning model used to map galaxy population statistics to cosmological parameters.

\subsection{Summary statistics}\label{sec:method_summary_statistics}

We use three main diagnostics to test whether the galaxy sample derived from the rescaled merger trees is biased: the halo mass function (HMF), the stellar mass function (SMF), and the galaxy real-space two-point correlation function (2PCF;~\citealt{Peebles_1980}).
For benchmarking purposes, we evaluate these on paired simulations with identical initial conditions, suppressing cosmic variance in the comparison between rescaled and directly simulated populations. We further use the SMF and 2PCF as inputs to the cosmological parameter estimation described in \autoref{sec:method_parameter_estimation}.
The HMF is the number density of halos as a function of mass.
We compute it by binning halos by virial mass above $10^{11}~\Msunh$ in bins of $0.2~\mathrm{dex}$.
Similarly, the SMF is the number density of galaxies as a function of stellar mass, which is regulated both by structure formation and galaxy formation physics.
We compute it above a stellar mass of $10^9~\Msunh$ in bins of $0.2~\mathrm{dex}$.

When computing the 2PCF, we downsample galaxies to achieve a number density of $n = 0.002~h^3\,\mathrm{Mpc}^{-3}$, corresponding to \num{2000} objects in our $100~\Mpch$ boxes.
The 2PCF $\xi(r)$ quantifies the probability $\dd P$ of finding a pair of points separated by a distance $r$,
\begin{equation}
    \dd P = n \left[ 1 + \xi(r) \right] \dd V,
\end{equation}
where $n$ is the mean number density and $\dd V$ is the differential volume element.
We compute $\xi(r)$ using \texttt{Corrfunc}~\citep{Corrfunc} with the Landy--Szalay estimator~\citep{Landy_1993},
\begin{equation}
    \xi(r) = \frac{1}{\mathrm{RR}} \left[ \mathrm{DD} \left( \frac{n_R}{n_D} \right)^2 - 2\mathrm{DR} \left( \frac{n_R}{n_D} \right) + \mathrm{RR} \right],
\end{equation}
where $\mathrm{DD}$, $\mathrm{DR}$, and $\mathrm{RR}$ are the counts of data-data, data-random, and random-random pairs in bins of separation $r$, and $n_D$, $n_R$ are the number densities of data and random samples, respectively.
The random catalogue contains $100$ times as many points as the data sample.
We compute $\xi(r)$ in 20 logarithmically spaced bins spanning from 1 to $30~\Mpch$.

\subsection{Matching of halos between simulations}\label{sec:method_matching}

The paired simulations additionally enable a one-to-one matching of halos between the original and target cosmologies~\citep{Butsky_2016,Desmond_2017,Mitchell_2018,Cataldi_2021,Gabrielpillai_2022,Stiskalek_2024}, allowing us to assess the rescaling accuracy at the level of individual objects.
We perform a bijective halo matching at the target redshift $\zt$ between the rescaled halos and the halos in the paired $N$-body simulation run with the target cosmology.
For each rescaled halo $A$, we identify the halo $B$ in the target simulation that shares the largest number of particles with $A$.
This matching requires a consistent particle ID assignment across simulations, with IDs indicating the particle's Lagrangian position.
We then verify the reverse match, checking that halo $A$ also contains the largest number of particles from halo $B$.
Formally, halos $A$ and $B$ are matched if and only if
\begin{align}
    B &= \arg\max_{B'} N_{\rm shared}(A, B') \quad \text{and} \nonumber\\
    A &= \arg\max_{A'} N_{\rm shared}(A', B),
\end{align}
where $N_{\rm shared}(A, B)$ denotes the number of particles shared between halos $A$ and $B$.
This procedure is repeated for all halos in the original simulation.
Such matched halos enable the study of how individual halo properties respond to changes in cosmology.

We now define the bias in individual halo (or galaxy) properties.
Let $x^\prime_i$ denote the rescaled property of the $i$\textsuperscript{th} halo (or galaxy) in cosmology $\Lt$, obtained by applying the rescaling algorithm to the original halo.
We compare this to the matched halo property $\tilde{x}^\prime_i$, obtained by directly simulating the matched halo in cosmology $\Lt$ with the same initial conditions.
We define the relative bias as
\begin{equation}\label{eq:relative_bias}
    \mathrm{Relative~bias}
    \equiv
    \frac{x^\prime_i - \tilde{x}^\prime_i}{\tilde{x}^\prime_i}.
\end{equation}

\subsection{Parameter estimation}\label{sec:method_parameter_estimation}

We use the rescaling procedure to generate training sets across multiple cosmologies from a limited number of $N$-body simulations.
These rescaled samples are then used to train cosmological parameter estimators based on the SMF and 2PCF, and the trained models are tested against samples obtained directly from $N$-body simulations to assess potential biases.
We use a multilayer perceptron (MLP) to jointly predict $\Om$ and $\sigma_8$.
We set aside 150 simulations as a test set. These are never used for training or rescaling. The training set comprises either the remaining $N$-body simulations or rescaled simulations generated from a subset of them; in both cases, 20\% is reserved as a validation set for early stopping.
As described in \autoref{sec:rescaled_suites}, the astrophysical parameters of \SC\ also vary across the training set, sampled on a Latin hypercube, so the MLP must implicitly marginalise over them.
The MLP is implemented using \texttt{flax}\footnote{\url{https://flax.readthedocs.io/en/stable/}} and consists of three fully connected layers with 64, 32, and 16 hidden units, respectively, each followed by a Gaussian error linear unit (GELU) activation function~\citep{Hendrycks_2016}.
The final output layer has two units corresponding to $\Om$ and $\sigma_8$.
Standard scaling is applied to the input features.
The network is trained with a batch size of 32, a dropout rate of 0.05~\citep{Gal_2015}, and a cosine learning rate schedule~\citep{Loshchilov_2016}.
The loss function consists of the mean squared error augmented with a bias penalty term,
\begin{align}
    \mathcal{L} &= \frac{1}{N} \sum_{i=1}^{N} \sum_{j} (\hat{\theta}_{i,j} - \theta_{{\rm true},i,j})^2 \nonumber\\
    &\quad+ \lambda_{\rm bias} \sum_{j} \left(\frac{1}{N} \sum_{i=1}^{N} (\hat{\theta}_{i,j} - \theta_{{\rm true},i,j})\right)^2,
\end{align}
where $i$ indexes the samples from 1 to $N$, $j$ indexes the two parameters ($\Om$ and $\sigma_8$), and we set $\lambda_{\rm bias} = 0.01$.
This penalty suppresses systematic bias in the predicted parameters.
Training is terminated when the validation loss does not improve for more than 150 epochs, and we select the neural network weights that achieved the lowest validation loss during training.
We do not perform hyperparameter optimisation but verify that the presented results are robust to hyperparameter choices.
The MLP provides point estimates of the target parameters $\hat{\bm{\theta}}$.
To assess the accuracy of the machine learning model, we compute the root mean square error (RMSE) separately for $\Om$ and $\sigma_8$.
Given a prediction for the $i$\textsuperscript{th} sample, denoted $\hat{\theta}_i$, and the true value $\theta_{\mathrm{true},\,i}$, the RMSE is defined as a sample average
\begin{equation}
    \mathrm{RMSE} \equiv \sqrt{\frac{1}{N} \sum_{i=1}^{N} (\hat{\theta}_i - \theta_{\mathrm{true},\,i})^2}.
\end{equation}
To improve robustness, we use an ensemble of 50 neural networks, each trained independently with different random weight initialisations.
When making predictions on new data, we compute a weighted average over all networks, where the weight for each network is given by the inverse of its validation loss.
This procedure upweights better-performing models and provides more stable predictions.


\section{Rescaling validation}\label{sec:results}

In \autoref{sec:NFW_scaling_correction_test}, we calibrate the NFW rescaling-correction parameter.
In \autoref{sec:results_individual_objects}, we test how well the rescaling algorithm reproduces the properties of individual halos and of galaxies generated from the rescaled merger trees.
In \autoref{sec:results_galaxy_population}, we compare the rescaled HMF, SMF and galaxy real-space 2PCF.

\vspace{1em}

We refer to halos rescaled from an original cosmology to the target cosmology as ``rescaled'' halos, and to halos simulated directly in the target cosmology as ``simulated'' halos.
When testing the rescaling algorithm, we use paired simulations with identical initial conditions but different cosmological parameters, enabling a direct comparison of rescaled and simulated halo properties.
Similarly, we refer to \SC\ galaxies generated based on the rescaled dark matter merger trees as ``rescaled galaxies,'' even though the rescaling algorithm is applied to the underlying dark matter halos.
We focus on rescaling in $\Om$ and $\sigma_8$, since these are the cosmological parameters sampled in \CS, and leave varying the other parameters for future work.\footnote{We note that in the \Bacco\ suite the rescaling is applied to $\Omega_{\rm b}$, $n_{\rm s}$, $h$, $M_{\nu}$, $w_0$ and $w_a$ as well~\citep{Angulo_2021}.}

The results presented in \autoref{sec:results_individual_objects} and \autoref{sec:results_galaxy_population} are for two steps in cosmological parameters: (i) $\sigma_8 = 0.8 \rightarrow 0.85$ while $\Om = 0.3$, and (ii) $\Om = 0.25 \rightarrow 0.3$ while $\sigma_8 = 0.8$, though we also tested other steps in cosmology.
Here we use the forward rescaling direction: we fix $\zo = 0$ and solve for $s$ and $\zt$. For these steps, the original cosmology is mapped from $\zo = 0$ to $\zt = 0.12,\,0.30$, respectively (and $s = 1.0,\,0.79$).
When presenting comparisons, the target simulation is always taken to be at redshift $\zt$.
We present results for arbitrary steps in cosmological parameters later in \autoref{sec:results_suite_validation}, in the context of the SMF and the galaxy 2PCF.

\subsection{Calibration of the NFW rescaling correction}\label{sec:NFW_scaling_correction_test}

The NFW rescaling correction introduced in \autoref{sec:rescaling_halo_catalogues} depends on a free parameter $\alpha$ in \autoref{eq:A_param}.
To calibrate it, we use paired $N$-body simulations and compare matched halos identified through the bijective particle-sharing procedure described in \autoref{sec:method_matching}.
For each pair, we apply the rescaling algorithm and measure the mean absolute relative bias in the rescaled halo virial mass $\Mvir$ relative to the simulated virial mass in the target cosmology (\autoref{eq:relative_bias}).
We scan over 20 values of $\alpha$ uniformly distributed between 0 and 1.5 and denote by $\hat{\alpha}$ the value that minimises the mean absolute bias.
\autoref{tab:alpha_calibration} presents the results for three representative cosmological steps and three mass definitions: $\Mvir$, $M_{\rm 200c}$, and $M_{\rm 200b}$.
The optimal $\alpha$ ranges from approximately $0.8$ to $1.2$, with mean absolute biases below $0.25\%$ for all mass definitions.
Without the correction ($\alpha = 0$), biases reach up to ${\sim}11\%$, demonstrating the importance of this term.
Based on these results, we adopt $\alpha = 1$ for all subsequent analyses, which provides good performance across the range of cosmological steps and mass definitions considered, yielding only marginally larger biases than $\hat{\alpha}$.
In principle, $\alpha$ could also depend on halo mass and redshift.
We have verified that $\alpha \approx 1$ remains a good approximation across the halo mass range resolved in \CS\ and at the redshifts probed here, but a mass- or redshift-dependent $\alpha$ may be required for substantially different regimes.
More broadly, the accuracy of the rescaling is governed by differences in the growth rate and power spectrum slope between the original and target cosmologies~\citep{Ondaro-Mallea_2022}, which suggests that extending the method to parameters that modify these quantities---such as $n_{\rm s}$ or dynamical dark energy---would require re-calibrating $\alpha$ as a function of these quantities.
This dependence also explains the trend in \autoref{fig:bias_in_mvir_stepsize}: larger steps in $\sigma_8$ produce larger differences in growth rate, increasing the rescaling error.

\begin{table*}
\centering
\begin{tabular*}{\textwidth}{@{\extracolsep{\fill}}ccc ccc @{\hspace{1.5em}} ccc @{\hspace{1.5em}} ccc@{}}
    \hline
    \multirow{2}{*}{$\Om \!\to\! \Om^\prime$} & \multirow{2}{*}{$\sigma_8 \!\to\! \sigma_8^\prime$} & \multirow{2}{*}{$\zt$} & \multicolumn{3}{c}{$\hat{\alpha}$} & \multicolumn{3}{c}{Bias ($\alpha = 0$)} & \multicolumn{3}{c}{Bias ($\hat{\alpha}$)} \\
    \cline{4-6} \cline{7-9} \cline{10-12}
    & & & $\Mvir$ & $M_{\rm 200c}$ & $M_{\rm 200b}$ & $\Mvir$ & $M_{\rm 200c}$ & $M_{\rm 200b}$ & $\Mvir$ & $M_{\rm 200c}$ & $M_{\rm 200b}$ \\
    \hline
    $0.25 \!\to\! 0.30$ & $0.80 \!\to\! 0.85$ & 0.42 & 1.03 & 1.11 & 0.79 & $9.71\%$ & $11.18\%$ & $7.01\%$ & $0.22\%$ & $0.01\%$ & $0.13\%$ \\
    $0.30 \!\to\! 0.30$ & $0.80 \!\to\! 0.90$ & 0.23 & 1.11 & 1.11 & 0.79 & $4.81\%$ & $5.68\%$ & $3.35\%$ & $0.10\%$ & $0.19\%$ & $0.09\%$ \\
    $0.30 \!\to\! 0.30$ & $0.80 \!\to\! 0.85$ & 0.12 & 1.11 & 1.18 & 0.87 & $2.80\%$ & $3.38\%$ & $1.89\%$ & $0.08\%$ & $0.08\%$ & $0.07\%$ \\
    \hline
\end{tabular*}
\caption{
    Mean absolute relative bias in rescaled halo mass for three cosmological steps and three mass definitions.
    For each step, we report the optimal $\hat{\alpha}$, the bias without the NFW correction ($\alpha = 0$), and the bias at $\hat{\alpha}$.
    The correction suppresses biases of up to ${\sim}11\%$ to below $0.25\%$ in all cases.
    The optimal $\hat{\alpha}$ depends on both the cosmological step and the mass definition: it is approximately $1.0$--$1.1$ for $\Mvir$ and $M_{\rm 200c}$, but lower ($0.8$--$0.9$) for $M_{\rm 200b}$.
    We adopt $\alpha = 1$ for all mass definitions, which yields only marginally larger biases than $\hat{\alpha}$. In the remainder of this work, we use $\Mvir$.
}
\label{tab:alpha_calibration}
\end{table*}

\subsection{Validation on individual objects}\label{sec:results_individual_objects}

\begin{figure*}
    \centering
    \includegraphics[width=\textwidth]{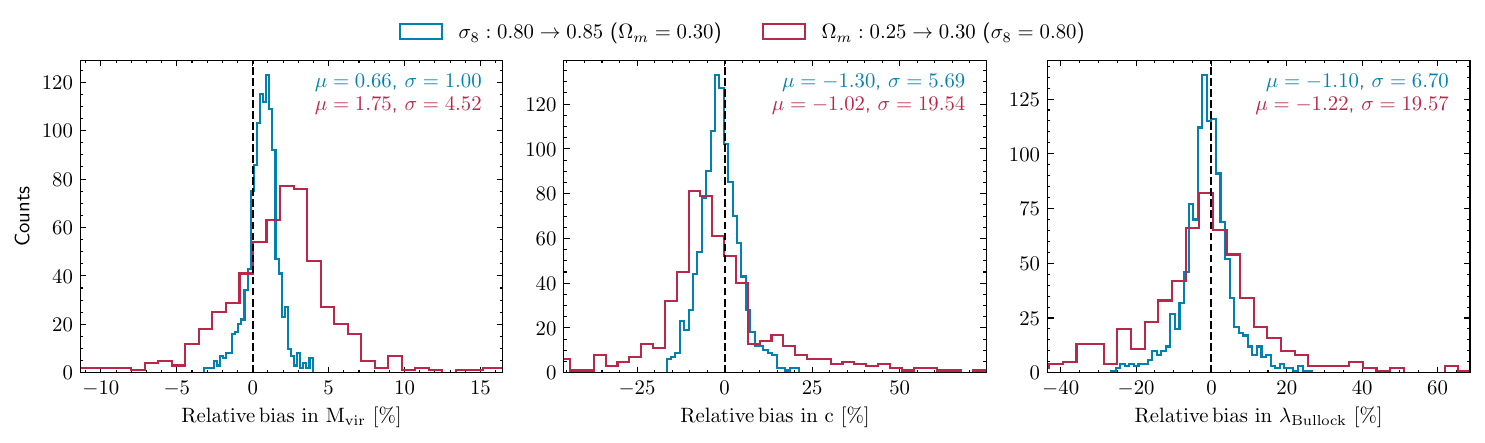}
    \caption{Relative bias in halo virial mass $\Mvir$, concentration $c$, and spin $\spin$ between individual rescaled and simulated halos for two cosmological steps: (i) $\sigma_8 = 0.8 \rightarrow 0.85$ at fixed $\Om = 0.3$, and (ii) $\Om = 0.25 \rightarrow 0.3$ at fixed $\sigma_8 = 0.8$, at the target redshifts $z = 0.12$ and $z = 0.3$, respectively.
    In the top right, we indicate the mean $\mu$ and standard deviation $\sigma$ of the distributions.
    The distributions are consistent with zero systematic bias in all three properties.}
    \label{fig:bias_in_halo_properties}
\end{figure*}

We now further validate the rescaling algorithm by comparing individual rescaled halos to their matched counterparts that were directly simulated in the target cosmology.
In \autoref{fig:bias_in_halo_properties}, we present the relative bias in $\Mvir$, $c$, and $\spin$ between matched rescaled and simulated halos.
We find that the rescaling algorithm accurately reproduces all three halo properties, with no significant systematic bias in any of them.
The same holds for additional halo properties including $\Vmax$, $\Vrms$, and alternative mass definitions ($M_{\rm 200b}$ and $M_{\rm 200c}$), which show similarly good agreement.
For the shift in $\sigma_8$, $\Mvir$ is recovered to within $1\%$ on average without systematic bias, while for the shift in $\Om$ it is recovered to within $5\%$ with a possible average systematic bias below $2\%$.
The per-halo scatter of the $\sigma_8$ step is comparable to the run-to-run stochasticity of the simulations themselves~\citep{Trenti_2010,Genel_2019}, which we measure by re-running an identical simulation on a different number of compute nodes: the changed order of floating-point operations scatters matched $\Mvir$ without systematic bias, from both the $N$-body gravity solver and the \texttt{Rockstar} halo finder.
Without the NFW corrections introduced in \autoref{sec:rescaling_halo_catalogues} and tuned in \autoref{sec:NFW_scaling_correction_test}, the average systematic bias for the two steps would be $2\%$ and $10\%$, respectively.
Both $c$ and $\spin$ are rescaled without systematic bias, with average spreads of about $5\%$ and $20\%$ for the steps in $\sigma_8$ and $\Om$, respectively.
The absence of systematic bias in $c$ is partly by construction, as we enforce the correct mass-concentration relation; the corresponding behaviour for $\spin$ reflects its weak dependence on mass and redshift in \LCDM~\citep{Bett_2007}.

\begin{figure*}
    \centering
    \includegraphics[width=\textwidth]{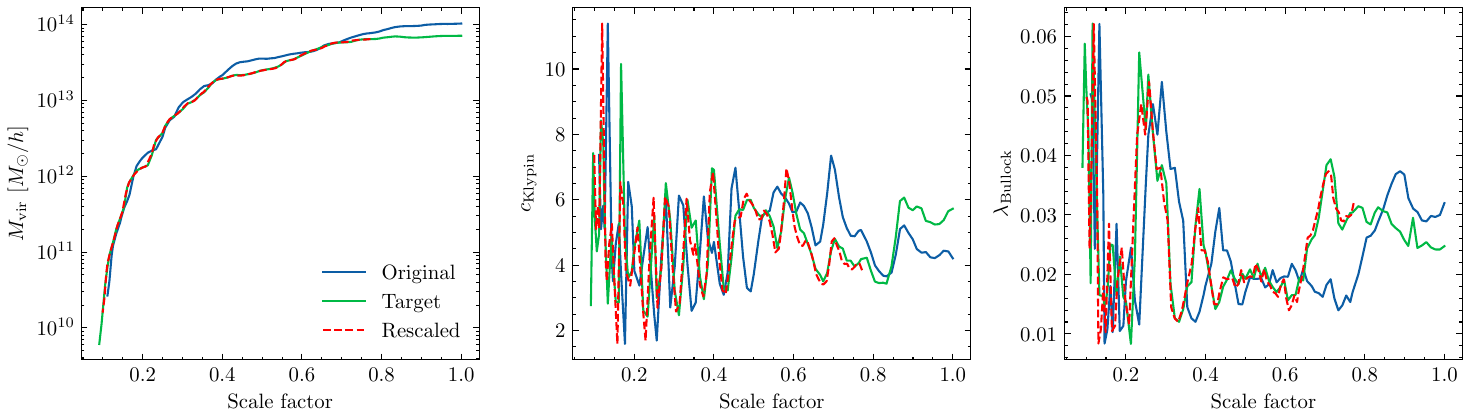}
    \caption{Example evolution of virial mass, concentration, and spin along the main progenitor branch of a single matched halo for a cosmological step $\Om = 0.25 \rightarrow 0.3$ at fixed $\sigma_8 = 0.8$.
    Solid blue lines show the evolution in the original cosmology, solid green lines show the same halo evolved in the target cosmology, and dotted red lines show the result of applying the rescaling procedure to the original halo.
    The rescaled halo closely reproduces the true evolution in the target cosmology across all redshifts for all three properties, confirming the accuracy of the method beyond the single redshift shown in \autoref{fig:bias_in_halo_properties}.
    }
    \label{fig:mpb_rescaled_example}
\end{figure*}

The results above are for the two cosmological steps described earlier.
In these, the $z = 0$ snapshot in the original cosmology maps to $\zt = 0.12$ and $\zt = 0.3$, respectively, in the target cosmology.
At higher redshifts, where the Universe is matter-dominated, the rescaling becomes more accurate than at $z = 0$: as typical \LCDM\ cosmologies then approach the Einstein--de Sitter regime, their growth histories converge, so the cosmology-dependence of the non-linear evolution that the rescaling cannot capture is smaller~\citep{Contreras_2020}.
In \autoref{fig:mpb_rescaled_example}, we show the evolution of $\Mvir$, $c$, and $\spin$ along the main progenitor branch of a single halo for the step in $\Om$ at fixed $\sigma_8$.
We compare the evolution of a matched halo in the original and target cosmologies to that of a rescaled halo, demonstrating that the rescaled evolution closely reproduces the behaviour in the target cosmology.

\begin{figure}
    \centering
    \includegraphics[width=\columnwidth]{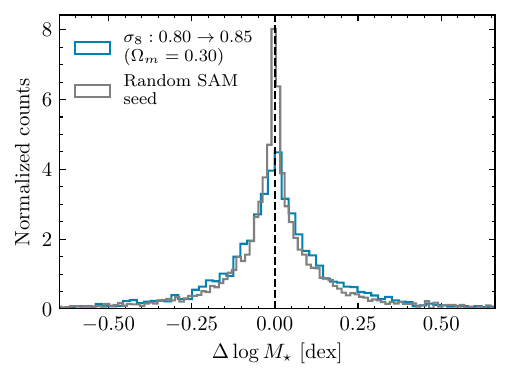}
    \caption{Logarithmic difference in $M_\star$ predicted by \SC\ from rescaled versus paired simulation merger trees for a step $\sigma_8 = 0.8 \rightarrow 0.85$ at $\Om = 0.3$ (shown in blue).
    The grey histogram shows the distribution of $M_\star$ differences when varying the random seed
    in \SC\ for the same halos (i.e., without any change in cosmology).
    Only halos with $\Mvir > 10^{12}~\Msunh$ are shown.
    }
    \label{fig:bias_in_mstar}
\end{figure}

Next, in \autoref{fig:bias_in_mstar}, we show the logarithmic difference in $M_\star$ predicted by \SC\ from rescaled versus paired simulation merger trees for the steps in $\Om$ and $\sigma_8$.
To obtain this comparison, we take two paired simulations at different cosmologies, scale the merger trees from the original to the target cosmology, apply the fiducial \SC\ to the rescaled trees, and compare the resulting $M_\star$ with that obtained by applying \SC\ to the paired simulation directly in the target cosmology, varying the random seed that determines the initial satellite orbits.
We find that the rescaled and random seed distributions are comparable, as shown in the figure, with standard deviations of $0.22$ and $0.20~\mathrm{dex}$, respectively, and both exhibit no systematic bias.
Unlike for halo properties, a dominant source of dispersion in stellar mass arises from the treatment of satellite orbits in \SC, as described in Section~\ref{sec:data}.
As a result, each dark matter halo has a distribution of plausible galaxy properties.
We show the difference only for halos with mass $\Mvir > 10^{12}~\Msunh$, thus selecting objects with well-resolved formation histories containing on the order of $10^4$ particles at $z = 0$.

\begin{figure}
    \centering
    \includegraphics[width=\columnwidth]{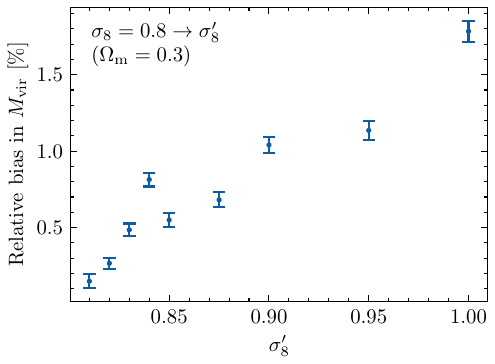}
    \caption{
        Bias in rescaled $\Mvir$ as a function of the target $\sigma_8$ for steps from $\sigma_8 = 0.8$ at fixed $\Om = 0.3$, computed for halos with $\Mvir > 10^{12}~\Msunh$.
        The bias increases with step size but remains modest, reaching only $1.8\%$ for the largest step to $\sigma_8 = 1.0$.
        Error bars represent $1\sigma$ bootstrap uncertainties.
    }
    \label{fig:bias_in_mvir_stepsize}
\end{figure}

The tests so far cover a few cosmological steps and show that the rescaling introduces no appreciable bias and modest scatter between the true and rescaled halo or galaxy properties.
The rescaling error nonetheless grows with the step size in $\Om$ and $\sigma_8$.
An illustration of the expected rescaling error in the matter power spectrum when directly rescaling the particle snapshots is provided in Fig.~3 of~\citet{Contreras_2020}.
Therefore, in \autoref{fig:bias_in_mvir_stepsize}, we show the bias in the rescaled $\Mvir$ as a function of steps in $\sigma_8$ from $\sigma_8 = 0.8$ to target values $\sigma_8^\prime$ up to 1.0, computed for halos with $\Mvir > 10^{12}~\Msunh$ matched at the target redshift to which the $z = 0$ snapshot of the $\sigma_8 = 0.8$ simulation is mapped.
As expected, the bias increases with step size, but the growth remains modest: even for the shift from $\sigma_8 = 0.8$ to $\sigma_8 = 1.0$, the bias is only $1.8\%$.

\subsection{Galaxy population summary statistics}\label{sec:results_galaxy_population}

We now examine summary statistics of the rescaled halo and galaxy populations: specifically, the HMF, SMF, and the real-space galaxy 2PCF, considering either a step in $\sigma_8$ at fixed $\Om$ or vice versa.
We compare two $N$-body simulations evolved with identical initial phases but under different cosmologies, thereby eliminating cosmic variance.
Since the rescaling also alters the simulation box size, we run the target simulations with a box size of $s L$, where $s$ is the rescaling factor and $L$ is the original box size.

\begin{figure*}
    \centering
    \includegraphics[width=\textwidth]{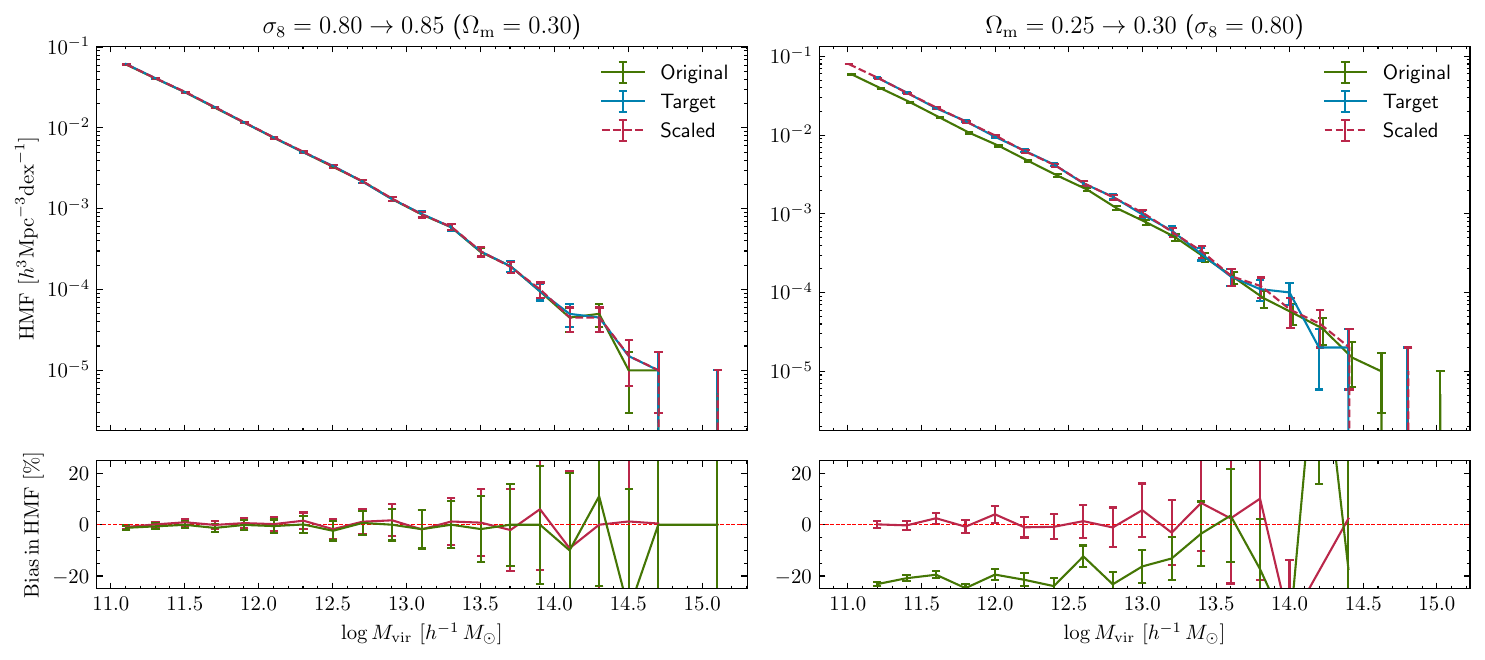}
    \caption{Comparison of the HMFs for steps in $\sigma_8$ (left; $\zt = 0.12$) and $\Om$ (right; $\zt = 0.3$).
    Top panels show the original cosmology at $z = 0$ (green), the target cosmology at $z = \zt$ (blue), and the rescaled result at $z = \zt$ (red); bottom panels show ratios to the target.
    For a step in $\sigma_8$ at fixed $\Om$, the HMFs already agree closely without rescaling, reflecting the near-equivalence to a time shift; rescaling mildly improves agreement at the high-mass end.
    For a step in $\Om$, the unscaled HMFs differ substantially, but the rescaled HMF reproduces the target with no systematic bias.
    Error bars show $1\sigma$ Poisson uncertainties.
    }
    \label{fig:HMF_comparison}
\end{figure*}

In \autoref{fig:HMF_comparison}, we compare the HMFs.
We demonstrate that there is no systematic bias for either step, with random scatter of $3\%$ and $10\%$ for the steps in $\sigma_8$ and $\Om$, respectively; this scatter is driven primarily by the highest-mass bins and remains consistent with Poisson uncertainties.
Error bars shown are $1\sigma$ Poisson uncertainties; for the ratio plot, only the Poisson error from the target HMF is propagated.
The plot shows the ``original'' HMF at cosmology $\Lo$ and evaluated at $z = 0$; the ``target'' HMF, simulated directly at cosmology $\Lt$ and shown at redshift $\zt$; and the ``rescaled'' HMF obtained by applying the rescaling to the dark matter halos at $z = \zt$.
As expected, for the step in $\sigma_8$, once redshifts are matched, there is only minimal difference between the original and target HMFs, except possibly at the high-mass end.
By contrast, for the step in $\Om$ the difference is substantial, primarily because halo masses must also be rescaled by the ratio of matter densities.

\begin{figure}
    \centering
    \includegraphics[width=\columnwidth]{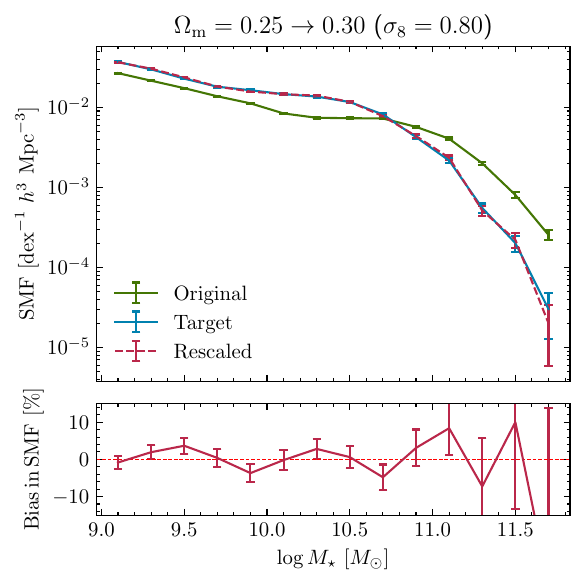}
    \caption{
        Comparison of the SMFs for a step $\Om = 0.25 \rightarrow 0.3$ at $\sigma_8 = 0.8$.
        The top panel shows the original cosmology at $z = 0$ (green), the target cosmology at $z = 0.3$ (blue), and the rescaled result at $z = 0.3$ (red); the bottom panel shows the ratio of rescaled to target.
        The rescaled SMF, derived from rescaled merger trees processed with \SC, closely matches the target.
        Error bars indicate $1\sigma$ Poisson uncertainties.
        }
    \label{fig:SMF_comparison}
\end{figure}

In \autoref{fig:SMF_comparison}, we compare the rescaled SMFs obtained by running the rescaled dark matter merger trees through the \SC\ model.
We show the step in $\Om = 0.25 \rightarrow 0.3$ at fixed $\sigma_8 = 0.8$ (which has a larger difference in the HMF; \autoref{fig:HMF_comparison}). Results for the $\sigma_8$ step are similar.
For reference, we also show the original SMF at $z = 0$.
The SMF derived from the rescaled merger trees agrees well with that from the paired simulation in the target cosmology, showing no systematic bias and a scatter consistent with Poisson noise.
We also verify that other population statistics remain unbiased: the star-formation rate density, gas mass function, and black hole mass function are all similarly unbiased; for brevity, we show only the SMF.
We further examine the distribution of stellar mass at fixed halo mass and find that the rescaling yields distributions consistent with those from simulations run directly in the target cosmology.

\begin{figure}
    \centering
    \includegraphics[width=\columnwidth]{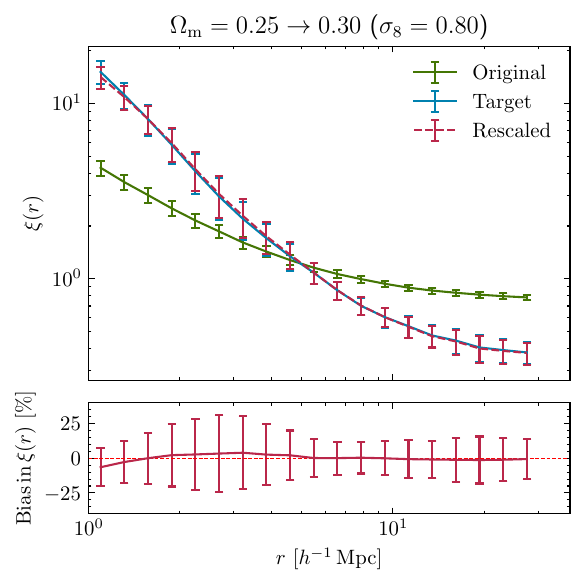}
    \caption{
        Comparison of the galaxy real-space 2PCF for a step $\Om = 0.25 \rightarrow 0.3$ at $\sigma_8 = 0.8$ for galaxies with $M_\star > 10^9~M_\odot$.
        The top panel shows the original cosmology at $z = 0$ (green), the target cosmology at $z = 0.3$ (blue), and the rescaled result at $z = 0.3$ (red); the bottom panel shows the ratio of rescaled to target.
        The rescaled 2PCF closely matches the target, with deviations of order $5\%$ on the smallest scales and ${\sim}1\%$ on larger scales, both subdominant to the statistical uncertainties.
        Error bars indicate $1\sigma$ jackknife uncertainties.
        }
    \label{fig:2PCF_comparison}
\end{figure}

Lastly, in \autoref{fig:2PCF_comparison}, we compare the galaxy real-space 2PCF for a step $\Om = 0.25 \rightarrow 0.3$ at fixed $\sigma_8$ for galaxies with $M_{\star} > 10^9~M_\odot$.
For reference, we also show the original 2PCF at $z = 0$.
The rescaled 2PCF closely matches that from the paired simulation in the target cosmology.
For this particular step, we find deviations of order $5\%$ on the smallest scales ($r \sim 1~\Mpch$) and of order $1\%$ on larger scales; both are substantially smaller than typical jackknife uncertainties on the 2PCF.
The modest deviations at small radii likely combine limitations of the rescaling algorithm in the strongly non-linear one-halo regime with the stochastic orbit prescription of the \SC\ orphan model, to which one-halo clustering is particularly sensitive.
Both contributions remain subdominant to the statistical uncertainties.
Selecting galaxies by a fixed abundance threshold rather than a fixed $M_\star$ threshold yields near-identical results.

\section{Rescaled parameter estimation}\label{sec:parameter_estimation}

Having validated the rescaling algorithm on individual objects and population statistics, we now apply it to cosmological parameter estimation.
In \autoref{sec:results_suite_validation}, we assess whether a suite of rescaled simulations can replace direct $N$-body simulations for training neural networks to infer $\Om$ and $\sigma_8$.
In \autoref{sec:results_smaller_LH_suite}, we explore how many base $N$-body simulations are needed to achieve a given level of accuracy.
In \autoref{sec:results_mlp_convergence}, we examine the convergence of the parameter estimation as the training set size increases.

\subsection{Validation of a suite of rescaled simulations}\label{sec:results_suite_validation}
We now assess whether the rescaling algorithm can generate training sets for cosmological parameter estimation when applied to a limited number of $N$-body simulations, and evaluate the precision and accuracy of the resulting parameter estimates.
We consider a simple set-up in which we generate \num{1000} rescaled simulations from a pool of \num{750} base $N$-body simulations, leaving \num{150} simulations aside for testing.
Because of the way the rescaled suite is constructed, a given base simulation may be used multiple times, particularly since the absence of future snapshots in \CS\ restricts us to rescaling towards cosmologies in which structure formation proceeds more slowly; in practice, \num{539} of the \num{750} available base simulations are used.
The \num{1000} rescaled simulations form the training pool, from which 20\% is held out as a validation set for early stopping, following the procedure in \autoref{sec:method_parameter_estimation}.
We train the MLP ensemble on the rescaled SMFs and 2PCFs and test it on the \num{150} held-out $N$-body simulations not used in the rescaling procedure.
For comparison, we train the same MLP ensemble on \num{750} \CS\ simulations and evaluate it on the same held-out set of \num{150} simulations.

The model trained on rescaled simulations achieves comparable performance to one trained exclusively on $N$-body simulations.
\autoref{tab:parameter_estimation} summarises the RMSE and bias for both models across the two observables and cosmological parameters.
The rescaled model yields RMSE values that closely match those of the $N$-body-only model, with differences typically within the uncertainties.
Moreover, in all cases, the bias remains consistent with zero and is substantially smaller than the RMSE.
We also verify this by visual inspection of the true versus predicted scatter plots.

\begin{table}
    \centering
    \begin{tabular}{llcc}
    \hline
    Observable & Parameter & Model & RMSE \\
    \hline
    \multirow{4}{*}{SMF} & \multirow{2}{*}{$\Omega_{\rm m}$} & $N$-body only & $0.012 \pm 0.001$ \\
    & & Rescaled & $0.013 \pm 0.002$ \\
    \cline{2-4}
    & \multirow{2}{*}{$\sigma_8$} & $N$-body only & $0.048 \pm 0.004$ \\
    & & Rescaled & $0.052 \pm 0.004$ \\
    \hline
    \multirow{4}{*}{2PCF} & \multirow{2}{*}{$\Omega_{\rm m}$} & $N$-body only & $0.048 \pm 0.003$ \\
    & & Rescaled & $0.048 \pm 0.003$ \\
    \cline{2-4}
    & \multirow{2}{*}{$\sigma_8$} & $N$-body only & $0.068 \pm 0.003$ \\
    & & Rescaled & $0.067 \pm 0.004$ \\
    \hline
    \end{tabular}
    \caption{Results of cosmological parameter estimation for models trained only on $N$-body simulations versus only on rescaled simulations, evaluated on \num{150} held-out SAM galaxy populations derived from direct $N$-body simulations. The $N$-body-only model was trained on \num{750} simulations; the rescaled model was trained on \num{1000} rescaled simulations generated from \num{539} unique base simulations (out of \num{750} available). RMSE values are quoted as mean $\pm$ standard deviation over bootstrap realisations.}
    \label{tab:parameter_estimation}
\end{table}

\subsection{How many simulations are needed?}\label{sec:results_smaller_LH_suite}

Having shown that \num{1000} rescaled simulations from \num{539} base simulations match the performance of \num{750} $N$-body simulations, we now examine how the goodness of fit depends on the number of available base simulations, mimicking scenarios with limited computational budgets.

We fix the initial base set to the 15 simulations at $\sigma_8 = 1$ that span $\Om \in [0.15, 0.5]$ uniformly, and augment it with a varying number of additional \CS\ simulations.
We consider adding 1, 10, 100, and 250 further simulations, randomly sampled from the \CS\ suite, which follows a Latin hypercube sampling in $\Om$ and $\sigma_8$.
When generating a suite of \num{1000} rescaled simulations from each pool (i.e., the added simulations plus the 15 at $\sigma_8 = 1$), the rescaling procedure draws from 16, 25, 115, and 265 unique base simulations, respectively.

Before examining the parameter estimation performance, we validate that the rescaled samples are statistically equivalent to samples drawn directly from $N$-body simulations using the \texttt{PQMass} likelihood-free comparison method~\citep{Lemos_2024}.
\texttt{PQMass} operates on multi-dimensional sample spaces by dividing them into non-overlapping regions and applying $\chi^2$ tests to the number of samples falling within each region, yielding a $p$-value that measures the probability that two sets of samples are drawn from the same multinomial distribution.
We divide the space into 50 regions.

\begin{figure*}
    \centering
    \includegraphics[width=\textwidth]{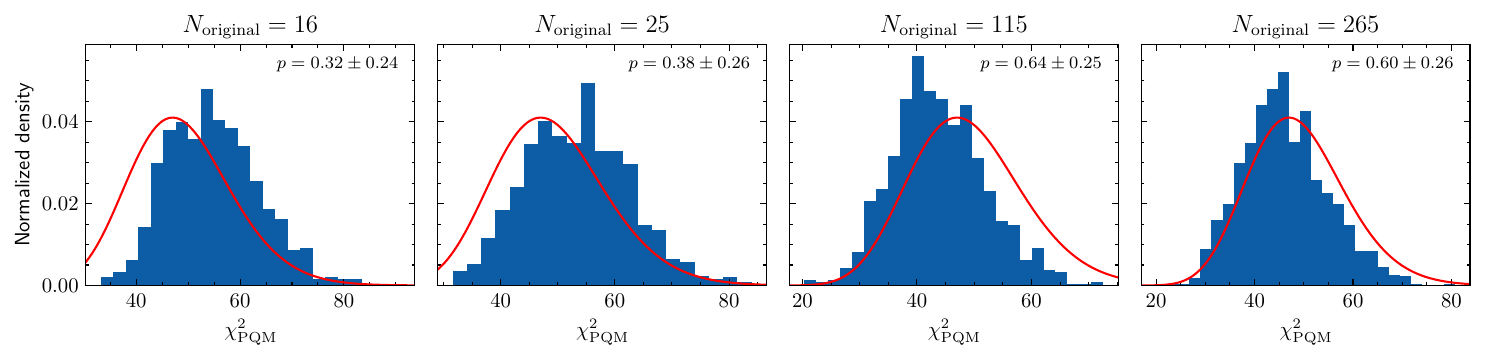}
    \caption{
        \texttt{PQMass} validation of rescaled SMFs to test whether they are statistically consistent with SMFs constructed directly on top of $N$-body simulations in \CS.
        Each panel shows the distribution of $\chi^2_{\rm PQM}$ values for $N_{\rm rescaled}=1000$ realisations generated from $N_{\rm sims}=16,\,25,\,115$, and $265$ base simulations, compared to the expected $\chi^2$ distribution with 49 degrees of freedom (red), corresponding to a partitioning of the parameter space into 50 regions.
        The null hypothesis is that the rescaled and direct SMFs are drawn from the same distribution.
        The corresponding $p$-values, shown in each panel, are consistent with the expected value of $0.5$, indicating no evidence against the null hypothesis; values significantly below $0.05$ would indicate rejection.
        In all cases, the rescaled and true SMFs are statistically consistent.
        }
    \label{fig:PQM_SMF}
\end{figure*}

In \autoref{fig:PQM_SMF}, we show the results of applying \texttt{PQMass} to compare the distribution of rescaled SMFs with the distribution of true SMFs from the \CS\ suite.
In all cases, the null hypothesis that the two distributions are the same is not rejected,
with the distribution of $\chi^2_{\rm PQM}$ values consistent with the expected $\chi^2$ distribution for the appropriate number of degrees of freedom.
Even the most extreme case with only $N_{\rm sims} = 16$ base simulations yields rescaled SMFs statistically consistent with those from direct simulations.
We obtain similar results when applying \texttt{PQMass} to the galaxy 2PCF. The rescaled and direct simulation samples are statistically indistinguishable to a high degree of confidence when using these summary statistics.

\begin{figure*}
    \centering
    \textbf{Increasing the number of base simulations used for rescaling}
    \vspace{-0.1em}

    \begin{subfigure}{0.48\textwidth}
        \centering
        \includegraphics[width=\textwidth]{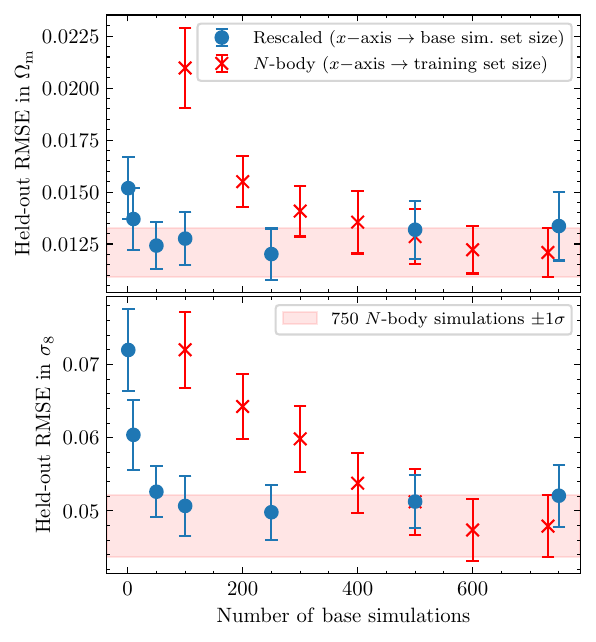}
        \caption{Models \textbf{trained on the SMF.}}
        \label{fig:SMF_vary_base}
    \end{subfigure}
    \hfill
    \begin{subfigure}{0.48\textwidth}
        \centering
        \includegraphics[width=\textwidth]{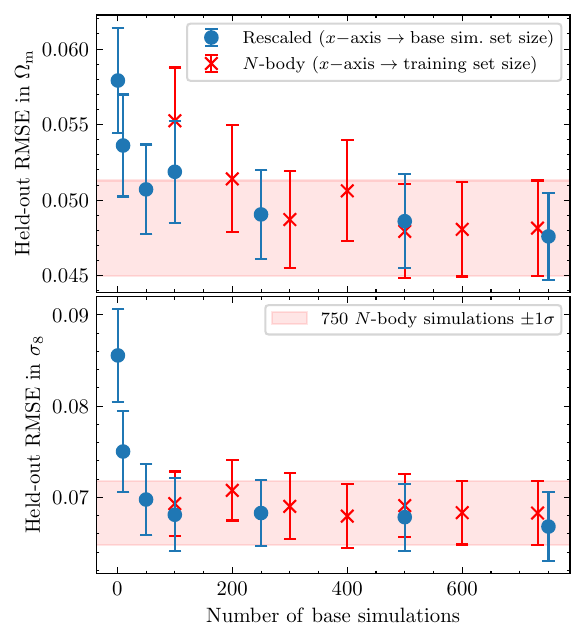}
        \caption{Models \textbf{trained on the galaxy 2PCF}.}
        \label{fig:2PCF_vary_base}
    \end{subfigure}
    \caption{
        RMSE in $\Om$ and $\sigma_8$ as a function of the number of base $N$-body simulations.
        Blue points: models trained on a fixed set of \num{1000} rescaled simulations generated from a varying number of base simulations.
        Red points: models trained directly on $N$-body simulations without rescaling, where the training set size equals the number on the $x$-axis.
        The RMSE is evaluated on \num{150} held-out SAM galaxy catalogues from direct $N$-body simulations.
        Rescaling from as few as 64 base simulations attains performance comparable to training on \num{750} direct $N$-body simulations.
        Error bars show $1\sigma$ bootstrap uncertainties; the red shaded band shows the RMSE $\pm 1\sigma$ when training on all \num{750} $N$-body simulations.
        }
    \label{fig:vary_base}
\end{figure*}

In \autoref{fig:SMF_vary_base}, we show the RMSE in $\Om$ and $\sigma_8$ as a function of the number of base simulations used to generate \num{1000} rescaled simulations for training (blue points), evaluated on \num{150} held-out $N$-body simulations.
For $\Om$, even models trained with as few as 16 or 25 base simulations achieve RMSE values comparable to those trained on the full set of \num{750} $N$-body simulations (red band).
For $\sigma_8$, the RMSE of the rescaled models converges once roughly 64 base simulations are used, matching the performance of the \num{750} $N$-body simulation model.

These results show that only a small number of base simulations is required to generate a sufficiently large training set through rescaling.

In \autoref{fig:2PCF_vary_base}, we show the corresponding results for the galaxy 2PCF, which lead to similar overall conclusions.
The first difference is that, with our choice of binning and number density, the RMSE in both $\Om$ and $\sigma_8$ is higher (approximately 0.05 and 0.07, respectively, compared to 0.01 and 0.05 for the SMF).
The second difference is that increasing the number of training $N$-body simulations (red points) yields only marginal improvement in $\Om$ and essentially none in $\sigma_8$ (without applying rescaling).
This is likely driven, at least in part, by Poisson noise in the 2PCF: we downsample the catalogue to a number density of $n=0.002~h^3\,\mathrm{Mpc}^{-3}$, corresponding to only $2000$ galaxies in a $100~\Mpch$ box, whereas the SMF uses the full catalogue.

We note, however, that these conclusions rest on the summary statistics considered here (SMF and 2PCF) and may not generalise to all possible observables. In particular, field-level inferences are likely more sensitive to the sampling of initial conditions.

\subsection{Convergence of the MLP}\label{sec:results_mlp_convergence}

Next, we examine the convergence of the MLP as the size of the training set increases, assessing whether the addition of cheaply generated rescaled simulations can improve performance on held-out $N$-body simulations.
A related question was addressed by~\citet{Bairagi_2025}, who studied simulation-based cosmological inference on power spectra from $N$-body simulations and found that as many as \num{32000} simulations were required to achieve convergence in cosmological inference and reach an asymptotic limit.
The \CS\ analysis of~\citet{Perez_2023} relies on \num{1000} $N$-body simulations, a fraction of which is held out for testing, and is therefore potentially far from convergence, although their setting differs from that of~\citet{Bairagi_2025} as they used the SMF, 2PCF, count-in-cells, and the void probability function rather than power spectra.

\begin{figure*}
    \centering
    \textbf{Increasing training set size through rescaling from \num{750} plausible base simulations}
    \vspace{-0.1em}

    \begin{subfigure}{0.48\textwidth}
        \centering
        \includegraphics[width=\textwidth]{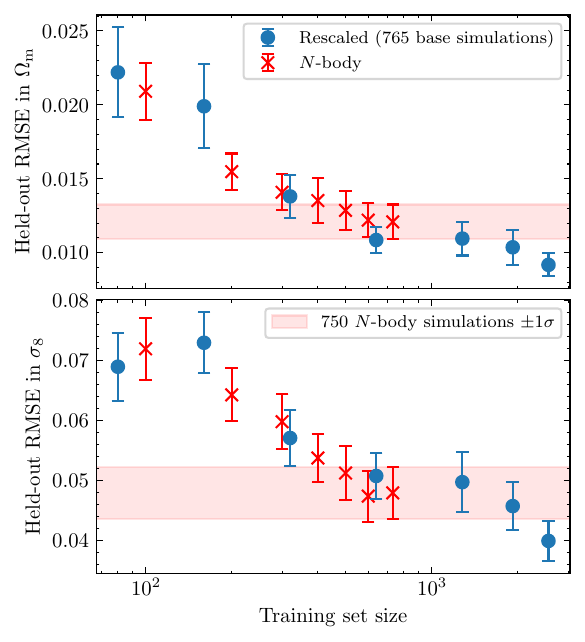}
        \caption{Models \textbf{trained on the SMF.}}
        \label{fig:SMF_vary_rescaled}
    \end{subfigure}
    \hfill
    \begin{subfigure}{0.48\textwidth}
        \centering
        \includegraphics[width=\textwidth]{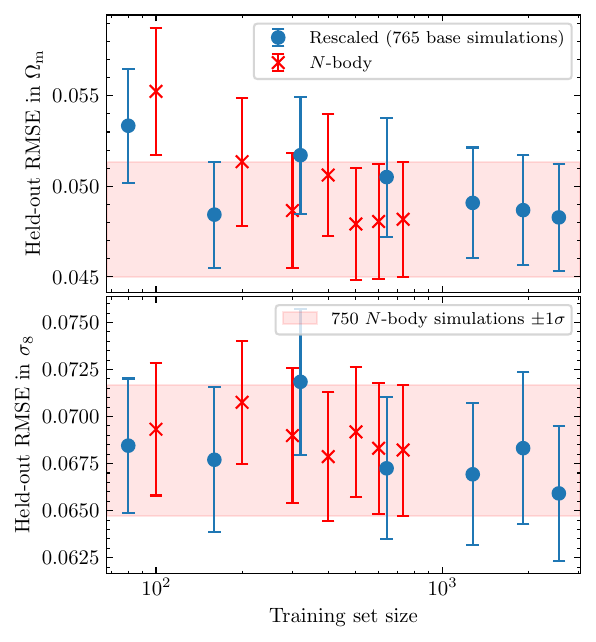}
        \caption{Models \textbf{trained on the galaxy 2PCF}.}
        \label{fig:2PCF_vary_rescaled}
    \end{subfigure}
    \caption{
        RMSE in $\Om$ and $\sigma_8$ as a function of training-set size.
        Blue points: models trained on rescaled simulations generated from a fixed pool of \num{750} base $N$-body simulations, with the number of rescaled realisations varying along the $x$-axis.
        Red points: models trained directly on $N$-body simulations without rescaling.
        The RMSE is evaluated on \num{150} held-out SAM galaxy populations built on direct $N$-body simulations.
        For the SMF, rescaling to \num{3200} simulations yields lower RMSE than training on \num{750} direct $N$-body simulations.
        Error bars show $1\sigma$ bootstrap uncertainties; the red shaded band shows the RMSE $\pm 1\sigma$ when training on all \num{750} $N$-body simulations.
        }
    \label{fig:vary_rescaled}
\end{figure*}

We present these results in \autoref{fig:vary_rescaled}, which shows the RMSE in $\Om$ and $\sigma_8$ as a function of the number of training simulations for models trained using rescaled simulations, where the number of base $N$-body simulations is fixed to \num{750} but between 100 and \num{3200} rescaled realisations are generated.
We find that the trend in RMSE as a function of training-set size is similar for rescaled simulations and direct $N$-body simulations when considering the SMF and $\Om$, with both decreasing at comparable rates.
More promisingly, we observe a marginal improvement when rescaling to \num{1600} simulations and a substantial improvement when rescaling to \num{3200}, for which the RMSE in $\Om$ is $0.0092 \pm 0.0008$ compared to $0.0121 \pm 0.0012$ when training exclusively on the \num{750} base $N$-body simulations.
We find an analogous behaviour for $\sigma_8$, where training on \num{3200} rescaled simulations yields an RMSE of $0.0399 \pm 0.0034$ compared to $0.0480 \pm 0.0043$ for models trained only on $N$-body simulations.

For the 2PCF, the trends are less pronounced, and increasing the number of $N$-body simulations yields only marginal improvements on the held-out set. Rescaling to \num{3200} simulations offers no significant improvement over directly training on \num{750} $N$-body simulations.
As before, this is likely driven by the larger Poisson noise in the 2PCF, given our choice of binning and number density.

\begin{figure*}
    \centering
    \textbf{Increasing training set size through rescaling from 25 plausible base simulations}
    \vspace{-0.1em}

    \begin{subfigure}{0.48\textwidth}
        \centering
        \includegraphics[width=\textwidth]{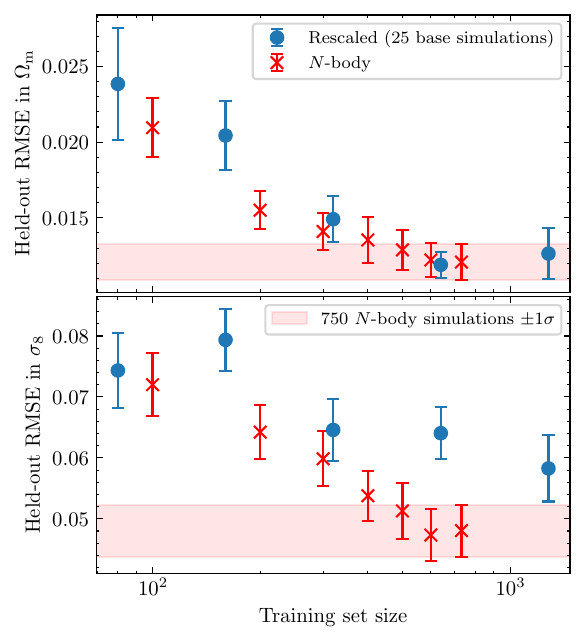}
        \caption{Results for models \textbf{trained on the SMF.}}
        \label{fig:SMF_vary_base_small}
    \end{subfigure}
    \hfill
    \begin{subfigure}{0.48\textwidth}
        \centering
        \includegraphics[width=\textwidth]{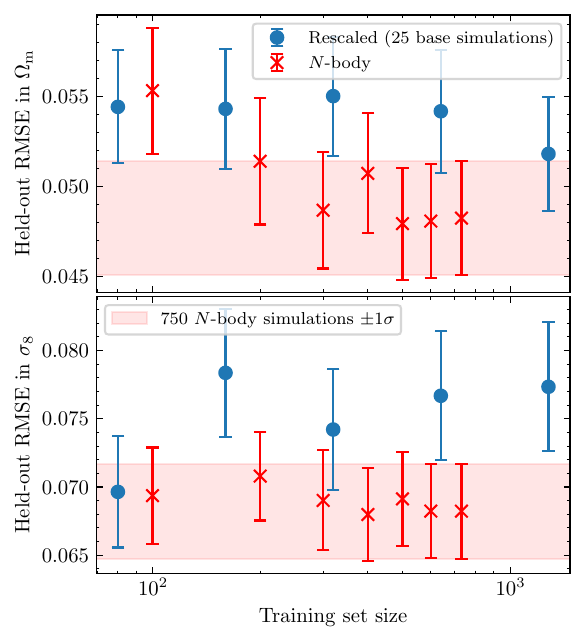}
        \caption{Results for models \textbf{trained on the galaxy 2PCF}.}
        \label{fig:2PCF_vary_base_small}
    \end{subfigure}
    \caption{
        RMSE in $\Om$ and $\sigma_8$ as a function of training-set size.
        Blue points: models trained on rescaled simulations generated from a fixed pool of only 25 base $N$-body simulations (10 from \CS\ plus 15 at $\sigma_8 = 1$), with the number of rescaled realisations varying along the $x$-axis.
        Red points: models trained directly on $N$-body simulations without rescaling.
        The RMSE is evaluated on \num{150} held-out SAM galaxy populations from direct $N$-body simulations.
        For the SMF, rescaling to \num{1600} simulations from only 25 base simulations matches the performance of training on \num{750} direct $N$-body simulations.
        Error bars show $1\sigma$ bootstrap uncertainties; the red shaded band shows the RMSE $\pm 1\sigma$ when training on all \num{750} $N$-body simulations.
        }
    \label{fig:vary_base_small}
\end{figure*}

In~\autoref{fig:vary_base_small}, we further examine an extreme scenario with only 25 base $N$-body simulations---10 from \CS\ plus 15 at $\sigma_8 = 1$---rescaled to generate up to \num{1600} realisations.
For the SMF, when rescaling to \num{1600} simulations we find that the RMSE in $\Om$ ($0.0126 \pm 0.0017$) matches the performance of training on \num{750} direct $N$-body simulations (\autoref{tab:parameter_estimation}).
For $\sigma_8$, the RMSE ($0.0582 \pm 0.0054$) remains approximately 21\% worse than the \num{750} $N$-body baseline (\autoref{tab:parameter_estimation}), though the trend may indicate further improvement with additional rescaled training data.
For the 2PCF, the RMSE in $\Om$ ($0.0518 \pm 0.0032$) remains approximately 8\% worse than the baseline and may improve with more training data, whereas for $\sigma_8$ the RMSE ($0.0773 \pm 0.0048$) is consistently 14\% worse than the baseline with no improvement trend, likely reflecting the greater sensitivity of the 2PCF to the limited number of initial conditions.
In \autoref{app:medium_base}, we show that increasing the number of base simulations to 79 (64 from \CS\ plus 15 at $\sigma_8 = 1$) largely eliminates the residual gap in 2PCF performance and yields SMF performance comparable to the \num{750}-base case.

\subsection{Cosmological coverage versus phase diversity}\label{sec:results_phase_mixing}

The held-out RMSE of the rescaled suites has two distinct error contributions: a rescaling error, which decreases as the Latin hypercube of base cosmologies becomes denser, and a cosmic-variance contribution, which decreases as the number of independent phase realisations across the base simulations grows.
The two are intertwined in our suite design, because each base simulation contributes both an independent phase realisation and a distinct cosmology.

We disentangle the two contributions by constructing a series of base-simulation suites with the same number of simulations and the same cosmologies, varying only the number of independent phase realisations.
As a reference, we take the 79-simulation base set of~\autoref{app:medium_base} (64 from \CS\ plus 15 at $\sigma_8 = 1$), in which each simulation has a distinct cosmology and an independent initial-condition phase.
We compare it to a set of phase-mixing suites that have the same 79 cosmologies but vary in their number of independent phase realisations, $N_{\rm ICs}$, from $1$ to $79$.
At $N_{\rm ICs} = 79$, every simulation carries an independent phase and the suite coincides with the reference set; at $N_{\rm ICs} = 1$, all 79 simulations share a single seed phase.
Intermediate values are mixtures: $N_{\rm ICs}$ simulations carry independent phases while the remaining $79 - N_{\rm ICs}$ share the seed phase.
We consider $N_{\rm ICs} \in \{1, 16, 26, 36, 46, 56, 66, 79\}$.
The 15 simulations at $\sigma_8 = 1$ are randomised first, so all suites with $N_{\rm ICs} \geq 16$ have these high-$\sigma_8$ simulations with independent phases.
For each $N_{\rm ICs}$, we generate \num{1000} rescaled simulations from the base set and train the MLP ensemble of \autoref{sec:method_parameter_estimation}, evaluated on the same \num{150} held-out $N$-body simulations used throughout this paper.

\begin{figure*}
    \centering
    \includegraphics[width=\textwidth]{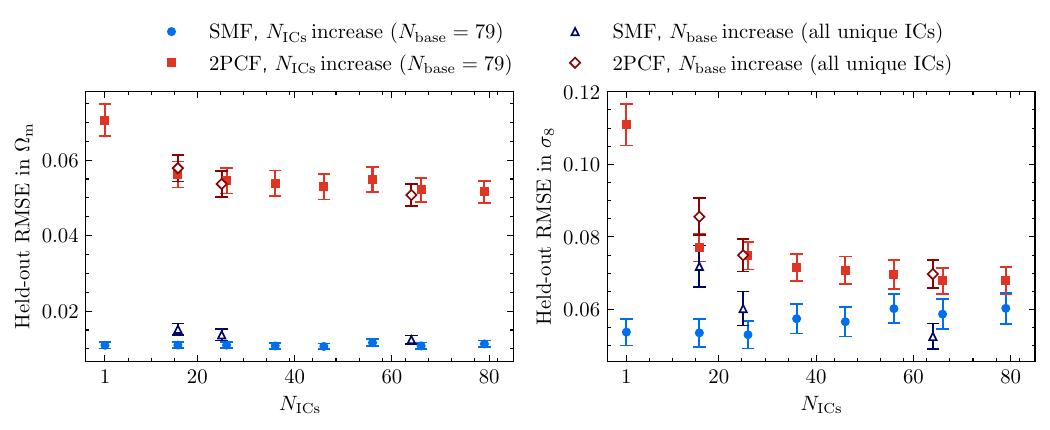}
    \caption{
        Held-out RMSE in $\Om$ (left) and $\sigma_8$ (right) as a function of the number $N_{\rm ICs}$ of unique initial-condition phases in the 79-simulation base set, shown as filled markers.
        Open markers show the reference runs of \autoref{sec:results_smaller_LH_suite}, in which all $N_{\rm base}$ simulations have independent phases, so $N_{\rm ICs} = N_{\rm base}$.
        Error bars give $1\sigma$ bootstrap uncertainties.
        The SMF RMSE is set entirely by rescaling error: cosmic variance makes no measurable contribution.
        The 2PCF RMSE is dominated by cosmic variance, saturating after ${\sim}16$ unique phases, with the rescaling-error contribution subdominant.
        }
    \label{fig:phase_mixing}
\end{figure*}

\autoref{fig:phase_mixing} shows the resulting held-out RMSE as a function of $N_{\rm ICs}$.
The SMF-trained models are insensitive to the random initial-condition sampling.
The 2PCF-trained models, by contrast, show a single large step between $N_{\rm ICs} = 1$ and $N_{\rm ICs} = 16$, a ${\sim}20$--$30\%$ reduction in RMSE, followed by a slow plateau.
In~\autoref{fig:phase_mixing} we also overlay the smaller-base-suite runs of~\autoref{sec:results_smaller_LH_suite}, in which $N_{\rm base} = N_{\rm ICs}$ simulations all have independent phases.
At a given $N_{\rm ICs}$, these runs match the corresponding phase-mixing suite
in the number of independent phase realisations but draw on only $N_{\rm base} = N_{\rm ICs}$ base cosmologies rather than 79; the rescaling therefore takes larger steps in $\Om$ and $\sigma_8$ and incurs more rescaling error.
The gap between them isolates this rescaling-error contribution at fixed $N_{\rm ICs}$.
This gap is significant only for the SMF-trained models with $N_{\rm ICs} = 16$, whereas in all other cases with higher $N_{\rm ICs}$ the gap is negligible.
Thus, for the SMF, the cosmic-variance contribution is negligible; the larger error at small $N_{\rm base}$ in~\autoref{fig:vary_base} is driven by rescaling error.
The 2PCF, by contrast, is more sensitive to the number of independent initial conditions, and is in a regime where the rescaling-error contribution is subdominant.

\begin{figure*}
    \centering
    \includegraphics[width=\textwidth]{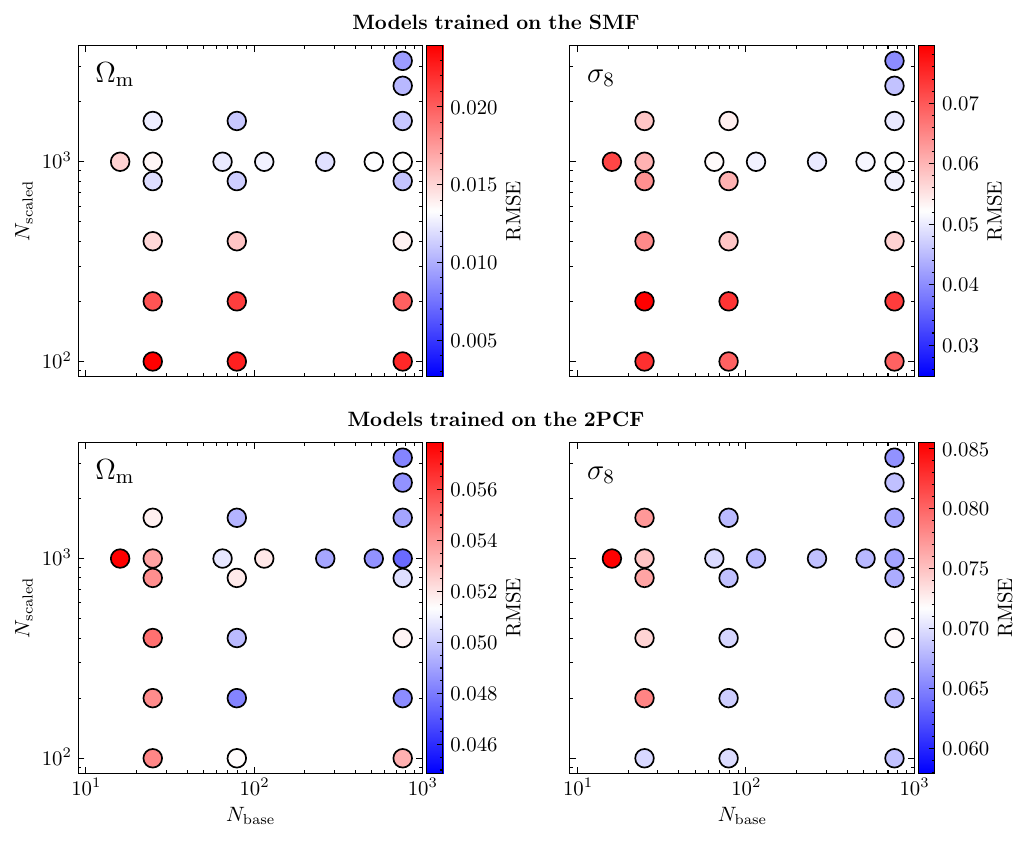}
    \caption{
        RMSE in $\Om$ and $\sigma_8$ as a function of the number of base $N$-body simulations ($x$-axis) and rescaled realisations ($y$-axis), for models trained on the SMF (top) and 2PCF (bottom), evaluated on \num{150} held-out SAM galaxy catalogues derived from direct $N$-body simulations.
        Marker colour encodes the bootstrap mean RMSE; the divergent colour map is centred on the mean RMSE $+ 1\sigma$ of the $N$-body baseline (trained on \num{750} simulations), chosen to visually balance the colour scale, so that blue markers indicate performance within $1\sigma$ of or better than the baseline and red markers indicate worse.
    }
    \label{fig:RMSE_2Dgrid}
\end{figure*}

\autoref{fig:RMSE_2Dgrid} summarises the parameter-estimation performance across all combinations of the number of base and rescaled simulations.
Each marker represents a single trained model, with colour encoding the bootstrap mean RMSE and a red edge indicating that the model matches the $N$-body baseline within $1\sigma$ or exceeds it.
Across all four panels, models with ${\sim}64$ or more base simulations and ${\sim}\num{1000}$ rescaled realisations match or exceed the performance of the \num{750}-simulation $N$-body baseline for both the SMF and 2PCF, with further improvements as the number of rescaled realisations increases.
Performance when training on the SMF to predict $\Om$ may continue to improve beyond the \num{3200} rescaled realisations tested here, but we do not explore this further.

We stress that these base simulations are randomly drawn from the \CS\ Latin hypercube, which was not designed with rescaling in mind. Each target cosmology is matched to the base simulation that minimises the expected rescaling error subject to constraints on the scale factor and length rescaling, but the pool of available bases is not itself optimised. A purpose-built suite with base cosmologies positioned to minimise the rescaling error across the target parameter space, as done for the \Bacco\ suite~\citep{Contreras_2020}, would reduce the number of required base simulations further still.

\section{Discussion}\label{sec:discussion}

In \autoref{sec:discussion_applications}, we examine the computational benefits of rescaling relative to running new $N$-body simulations.
In \autoref{sec:discussion_optimal_placement}, we discuss strategies for optimal placement of base simulations.
In \autoref{sec:discussion_limitations}, we outline limitations and systematic uncertainties, and in \autoref{sec:discussion_future_work}, we identify directions for future work.

\subsection{Computational savings}\label{sec:discussion_applications}

We have demonstrated that a small number of base simulations can be used to generate large training sets for cosmological parameter estimation from the SMF and 2PCF.
In \autoref{sec:results_smaller_LH_suite}, we showed that as few as 64 base simulations suffice to match the parameter-estimation performance of models trained on \num{750} full $N$-body realisations, representing more than an order-of-magnitude reduction in the required number of expensive simulations.
Although rescaling introduces systematic errors that grow with the step size in cosmological parameters, these errors remain subdominant to statistical uncertainties for the summary statistics considered here.
The computational cost of rescaling is negligible: rescaling a halo catalogue with $\mathcal{O}(\num{10000})$ objects takes a few seconds on a single CPU, and one run of \SC\ takes ${\sim}8$~CPUh for this setup, whereas each \CS\ $N$-body simulation required several thousand~CPUh.
The dominant cost is therefore the $N$-body simulations themselves, and rescaling reduces the number required from ${\sim}750$ to ${\sim}64$, and potentially fewer, while each additional rescaled realisation adds only ${\sim}8$~CPUh of SAM runtime. Doubling the \CS\ box size to $200~\Mpch$ at fixed mass resolution would increase the cost of each $N$-body simulation by roughly an order of magnitude, making rescaling-like approaches a necessity.

The \Bacco\ suite~\citep{Angulo_2021} similarly uses rescaling, modelling the non-linear matter power spectrum from only three $N$-body simulations, while \CS~\citep{Perez_2023} reaches larger volumes than hydrodynamic \texttt{CAMELS}~\citep{CAMELS_2021}.
We combine these approaches, exploiting the low cost of SAMs while reducing the $N$-body burden and enabling expansion to larger volumes or additional cosmological parameters.
By rescaling merger trees and rerunning the SAM, one can generate galaxy catalogues spanning cosmology and astrophysics without the expense of thousands of $N$-body realisations.

\subsection{Optimal placement of base simulations}\label{sec:discussion_optimal_placement}

The cosmological rescaling reuses existing $N$-body simulations, but the choice of which cosmologies to simulate and rescale from is flexible.
For example, the \Bacco\ suite runs simulations at $\sigma_8 = 0.9$, a value selected to minimise the theoretical rescaling error~\citep{Contreras_2020}, well above the \textit{Planck} value of $0.811 \pm 0.006$~\citep{Planck_2020}, exploiting faster structure formation at high $\sigma_8$ to scale to cosmologies with slower structure formation.

In contrast, \CS\ samples $\Om$ and $\sigma_8$ uniformly, yielding representative coverage but suboptimal placement for rescaling.
Since we do not have future snapshots at $z < 0$, we were restricted to rescaling towards cosmologies with slower structure formation (generally lower $\sigma_8$ or $\Om$).
Therefore, to improve the coverage at high $\sigma_8$, we included 15 simulations at $\sigma_8 = 1$ as high-$\sigma_8$ anchors.
However, optimising the choice of cosmological parameter space locations
to simulate for rescaling efficiency reduces general-purpose applicability, and the optimal solution is problem-dependent.

As noted in~\autoref{sec:results_smaller_LH_suite}, our base simulations are randomly drawn from the \CS\ Latin hypercube rather than optimally placed for rescaling, and a dedicated suite design would reduce the required number further.

Active learning could address this.
Rescaling errors increase with step size in cosmological parameters, creating a trade-off between the number of base simulations and the accuracy of rescaled outputs.
One could begin with a few simulations placed at high $\sigma_8$ or uniformly across parameter space, and iteratively add new base simulations to reduce rescaling errors where they are largest.

\subsection{Limitations and systematic uncertainties}\label{sec:discussion_limitations}

A key limitation is that reducing the number of base simulations also reduces the number of unique initial conditions, potentially making cosmic variance a limiting factor.
While as few as 64 base simulations suffice for converged parameter estimation from the 2PCF and SMF---matching performance with \num{750}-simulation training sets---they share a limited pool of initial-condition phases.
In \autoref{sec:results_phase_mixing}, we disentangle these two contributions: rescaling error dominates the SMF RMSE for far fewer than 64 base simulations, whereas cosmic variance dominates the 2PCF RMSE, which saturates beyond ${\sim}16$ unique phases.
A direct consequence is that a future suite optimised for rescaling could retain the LH coverage while recycling a small pool of independent phases across cosmological positions, reducing the number of $N$-body realisations required for the base set.
For applications requiring many realisations at the same or similar cosmologies (e.g.,~field-level inference), this limited phase diversity becomes a fundamental constraint: rescaling cannot replace the need for independent realisations when phase information matters.
The applicability of rescaling depends on target observables and the acceptable level of error.
Summary statistics such as the SMF, which average over many objects, are relatively insensitive to rescaling errors.

The original rescaling algorithm of~\citet{Angulo_2010} includes large-scale LPT corrections to adjust the amplitudes of linear density modes, ensuring that the rescaled simulation reproduces the correct large-scale power spectrum in the target cosmology.
As in~\citet{Ruiz_2011}, we omit these corrections because the \CS\ box size of $100~\cMpch$ does not sample the large-scale modes for which they are relevant.
For extensions to larger simulation volumes, where box sizes exceed several hundred $\Mpch$, LPT corrections would become necessary to accurately capture large-scale clustering.
While not tested here, this would primarily impact the 2PCF, with a subdominant effect on the SMF.

Since \CS\ lacks snapshots at $z < 0$, we could only scale to higher redshifts and to cosmologies with generally slower structure formation (lower $\sigma_8$ or $\Om$).
Ideally, a suite designed for rescaling would include future snapshots to enable rescaling in both directions.
Nevertheless, as we demonstrated, this restriction does not pose a major limitation for the parameter ranges and observables considered here even when rescaling from only a very small number of base simulations.

Finally, our tests are limited to specific observables---the SMF and 2PCF---and to the \CS\ simulation suite.
Whether these results extend to higher-order clustering or field-level inference of SAM-modelled galaxy properties remains to be tested.
We expect rescaling to perform well for observables that average over many objects, but field-level or higher-order statistics may be more sensitive to the limited number of independent initial conditions.

\subsection{Future work}\label{sec:discussion_future_work}

An immediate extension is applying rescaling to larger simulation volumes matching modern survey footprints.
Even SDSS~\citep{Strauss_2002} spans approximately $(1~\mathrm{Gpc}/h)^3$, while DESI~\citep{Levi_2019}, Euclid~\citep{Euclid}, and the Vera C. Rubin Observatory~\citep{Zeljko_2019} will probe substantially larger volumes.
Modelling such surveys requires box sizes of at least several hundred $\Mpch$, where the cost of individual realisations grows rapidly and rescaling offers significant advantages.

In this work, we compute the galaxy 2PCF in real space only.
A natural extension is to validate the rescaling for redshift-space clustering, which is the primary observable for spectroscopic surveys.
Since the rescaling also transforms halo velocities, the rescaled merger trees contain the information needed to construct redshift-space catalogues.
The \Bacco\ project has demonstrated that rescaled simulations accurately reproduce the redshift-space power spectrum of dark matter and halos~\citep{ibanez_2023}; we therefore do not anticipate any difficulty in extending our approach to redshift space, but leave explicit validation for future work.

Another direction is expanding the cosmological parameter space beyond $\Om$ and $\sigma_8$ to include $H_0$, $n_{\rm s}$, $\sum m_\nu$, and $w$.
Uniform sampling in six dimensions would require a prohibitively large number of simulations; for comparison, even the ${\sim}\num{32000}$ simulations of the BSQ suite do not yield fully converged power spectrum inference~\citep{Bairagi_2025}.
Rescaling might alleviate this curse of dimensionality.

A natural application is extending the \CS\ suite itself.
The current suite consists of $100~\Mpch$ boxes sampling only $\Om$ and $\sigma_8$; rescaling would enable extension to larger volumes and denser sampling of a larger cosmological parameter space without requiring a correspondingly large number of $N$-body simulations.
This would yield a next-generation \CS\ suite capable of modelling galaxy populations for upcoming surveys, retaining the computational efficiency of SAM-based galaxy formation.
Target observables could include galaxy luminosity functions and colours, extending inference beyond the summary statistics considered here~\citep{Lovell_2025}. Based on our SMF results, such observables would likely be insensitive to the limited number of initial conditions.

Because summary statistics such as the SMF and 2PCF average over many objects, the rescaled suite is particularly well suited for training emulators of these quantities as continuous functions of both cosmological and astrophysical parameters---analogous to the \Bacco\ emulators for the matter power spectrum and halo statistics~\citep{Contreras_2020,Angulo_2021,Zennaro_2023}.
Such emulators would provide rapid predictions at arbitrary parameter values, enabling explicit likelihood analyses and efficient calibration of SAM parameters against observed galaxy statistics.

The parameter estimation framework presented here uses simple neural network regression, though it is readily extensible to simulation-based inference; neither approach requires gradients of the forward model.
However, the rescaling procedure is a straightforward algebraic operation amenable to automatic differentiation.
One could implement differentiable rescaling on top of existing merger trees and propagate gradients through a differentiable SAM such as \texttt{Sapphire}~\citep{Pandya_2026}.
This would enable gradient-based inference methods such as Hamiltonian Monte Carlo~\citep[HMC;][]{Hoffman_2014}, which could improve sampling efficiency for joint cosmological and astrophysical parameter estimation.

A potential longer-term goal is extending rescaling beyond SAM post-processing to hydrodynamical simulations.
Our approach directly targets galaxy observables by re-running the relatively inexpensive SAM on rescaled merger trees.
Extending this to hydrodynamical simulations would be more challenging, as baryonic physics operates predominantly on non-linear scales.
However, one could target specific observables using simple mappings---or machine learning---analogous to the NFW rescaling correction we introduced for halo mass.
An alternative is to bypass the SAM altogether and emulate galaxy properties directly from merger trees, training on small-volume hydrodynamical simulations and evaluating on large-volume rescaled halo catalogues~\citep[e.g., the \texttt{Mangrove} graph neural network;][]{Jespersen_2022}.

\section{Conclusion}\label{sec:conclusion}

We have extended cosmological rescaling algorithms to operate directly on halo merger trees---rather than particle snapshots or halo catalogues---and shown that this technique can be used to efficiently generate large training sets of semi-analytic galaxy populations.
By rescaling merger trees from the \CS\ suite and rerunning the \SC\ model, we produce galaxy catalogues spanning a range of cosmologies at a fraction of the cost of running full $N$-body simulations.

We introduced a novel NFW rescaling correction, parametrised by a single parameter, that suppresses systematic bias in rescaled halo masses for steps in both $\Om$ and $\sigma_8$.
Other halo properties (concentration, spin) are likewise reproduced without systematic bias across all redshifts.
Galaxy stellar masses (and other properties such as gas mass and black hole mass) from rescaled merger trees exhibit scatter comparable to that from varying the random seed in \SC, confirming that rescaling errors are subdominant to stochasticity originating from the treatment of satellite galaxies in \SC.
Similarly, population statistics---the HMF, SMF, and galaxy 2PCF---from rescaled trees closely match those from direct simulations.

For cosmological parameter estimation from the SMF and 2PCF, models trained on rescaled simulations match the performance of models trained on full $N$-body suites.
As few as 64 base $N$-body simulations---and likely even fewer depending on error tolerance---when used to generate \num{1000} rescaled training samples, match the parameter-estimation accuracy of \num{750} $N$-body simulations. \autoref{fig:RMSE_2Dgrid} summarises this across all tested combinations of the number of base and rescaled simulations: for both the SMF and 2PCF, models with ${\sim}64$ base simulations and ${\sim}\num{1000}$ rescaled realisations reach the $N$-body baseline, and performance continues to improve with additional rescaled training data.
With 79 base simulations (64 from the Latin hypercube and 15 high-$\sigma_8$ anchors), the residual gap in 2PCF performance observed with fewer bases is largely eliminated (\autoref{app:medium_base}).
Generating yet larger training sets via rescaling further improves inference precision: \num{3200} realisations generated from \num{750} base simulations reduce the RMSE in $\Om$ when trained on the SMF by ${\sim}25\%$ relative to training on \num{750} $N$-body simulations alone.
Since our base simulations are randomly drawn from the \CS\ Latin hypercube rather than optimally placed for rescaling, the number of required bases could be reduced further with a dedicated suite design.
In effect, rescaling acts as a physics-informed interpolation scheme between existing simulations, using known scaling relations to generate new training samples at intermediate cosmologies.
For \CS, rescaling merger trees on a single CPU takes less than a minute per realisation; \SC\ requires ${\sim}8$~CPUh per realisation; and each $N$-body simulation costs several thousand CPUh.
Merger-tree rescaling, combined with SAM post-processing, thus offers a practical route to the large simulation suites required for modern cosmological inference on galaxy observables.

\begin{acknowledgments}
Richard Stiskalek acknowledges financial support from the CCA Pre-doctoral Program, STFC Grant No. ST/X508664/1, the Snell Exhibition of Balliol College, Oxford, and a Hintze Fellowship at the Oxford Centre for Astrophysical Surveys, funded through generous support from the Hintze Family Charitable Foundation.
The Flatiron Institute is supported by the Simons Foundation.
Sergio Contreras acknowledges the support of the ``Ram\'{o}n y Cajal'' fellowship (RYC2023-043783-I).
\end{acknowledgments}

\bibliography{ref}{}

@ARTICLE{Ortega_2026,
       author = {{Ortega-Martinez}, Sara and {Angulo}, Raul E. and {Contreras}, Sergio and {Chaves-Montero}, Jon{\'a}s and {Zennaro}, Matteo and {Bose}, Sownak and {Hadzhiyska}, Boryana and {Hern{\'a}ndez-Aguayo}, C{\'e}sar and {Hernquist}, Lars and {Springel}, Volker},
        title = "{Cosmological constraints from the small scale clustering of Emission Line Galaxies}",
      journal = {arXiv e-prints},
         year = 2026,
        month = apr,
          eid = {arXiv:2604.19449},
        pages = {arXiv:2604.19449},
          doi = {10.48550/arXiv.2604.19449},
archivePrefix = {arXiv},
       eprint = {2604.19449},
 primaryClass = {astro-ph.CO},
       adsurl = {https://ui.adsabs.harvard.edu/abs/2026arXiv260419449O}
}

@ARTICLE{Mahony_2026,
       author = {{Mahony}, Constance and {Contreras}, Sergio and {Angulo}, Raul E. and {Alonso}, David and {Georgiou}, Christos and {Dvornik}, Andrej},
        title = "{Cosmological constraints from galaxy clustering and galaxy─galaxy lensing with extended SubHalo Abundance Matching}",
      journal = {\mnras},
         year = 2026,
        month = feb,
       volume = {545},
       number = {4},
          eid = {staf2143},
        pages = {staf2143},
          doi = {10.1093/mnras/staf2143},
archivePrefix = {arXiv},
       eprint = {2507.01601},
 primaryClass = {astro-ph.CO},
       adsurl = {https://ui.adsabs.harvard.edu/abs/2026MNRAS.545f2143M}
}

@ARTICLE{Ondaro-Mallea_2022,
       author = {{Ondaro-Mallea}, Lurdes and {Angulo}, Raul E. and {Zennaro}, Matteo and {Contreras}, Sergio and {Aric{\`o}}, Giovanni},
        title = "{Non-universality of the mass function: dependence on the growth rate and power spectrum shape}",
      journal = {\mnras},
         year = 2022,
        month = feb,
       volume = {509},
       number = {4},
        pages = {6077-6090},
          doi = {10.1093/mnras/stab3337},
archivePrefix = {arXiv},
       eprint = {2102.08958},
 primaryClass = {astro-ph.CO},
       adsurl = {https://ui.adsabs.harvard.edu/abs/2022MNRAS.509.6077O}
}

@ARTICLE{CAMELS_2021,
       author = {{Villaescusa-Navarro}, Francisco and {Angl{\'e}s-Alc{\'a}zar}, Daniel and {Genel}, Shy and {Spergel}, David N. and {Somerville}, Rachel S. and {Dave}, Romeel and {Pillepich}, Annalisa and {Hernquist}, Lars and {Nelson}, Dylan and {Torrey}, Paul and {Narayanan}, Desika and {Li}, Yin and {Philcox}, Oliver and {La Torre}, Valentina and {Maria Delgado}, Ana and {Ho}, Shirley and {Hassan}, Sultan and {Burkhart}, Blakesley and {Wadekar}, Digvijay and {Battaglia}, Nicholas and {Contardo}, Gabriella and {Bryan}, Greg L.},
        title = "{The CAMELS Project: Cosmology and Astrophysics with Machine-learning Simulations}",
      journal = {\apj},
         year = 2021,
        month = jul,
       volume = {915},
       number = {1},
          eid = {71},
        pages = {71},
          doi = {10.3847/1538-4357/abf7ba},
archivePrefix = {arXiv},
       eprint = {2010.00619},
 primaryClass = {astro-ph.CO},
       adsurl = {https://ui.adsabs.harvard.edu/abs/2021ApJ...915...71V}
}

@ARTICLE{Perez_2023,
       author = {{Perez}, Lucia A. and {Genel}, Shy and {Villaescusa-Navarro}, Francisco and {Somerville}, Rachel S. and {Gabrielpillai}, Austen and {Angl{\'e}s-Alc{\'a}zar}, Daniel and {Wandelt}, Benjamin D. and {Yung}, L.~Y. Aaron},
        title = "{Constraining Cosmology with Machine Learning and Galaxy Clustering: The CAMELS-SAM Suite}",
      journal = {\apj},
         year = 2023,
        month = sep,
       volume = {954},
       number = {1},
          eid = {11},
        pages = {11},
          doi = {10.3847/1538-4357/accd52},
archivePrefix = {arXiv},
       eprint = {2204.02408},
 primaryClass = {astro-ph.GA},
       adsurl = {https://ui.adsabs.harvard.edu/abs/2023ApJ...954...11P}
}

@ARTICLE{AREPO,
       author = {{Weinberger}, Rainer and {Springel}, Volker and {Pakmor}, R{\"u}diger},
        title = "{The AREPO Public Code Release}",
      journal = {\apjs},
         year = 2020,
        month = jun,
       volume = {248},
       number = {2},
          eid = {32},
        pages = {32},
          doi = {10.3847/1538-4365/ab908c},
archivePrefix = {arXiv},
       eprint = {1909.04667},
 primaryClass = {astro-ph.IM},
       adsurl = {https://ui.adsabs.harvard.edu/abs/2020ApJS..248...32W}
}

@software{CAMB,
       author = {{Lewis}, Antony and {Challinor}, Anthony},
        title = "{CAMB: Code for Anisotropies in the Microwave Background}",
 howpublished = {Astrophysics Source Code Library, record ascl:1102.026},
         year = 2011,
        month = feb,
          eid = {ascl:1102.026},
archivePrefix = {ascl},
       eprint = {1102.026},
       adsurl = {https://ui.adsabs.harvard.edu/abs/2011ascl.soft02026L}
}

@ARTICLE{Behroozi_2013,
       author = {{Behroozi}, Peter S. and {Wechsler}, Risa H. and {Wu}, Hao-Yi},
        title = "{The ROCKSTAR Phase-space Temporal Halo Finder and the Velocity Offsets of Cluster Cores}",
      journal = {\apj},
         year = 2013,
        month = jan,
       volume = {762},
       number = {2},
          eid = {109},
        pages = {109},
          doi = {10.1088/0004-637X/762/2/109},
archivePrefix = {arXiv},
       eprint = {1110.4372},
 primaryClass = {astro-ph.CO},
       adsurl = {https://ui.adsabs.harvard.edu/abs/2013ApJ...762..109B}
}

@ARTICLE{Behroozi_2013_cs,
       author = {{Behroozi}, Peter S. and {Wechsler}, Risa H. and {Wu}, Hao-Yi and {Busha}, Michael T. and {Klypin}, Anatoly A. and {Primack}, Joel R.},
        title = "{Gravitationally Consistent Halo Catalogs and Merger Trees for Precision Cosmology}",
      journal = {\apj},
         year = 2013,
        month = jan,
       volume = {763},
       number = {1},
          eid = {18},
        pages = {18},
          doi = {10.1088/0004-637X/763/1/18},
archivePrefix = {arXiv},
       eprint = {1110.4370},
 primaryClass = {astro-ph.CO},
       adsurl = {https://ui.adsabs.harvard.edu/abs/2013ApJ...763...18B}
}

@ARTICLE{Somerville_1999,
       author = {{Somerville}, Rachel S. and {Primack}, Joel R.},
        title = "{Semi-analytic modelling of galaxy formation: the local Universe}",
      journal = {\mnras},
         year = 1999,
        month = dec,
       volume = {310},
       number = {4},
        pages = {1087-1110},
          doi = {10.1046/j.1365-8711.1999.03032.x},
archivePrefix = {arXiv},
       eprint = {astro-ph/9802268},
 primaryClass = {astro-ph},
       adsurl = {https://ui.adsabs.harvard.edu/abs/1999MNRAS.310.1087S}
}

@ARTICLE{Somerville_2008,
       author = {{Somerville}, Rachel S. and {Hopkins}, Philip F. and {Cox}, Thomas J. and {Robertson}, Brant E. and {Hernquist}, Lars},
        title = "{A semi-analytic model for the co-evolution of galaxies, black holes and active galactic nuclei}",
      journal = {\mnras},
         year = 2008,
        month = dec,
       volume = {391},
       number = {2},
        pages = {481-506},
          doi = {10.1111/j.1365-2966.2008.13805.x},
archivePrefix = {arXiv},
       eprint = {0808.1227},
 primaryClass = {astro-ph},
       adsurl = {https://ui.adsabs.harvard.edu/abs/2008MNRAS.391..481S}
}

@ARTICLE{Somerville_2015,
       author = {{Somerville}, Rachel S. and {Popping}, Gerg{\"o} and {Trager}, Scott C.},
        title = "{Star formation in semi-analytic galaxy formation models with multiphase gas}",
      journal = {\mnras},
         year = 2015,
        month = nov,
       volume = {453},
       number = {4},
        pages = {4337-4367},
          doi = {10.1093/mnras/stv1877},
archivePrefix = {arXiv},
       eprint = {1503.00755},
 primaryClass = {astro-ph.GA},
       adsurl = {https://ui.adsabs.harvard.edu/abs/2015MNRAS.453.4337S}
}

@ARTICLE{Somerville_2021,
       author = {{Somerville}, Rachel S. and {Olsen}, Charlotte and {Yung}, L.~Y. Aaron and {Pacifici}, Camilla and {Ferguson}, Henry C. and {Behroozi}, Peter and {Osborne}, Shannon and {Wechsler}, Risa H. and {Pandya}, Viraj and {Faber}, Sandra M. and {Primack}, Joel R. and {Dekel}, Avishai},
        title = "{Mock light-cones and theory friendly catalogues for the CANDELS survey}",
      journal = {\mnras},
         year = 2021,
        month = apr,
       volume = {502},
       number = {4},
        pages = {4858-4876},
          doi = {10.1093/mnras/stab231},
archivePrefix = {arXiv},
       eprint = {2102.00108},
 primaryClass = {astro-ph.GA},
       adsurl = {https://ui.adsabs.harvard.edu/abs/2021MNRAS.502.4858S}
}

@ARTICLE{Gabrielpillai_2022,
       author = {{Gabrielpillai}, Austen and {Somerville}, Rachel S. and {Genel}, Shy and {Rodriguez-Gomez}, Vicente and {Pandya}, Viraj and {Yung}, L.~Y. Aaron and {Hernquist}, Lars},
        title = "{Galaxy formation in the Santa Cruz semi-analytic model compared with IllustrisTNG - I. Galaxy scaling relations, dispersions, and residuals at z = 0}",
      journal = {\mnras},
         year = 2022,
        month = dec,
       volume = {517},
       number = {4},
        pages = {6091-6111},
          doi = {10.1093/mnras/stac2297},
archivePrefix = {arXiv},
       eprint = {2111.03077},
 primaryClass = {astro-ph.GA},
       adsurl = {https://ui.adsabs.harvard.edu/abs/2022MNRAS.517.6091G}
}

@ARTICLE{Angulo_2010,
       author = {{Angulo}, R.~E. and {White}, S.~D.~M.},
        title = "{One simulation to fit them all - changing the background parameters of a cosmological N-body simulation}",
      journal = {\mnras},
         year = 2010,
        month = jun,
       volume = {405},
       number = {1},
        pages = {143-154},
          doi = {10.1111/j.1365-2966.2010.16459.x},
archivePrefix = {arXiv},
       eprint = {0912.4277},
 primaryClass = {astro-ph.CO},
       adsurl = {https://ui.adsabs.harvard.edu/abs/2010MNRAS.405..143A}
}

@ARTICLE{Ruiz_2011,
       author = {{Ruiz}, Andr{\'e}s. N. and {Padilla}, Nelson D. and {Dom{\'\i}nguez}, Mariano J. and {Cora}, Sof{\'\i}a. A.},
        title = "{How accurate is it to update the cosmology of your halo catalogues?}",
      journal = {\mnras},
         year = 2011,
        month = dec,
       volume = {418},
       number = {4},
        pages = {2422-2434},
          doi = {10.1111/j.1365-2966.2011.19635.x},
archivePrefix = {arXiv},
       eprint = {1103.5074},
 primaryClass = {astro-ph.CO},
       adsurl = {https://ui.adsabs.harvard.edu/abs/2011MNRAS.418.2422R}
}

@ARTICLE{Mead_2014,
       author = {{Mead}, A.~J. and {Peacock}, J.~A.},
        title = "{Remapping dark matter halo catalogues between cosmological simulations}",
      journal = {\mnras},
         year = 2014,
        month = may,
       volume = {440},
       number = {2},
        pages = {1233-1247},
          doi = {10.1093/mnras/stu345},
archivePrefix = {arXiv},
       eprint = {1308.5183},
 primaryClass = {astro-ph.CO},
       adsurl = {https://ui.adsabs.harvard.edu/abs/2014MNRAS.440.1233M}
}

@ARTICLE{Navarro_1997,
       author = {{Navarro}, Julio F. and {Frenk}, Carlos S. and {White}, Simon D.~M.},
        title = "{A Universal Density Profile from Hierarchical Clustering}",
      journal = {\apj},
         year = 1997,
        month = dec,
       volume = {490},
       number = {2},
        pages = {493-508},
          doi = {10.1086/304888},
archivePrefix = {arXiv},
       eprint = {astro-ph/9611107},
 primaryClass = {astro-ph},
       adsurl = {https://ui.adsabs.harvard.edu/abs/1997ApJ...490..493N}
}

@ARTICLE{Bryan_1998,
       author = {{Bryan}, Greg L. and {Norman}, Michael L.},
        title = "{Statistical Properties of X-Ray Clusters: Analytic and Numerical Comparisons}",
      journal = {\apj},
         year = 1998,
        month = mar,
       volume = {495},
       number = {1},
        pages = {80-99},
          doi = {10.1086/305262},
archivePrefix = {arXiv},
       eprint = {astro-ph/9710107},
 primaryClass = {astro-ph},
       adsurl = {https://ui.adsabs.harvard.edu/abs/1998ApJ...495...80B}
}

@ARTICLE{Butsky_2016,
       author = {{Butsky}, Iryna and {Macci{\`o}}, Andrea V. and {Dutton}, Aaron A. and {Wang}, Liang and {Obreja}, Aura and {Stinson}, Greg S. and {Penzo}, Camilla and {Kang}, Xi and {Keller}, Ben W. and {Wadsley}, James},
        title = "{NIHAO project II: halo shape, phase-space density and velocity distribution of dark matter in galaxy formation simulations}",
      journal = {\mnras},
         year = 2016,
        month = oct,
       volume = {462},
       number = {1},
        pages = {663-680},
          doi = {10.1093/mnras/stw1688},
archivePrefix = {arXiv},
       eprint = {1503.04814},
 primaryClass = {astro-ph.GA},
       adsurl = {https://ui.adsabs.harvard.edu/abs/2016MNRAS.462..663B}
}

@ARTICLE{Desmond_2017,
       author = {{Desmond}, Harry and {Mao}, Yao-Yuan and {Wechsler}, Risa H. and {Crain}, Robert A. and {Schaye}, Joop},
        title = "{On the galaxy-halo connection in the EAGLE simulation}",
      journal = {\mnras},
         year = 2017,
        month = oct,
       volume = {471},
       number = {1},
        pages = {L11-L15},
          doi = {10.1093/mnrasl/slx093},
archivePrefix = {arXiv},
       eprint = {1612.01029},
 primaryClass = {astro-ph.GA},
       adsurl = {https://ui.adsabs.harvard.edu/abs/2017MNRAS.471L..11D}
}

@ARTICLE{Mitchell_2018,
       author = {{Mitchell}, Peter D. and {Lacey}, Cedric G. and {Lagos}, Claudia D.~P. and {Frenk}, Carlos S. and {Bower}, Richard G. and {Cole}, Shaun and {Helly}, John C. and {Schaller}, Matthieu and {Gonzalez-Perez}, Violeta and {Theuns}, Tom},
        title = "{Comparing galaxy formation in semi-analytic models and hydrodynamical simulations}",
      journal = {\mnras},
         year = 2018,
        month = feb,
       volume = {474},
       number = {1},
        pages = {492-521},
          doi = {10.1093/mnras/stx2770},
archivePrefix = {arXiv},
       eprint = {1709.08647},
 primaryClass = {astro-ph.GA},
       adsurl = {https://ui.adsabs.harvard.edu/abs/2018MNRAS.474..492M}
}

@ARTICLE{Cataldi_2021,
       author = {{Cataldi}, P. and {Pedrosa}, S.~E. and {Tissera}, P.~B. and {Artale}, M.~C.},
        title = "{Baryons shaping dark matter haloes}",
      journal = {\mnras},
         year = 2021,
        month = mar,
       volume = {501},
       number = {4},
        pages = {5679-5691},
          doi = {10.1093/mnras/staa3988},
archivePrefix = {arXiv},
       eprint = {2008.02404},
 primaryClass = {astro-ph.GA},
       adsurl = {https://ui.adsabs.harvard.edu/abs/2021MNRAS.501.5679C}
}

@ARTICLE{Stiskalek_2024,
       author = {{Stiskalek}, Richard and {Desmond}, Harry and {Devriendt}, Julien and {Slyz}, Adrianne},
        title = "{Evaluating the variance of individual halo properties in constrained cosmological simulations}",
      journal = {\mnras},
         year = 2024,
        month = nov,
       volume = {534},
       number = {4},
        pages = {3120-3132},
          doi = {10.1093/mnras/stae2292},
archivePrefix = {arXiv},
       eprint = {2310.20672},
 primaryClass = {astro-ph.CO},
       adsurl = {https://ui.adsabs.harvard.edu/abs/2024MNRAS.534.3120S}
}

@ARTICLE{Ludlow_2016,
       author = {{Ludlow}, Aaron D. and {Bose}, Sownak and {Angulo}, Ra{\'u}l E. and {Wang}, Lan and {Hellwing}, Wojciech A. and {Navarro}, Julio F. and {Cole}, Shaun and {Frenk}, Carlos S.},
        title = "{The mass-concentration-redshift relation of cold and warm dark matter haloes}",
      journal = {\mnras},
         year = 2016,
        month = aug,
       volume = {460},
       number = {2},
        pages = {1214-1232},
          doi = {10.1093/mnras/stw1046},
archivePrefix = {arXiv},
       eprint = {1601.02624},
 primaryClass = {astro-ph.CO},
       adsurl = {https://ui.adsabs.harvard.edu/abs/2016MNRAS.460.1214L}
}

@ARTICLE{Corrfunc,
       author = {{Sinha}, Manodeep and {Garrison}, Lehman H.},
        title = "{CORRFUNC - a suite of blazing fast correlation functions on the CPU}",
      journal = {\mnras},
         year = 2020,
        month = jan,
       volume = {491},
       number = {2},
        pages = {3022-3041},
          doi = {10.1093/mnras/stz3157},
archivePrefix = {arXiv},
       eprint = {1911.03545},
 primaryClass = {astro-ph.CO},
       adsurl = {https://ui.adsabs.harvard.edu/abs/2020MNRAS.491.3022S}
}

@ARTICLE{Landy_1993,
       author = {{Landy}, Stephen D. and {Szalay}, Alexander S.},
        title = "{Bias and Variance of Angular Correlation Functions}",
      journal = {\apj},
         year = 1993,
        month = jul,
       volume = {412},
        pages = {64},
          doi = {10.1086/172900},
       adsurl = {https://ui.adsabs.harvard.edu/abs/1993ApJ...412...64L}
}

@ARTICLE{Cranmer_2020,
       author = {{Cranmer}, Kyle and {Brehmer}, Johann and {Louppe}, Gilles},
        title = "{The frontier of simulation-based inference}",
      journal = {Proceedings of the National Academy of Science},
         year = 2020,
        month = dec,
       volume = {117},
       number = {48},
        pages = {30055-30062},
          doi = {10.1073/pnas.1912789117},
archivePrefix = {arXiv},
       eprint = {1911.01429},
 primaryClass = {stat.ML},
       adsurl = {https://ui.adsabs.harvard.edu/abs/2020PNAS..11730055C}
}

@ARTICLE{Greenberg_2019,
       author = {{Greenberg}, David S. and {Nonnenmacher}, Marcel and {Macke}, Jakob H.},
        title = "{Automatic Posterior Transformation for Likelihood-Free Inference}",
      journal = {arXiv e-prints},
         year = 2019,
        month = may,
          eid = {arXiv:1905.07488},
        pages = {arXiv:1905.07488},
          doi = {10.48550/arXiv.1905.07488},
archivePrefix = {arXiv},
       eprint = {1905.07488},
 primaryClass = {cs.LG},
       adsurl = {https://ui.adsabs.harvard.edu/abs/2019arXiv190507488G}
}

@ARTICLE{Alsing_2018,
       author = {{Alsing}, Justin and {Wandelt}, Benjamin and {Feeney}, Stephen},
        title = "{Massive optimal data compression and density estimation for scalable, likelihood-free inference in cosmology}",
      journal = {\mnras},
         year = 2018,
        month = jul,
       volume = {477},
       number = {3},
        pages = {2874-2885},
          doi = {10.1093/mnras/sty819},
archivePrefix = {arXiv},
       eprint = {1801.01497},
 primaryClass = {astro-ph.CO},
       adsurl = {https://ui.adsabs.harvard.edu/abs/2018MNRAS.477.2874A}
}

@ARTICLE{Papamakarios_2019,
       author = {{Papamakarios}, George and {Nalisnick}, Eric and {Jimenez Rezende}, Danilo and {Mohamed}, Shakir and {Lakshminarayanan}, Balaji},
        title = "{Normalizing Flows for Probabilistic Modeling and Inference}",
      journal = {arXiv e-prints},
         year = 2019,
        month = dec,
          eid = {arXiv:1912.02762},
        pages = {arXiv:1912.02762},
          doi = {10.48550/arXiv.1912.02762},
archivePrefix = {arXiv},
       eprint = {1912.02762},
 primaryClass = {stat.ML},
       adsurl = {https://ui.adsabs.harvard.edu/abs/2019arXiv191202762P}
}

@ARTICLE{Hendrycks_2016,
       author = {{Hendrycks}, Dan and {Gimpel}, Kevin},
        title = "{Gaussian Error Linear Units (GELUs)}",
      journal = {arXiv e-prints},
         year = 2016,
        month = jun,
          eid = {arXiv:1606.08415},
        pages = {arXiv:1606.08415},
          doi = {10.48550/arXiv.1606.08415},
archivePrefix = {arXiv},
       eprint = {1606.08415},
 primaryClass = {cs.LG},
       adsurl = {https://ui.adsabs.harvard.edu/abs/2016arXiv160608415H}
}

@ARTICLE{Angulo_2021,
       author = {{Angulo}, Raul E. and {Zennaro}, Matteo and {Contreras}, Sergio and {Aric{\`o}}, Giovanni and {Pellejero-Iba{\~n}ez}, Marcos and {St{\"u}cker}, Jens},
        title = "{The BACCO simulation project: exploiting the full power of large-scale structure for cosmology}",
      journal = {\mnras},
         year = 2021,
        month = nov,
       volume = {507},
       number = {4},
        pages = {5869-5881},
          doi = {10.1093/mnras/stab2018},
archivePrefix = {arXiv},
       eprint = {2004.06245},
 primaryClass = {astro-ph.CO},
       adsurl = {https://ui.adsabs.harvard.edu/abs/2021MNRAS.507.5869A}
}

@ARTICLE{Genel_2019,
       author = {{Genel}, Shy and {Bryan}, Greg L. and {Springel}, Volker and {Hernquist}, Lars and {Nelson}, Dylan and {Pillepich}, Annalisa and {Weinberger}, Rainer and {Pakmor}, R{\"u}diger and {Marinacci}, Federico and {Vogelsberger}, Mark},
        title = "{A Quantification of the Butterfly Effect in Cosmological Simulations and Implications for Galaxy Scaling Relations}",
      journal = {\apj},
         year = 2019,
        month = jan,
       volume = {871},
       number = {1},
          eid = {21},
        pages = {21},
          doi = {10.3847/1538-4357/aaf4bb},
archivePrefix = {arXiv},
       eprint = {1807.07084},
 primaryClass = {astro-ph.GA},
       adsurl = {https://ui.adsabs.harvard.edu/abs/2019ApJ...871...21G}
}

@ARTICLE{Contreras_2020,
       author = {{Contreras}, S. and {Angulo}, R.~E. and {Zennaro}, M. and {Aric{\`o}}, G. and {Pellejero-Iba{\~n}ez}, M.},
        title = "{3 per cent-accurate predictions for the clustering of dark matter, haloes, and subhaloes, over a wide range of cosmologies and scales}",
      journal = {\mnras},
         year = 2020,
        month = dec,
       volume = {499},
       number = {4},
        pages = {4905-4917},
          doi = {10.1093/mnras/staa3117},
archivePrefix = {arXiv},
       eprint = {2001.03176},
 primaryClass = {astro-ph.CO},
       adsurl = {https://ui.adsabs.harvard.edu/abs/2020MNRAS.499.4905C}
}

@BOOK{Peebles_1980,
       author = {{Peebles}, P.~J.~E.},
        title = "{The large-scale structure of the universe}",
         year = 1980,
       adsurl = {https://ui.adsabs.harvard.edu/abs/1980lssu.book.....P}
}

@ARTICLE{Mokeddem_2025,
       author = {{Mokeddem}, Rahima and {Bizarria}, Bruno B. and {Zhang}, Jiajun and {Hip{\'o}lito-Ricaldi}, Wiliam S. and {Wuensche}, Carlos Alexandre and {Abdalla}, Elcio and {Abdalla}, Filipe B. and {Queiroz}, Amilcar R. and {Villela}, Thyrso and {Wang}, Bin and {Feng}, Chang and {Gurj{\~a}o}, Edmar C. and {Marins}, Alessandro},
        title = "{Cosmological remapping for efficient generation of 21 cm intensity mapping mocks}",
      journal = {\jcap},
         year = 2026,
        month = jan,
       volume = {2026},
       number = {1},
          eid = {019},
        pages = {019},
          doi = {10.1088/1475-7516/2026/01/019},
archivePrefix = {arXiv},
       eprint = {2506.14588},
 primaryClass = {astro-ph.CO},
       adsurl = {https://ui.adsabs.harvard.edu/abs/2026JCAP...01..019M}
}

@ARTICLE{Gal_2015,
       author = {{Gal}, Yarin and {Ghahramani}, Zoubin},
        title = "{Dropout as a Bayesian Approximation: Representing Model Uncertainty in Deep Learning}",
      journal = {arXiv e-prints},
         year = 2015,
        month = jun,
          eid = {arXiv:1506.02142},
        pages = {arXiv:1506.02142},
          doi = {10.48550/arXiv.1506.02142},
archivePrefix = {arXiv},
       eprint = {1506.02142},
 primaryClass = {stat.ML},
       adsurl = {https://ui.adsabs.harvard.edu/abs/2015arXiv150602142G}
}

@ARTICLE{Loshchilov_2016,
       author = {{Loshchilov}, Ilya and {Hutter}, Frank},
        title = "{SGDR: Stochastic Gradient Descent with Warm Restarts}",
      journal = {arXiv e-prints},
         year = 2016,
        month = aug,
          eid = {arXiv:1608.03983},
        pages = {arXiv:1608.03983},
          doi = {10.48550/arXiv.1608.03983},
archivePrefix = {arXiv},
       eprint = {1608.03983},
 primaryClass = {cs.LG},
       adsurl = {https://ui.adsabs.harvard.edu/abs/2016arXiv160803983L}
}

@ARTICLE{Arico_2021,
       author = {{Aric{\`o}}, Giovanni and {Angulo}, Raul E. and {Contreras}, Sergio and {Ondaro-Mallea}, Lurdes and {Pellejero-Iba{\~n}ez}, Marcos and {Zennaro}, Matteo},
        title = "{The BACCO simulation project: a baryonification emulator with neural networks}",
      journal = {\mnras},
         year = 2021,
        month = sep,
       volume = {506},
       number = {3},
        pages = {4070-4082},
          doi = {10.1093/mnras/stab1911},
archivePrefix = {arXiv},
       eprint = {2011.15018},
 primaryClass = {astro-ph.CO},
       adsurl = {https://ui.adsabs.harvard.edu/abs/2021MNRAS.506.4070A}
}

@ARTICLE{Zennaro_2023,
       author = {{Zennaro}, Matteo and {Angulo}, Raul E. and {Pellejero-Ib{\'a}{\~n}ez}, Marcos and {St{\"u}cker}, Jens and {Contreras}, Sergio and {Aric{\`o}}, Giovanni},
        title = "{The BACCO simulation project: biased tracers in real space}",
      journal = {\mnras},
         year = 2023,
        month = sep,
       volume = {524},
       number = {2},
        pages = {2407-2419},
          doi = {10.1093/mnras/stad2008},
archivePrefix = {arXiv},
       eprint = {2101.12187},
 primaryClass = {astro-ph.CO},
       adsurl = {https://ui.adsabs.harvard.edu/abs/2023MNRAS.524.2407Z}
}

@ARTICLE{ibanez_2023,
       author = {{Pellejero Iba{\~n}ez}, Marcos and {Angulo}, Raul E. and {Zennaro}, Matteo and {St{\"u}cker}, Jens and {Contreras}, Sergio and {Aric{\`o}}, Giovanni and {Maion}, Francisco},
        title = "{The bacco simulation project: bacco hybrid Lagrangian bias expansion model in redshift space}",
      journal = {\mnras},
         year = 2023,
        month = apr,
       volume = {520},
       number = {3},
        pages = {3725-3741},
          doi = {10.1093/mnras/stad368},
archivePrefix = {arXiv},
       eprint = {2207.06437},
 primaryClass = {astro-ph.CO},
       adsurl = {https://ui.adsabs.harvard.edu/abs/2023MNRAS.520.3725P}
}

@INPROCEEDINGS{Lemos_2024,
       author = {{Lemos}, Pablo and {Sharief}, Sammy and {Malkin}, Nikolay and {Salhi}, Salma and {Stone}, Connor and {Perreault-Levasseur}, Laurence and {Hezaveh}, Yashar},
        title = "{PQMass: Probabilistic Assessment of the Quality of Generative Models using Probability Mass Estimation}",
    booktitle = {The Thirteenth International Conference on Learning Representations},
         year = 2025,
        month = jan,
          eid = {7567},
        pages = {7567},
          doi = {10.48550/arXiv.2402.04355},
archivePrefix = {arXiv},
       eprint = {2402.04355},
 primaryClass = {stat.ML},
       adsurl = {https://ui.adsabs.harvard.edu/abs/2025iclr.conf.7567L}
}

@ARTICLE{Bairagi_2025,
       author = {{Bairagi}, Anirban and {Wandelt}, Benjamin and {Villaescusa-Navarro}, Francisco},
        title = "{The BIG SOBOL SEQUENCE: How many simulations do we need for simulation-based inference in cosmology?}",
      journal = {\aap},
         year = 2025,
        month = nov,
       volume = {703},
          eid = {A301},
        pages = {A301},
          doi = {10.1051/0004-6361/202554602},
archivePrefix = {arXiv},
       eprint = {2503.13755},
 primaryClass = {astro-ph.CO},
       adsurl = {https://ui.adsabs.harvard.edu/abs/2025A&A...703A.301B}
}

@ARTICLE{Strauss_2002,
       author = {{Strauss}, Michael A. and {Weinberg}, David H. and {Lupton}, Robert H. and {Narayanan}, Vijay K. and {Annis}, James and {Bernardi}, Mariangela and {Blanton}, Michael and {Burles}, Scott and {Connolly}, A.~J. and {Dalcanton}, Julianne and {Doi}, Mamoru and {Eisenstein}, Daniel and {Frieman}, Joshua A. and {Fukugita}, Masataka and {Gunn}, James E. and {Ivezi{\'c}}, {\v{Z}}eljko and {Kent}, Stephen and {Kim}, Rita S.~J. and {Knapp}, G.~R. and {Kron}, Richard G. and {Munn}, Jeffrey A. and {Newberg}, Heidi Jo and {Nichol}, R.~C. and {Okamura}, Sadanori and {Quinn}, Thomas R. and {Richmond}, Michael W. and {Schlegel}, David J. and {Shimasaku}, Kazuhiro and {SubbaRao}, Mark and {Szalay}, Alexander S. and {Vanden Berk}, Dan and {Vogeley}, Michael S. and {Yanny}, Brian and {Yasuda}, Naoki and {York}, Donald G. and {Zehavi}, Idit},
        title = "{Spectroscopic Target Selection in the Sloan Digital Sky Survey: The Main Galaxy Sample}",
      journal = {\aj},
         year = 2002,
        month = sep,
       volume = {124},
       number = {3},
        pages = {1810-1824},
          doi = {10.1086/342343},
archivePrefix = {arXiv},
       eprint = {astro-ph/0206225},
 primaryClass = {astro-ph},
       adsurl = {https://ui.adsabs.harvard.edu/abs/2002AJ....124.1810S}
}

@INPROCEEDINGS{Levi_2019,
       author = {{Levi}, Michael and {Allen}, Lori E. and {Raichoor}, Anand and {Baltay}, Charles and {BenZvi}, Segev and {Beutler}, Florian and {Bolton}, Adam and {Castander}, Francisco J. and {Chuang}, Chia-Hsun and {Cooper}, Andrew and {Cuby}, Jean-Gabriel and {Dey}, Arjun and {Eisenstein}, Daniel and {Fan}, Xiaohui and {Flaugher}, Brenna and {Frenk}, Carlos and {Gonzalez-Morales}, Alma X. and {Graur}, Or and {Guy}, Julien and {Habib}, Salman and {Honscheid}, Klaus and {Juneau}, Stephanie and {Kneib}, Jean-Paul and {Lahav}, Ofer and {Lang}, Dustin and {Leauthaud}, Alexie and {Lusso}, Betta and {de la Macorra}, Axel and {Manera}, Marc and {Martini}, Paul and {Mao}, Shude and {Newman}, Jeffrey A. and {Palanque-Delabrouille}, Nathalie and {Percival}, Will J. and {Allende Prieto}, Carlos and {Rockosi}, Constance M. and {Ruhlmann-Kleider}, Vanina and {Schlegel}, David and {Seo}, Hee-Jong and {Song}, Yong-Seon and {Tarle}, Greg and {Wechsler}, Risa and {Weinberg}, David and {Yeche}, Christophe and {Zu}, Ying},
        title = "{The Dark Energy Spectroscopic Instrument (DESI)}",
    booktitle = {Bulletin of the American Astronomical Society},
         year = 2019,
       volume = {51},
        month = sep,
          eid = {57},
        pages = {57},
          doi = {10.48550/arXiv.1907.10688},
archivePrefix = {arXiv},
       eprint = {1907.10688},
 primaryClass = {astro-ph.IM},
       adsurl = {https://ui.adsabs.harvard.edu/abs/2019BAAS...51g..57L}
}

@ARTICLE{Euclid,
       author = {{Euclid Collaboration} and {Mellier}, Y. and {Abdurro'uf} and {Acevedo Barroso}, J.~A. and {Ach{\'u}carro}, A. and {Adamek}, J. and {Adam}, R. and {Addison}, G.~E. and {Aghanim}, N. and {Aguena}, M. and et al.},
        title = "{Euclid: I. Overview of the Euclid mission}",
      journal = {\aap},
         year = 2025,
        month = may,
       volume = {697},
          eid = {A1},
        pages = {A1},
          doi = {10.1051/0004-6361/202450810},
archivePrefix = {arXiv},
       eprint = {2405.13491},
 primaryClass = {astro-ph.CO},
       adsurl = {https://ui.adsabs.harvard.edu/abs/2025A&A...697A...1E}
}

@ARTICLE{Zeljko_2019,
       author = {{Ivezi{\'c}}, {\v{Z}}eljko and {Kahn}, Steven M. and {Tyson}, J. Anthony and {Abel}, Bob and {Acosta}, Emily and {Allsman}, Robyn and {Alonso}, David and {AlSayyad}, Yusra and {Anderson}, Scott F. and {Andrew}, John and et al.},
        title = "{LSST: From Science Drivers to Reference Design and Anticipated Data Products}",
      journal = {\apj},
         year = 2019,
        month = mar,
       volume = {873},
       number = {2},
          eid = {111},
        pages = {111},
          doi = {10.3847/1538-4357/ab042c},
archivePrefix = {arXiv},
       eprint = {0805.2366},
 primaryClass = {astro-ph},
       adsurl = {https://ui.adsabs.harvard.edu/abs/2019ApJ...873..111I}
}

@ARTICLE{Planck_2020,
       author = {{Planck Collaboration} and {Aghanim}, N. and {Akrami}, Y. and {Ashdown}, M. and {Aumont}, J. and {Baccigalupi}, C. and {Ballardini}, M. and {Banday}, A.~J. and {Barreiro}, R.~B. and {Bartolo}, N. and {Basak}, S. and {Battye}, R. and {Benabed}, K. and {Bernard}, J.-P. and {Bersanelli}, M. and {Bielewicz}, P. and {Bock}, J.~J. and {Bond}, J.~R. and {Borrill}, J. and {Bouchet}, F.~R. and {Boulanger}, F. and {Bucher}, M. and {Burigana}, C. and {Butler}, R.~C. and {Calabrese}, E. and {Cardoso}, J.-F. and {Carron}, J. and {Challinor}, A. and {Chiang}, H.~C. and {Chluba}, J. and {Colombo}, L.~P.~L. and {Combet}, C. and {Contreras}, D. and {Crill}, B.~P. and {Cuttaia}, F. and {de Bernardis}, P. and {de Zotti}, G. and {Delabrouille}, J. and {Delouis}, J.-M. and {Di Valentino}, E. and {Diego}, J.~M. and {Dor{\'e}}, O. and {Douspis}, M. and {Ducout}, A. and {Dupac}, X. and {Dusini}, S. and {Efstathiou}, G. and {Elsner}, F. and {En{\ss}lin}, T.~A. and {Eriksen}, H.~K. and {Fantaye}, Y. and {Farhang}, M. and {Fergusson}, J. and {Fernandez-Cobos}, R. and {Finelli}, F. and {Forastieri}, F. and {Frailis}, M. and {Fraisse}, A.~A. and {Franceschi}, E. and {Frolov}, A. and {Galeotta}, S. and {Galli}, S. and {Ganga}, K. and {G{\'e}nova-Santos}, R.~T. and {Gerbino}, M. and {Ghosh}, T. and {Gonz{\'a}lez-Nuevo}, J. and {G{\'o}rski}, K.~M. and {Gratton}, S. and {Gruppuso}, A. and {Gudmundsson}, J.~E. and {Hamann}, J. and {Handley}, W. and {Hansen}, F.~K. and {Herranz}, D. and {Hildebrandt}, S.~R. and {Hivon}, E. and {Huang}, Z. and {Jaffe}, A.~H. and {Jones}, W.~C. and {Karakci}, A. and {Keih{\"a}nen}, E. and {Keskitalo}, R. and {Kiiveri}, K. and {Kim}, J. and {Kisner}, T.~S. and {Knox}, L. and {Krachmalnicoff}, N. and {Kunz}, M. and {Kurki-Suonio}, H. and {Lagache}, G. and {Lamarre}, J.-M. and {Lasenby}, A. and {Lattanzi}, M. and {Lawrence}, C.~R. and {Le Jeune}, M. and {Lemos}, P. and {Lesgourgues}, J. and {Levrier}, F. and {Lewis}, A. and {Liguori}, M. and {Lilje}, P.~B. and {Lilley}, M. and {Lindholm}, V. and {L{\'o}pez-Caniego}, M. and {Lubin}, P.~M. and {Ma}, Y.-Z. and {Mac{\'\i}as-P{\'e}rez}, J.~F. and {Maggio}, G. and {Maino}, D. and {Mandolesi}, N. and {Mangilli}, A. and {Marcos-Caballero}, A. and {Maris}, M. and {Martin}, P.~G. and {Martinelli}, M. and {Mart{\'\i}nez-Gonz{\'a}lez}, E. and {Matarrese}, S. and {Mauri}, N. and {McEwen}, J.~D. and {Meinhold}, P.~R. and {Melchiorri}, A. and {Mennella}, A. and {Migliaccio}, M. and {Millea}, M. and {Mitra}, S. and {Miville-Desch{\^e}nes}, M.-A. and {Molinari}, D. and {Montier}, L. and {Morgante}, G. and {Moss}, A. and {Natoli}, P. and {N{\o}rgaard-Nielsen}, H.~U. and {Pagano}, L. and {Paoletti}, D. and {Partridge}, B. and {Patanchon}, G. and {Peiris}, H.~V. and {Perrotta}, F. and {Pettorino}, V. and {Piacentini}, F. and {Polastri}, L. and {Polenta}, G. and {Puget}, J.-L. and {Rachen}, J.~P. and {Reinecke}, M. and {Remazeilles}, M. and {Renzi}, A. and {Rocha}, G. and {Rosset}, C. and {Roudier}, G. and {Rubi{\~n}o-Mart{\'\i}n}, J.~A. and {Ruiz-Granados}, B. and {Salvati}, L. and {Sandri}, M. and {Savelainen}, M. and {Scott}, D. and {Shellard}, E.~P.~S. and {Sirignano}, C. and {Sirri}, G. and {Spencer}, L.~D. and {Sunyaev}, R. and {Suur-Uski}, A.-S. and {Tauber}, J.~A. and {Tavagnacco}, D. and {Tenti}, M. and {Toffolatti}, L. and {Tomasi}, M. and {Trombetti}, T. and {Valenziano}, L. and {Valiviita}, J. and {Van Tent}, B. and {Vibert}, L. and {Vielva}, P. and {Villa}, F. and {Vittorio}, N. and {Wandelt}, B.~D. and {Wehus}, I.~K. and {White}, M. and {White}, S.~D.~M. and {Zacchei}, A. and {Zonca}, A.},
        title = "{Planck 2018 results. VI. Cosmological parameters}",
      journal = {\aap},
         year = 2020,
        month = sep,
       volume = {641},
          eid = {A6},
        pages = {A6},
          doi = {10.1051/0004-6361/201833910},
archivePrefix = {arXiv},
       eprint = {1807.06209},
 primaryClass = {astro-ph.CO},
       adsurl = {https://ui.adsabs.harvard.edu/abs/2020A&A...641A...6P}
}

@article{Hoffman_2014,
  author  = {{Hoffman}, Matthew D. and {Gelman}, Andrew},
  title   = {The No-U-Turn Sampler: Adaptively Setting Path Lengths in Hamiltonian Monte Carlo},
  journal = {Journal of Machine Learning Research},
  year    = {2014},
  volume  = {15},
  number  = {47},
  pages   = {1593--1623},
  url     = {http://jmlr.org/papers/v15/hoffman14a.html}
}

@ARTICLE{Lovell_2025,
       author = {{Lovell}, Christopher C. and {Starkenburg}, Tjitske and {Ho}, Matthew and {Angl{\'e}s-Alc{\'a}zar}, Daniel and {Dav{\'e}}, Romeel and {Gabrielpillai}, Austen and {Iyer}, Kartheik G. and {Matthews}, Alice E. and {Roper}, William J. and {Somerville}, Rachel S. and {Sommovigo}, Laura and {Villaescusa-Navarro}, Francisco},
        title = "{Learning the Universe: cosmological and astrophysical parameter inference with galaxy luminosity functions and colours}",
      journal = {\mnras},
         year = 2025,
        month = dec,
       volume = {544},
       number = {4},
        pages = {3949-3979},
          doi = {10.1093/mnras/staf1888},
archivePrefix = {arXiv},
       eprint = {2411.13960},
 primaryClass = {astro-ph.GA},
       adsurl = {https://ui.adsabs.harvard.edu/abs/2025MNRAS.544.3949L}
}

@ARTICLE{Springel_2018,
       author = {{Springel}, Volker and {Pakmor}, R{\"u}diger and {Pillepich}, Annalisa and {Weinberger}, Rainer and {Nelson}, Dylan and {Hernquist}, Lars and {Vogelsberger}, Mark and {Genel}, Shy and {Torrey}, Paul and {Marinacci}, Federico and {Naiman}, Jill},
        title = "{First results from the IllustrisTNG simulations: matter and galaxy clustering}",
      journal = {\mnras},
         year = 2018,
        month = mar,
       volume = {475},
       number = {1},
        pages = {676-698},
          doi = {10.1093/mnras/stx3304},
archivePrefix = {arXiv},
       eprint = {1707.03397},
 primaryClass = {astro-ph.GA},
       adsurl = {https://ui.adsabs.harvard.edu/abs/2018MNRAS.475..676S}
}

@ARTICLE{Pillepich_2018,
       author = {{Pillepich}, Annalisa and {Springel}, Volker and {Nelson}, Dylan and {Genel}, Shy and {Naiman}, Jill and {Pakmor}, R{\"u}diger and {Hernquist}, Lars and {Torrey}, Paul and {Vogelsberger}, Mark and {Weinberger}, Rainer and {Marinacci}, Federico},
        title = "{Simulating galaxy formation with the IllustrisTNG model}",
      journal = {\mnras},
         year = 2018,
        month = jan,
       volume = {473},
       number = {3},
        pages = {4077-4106},
          doi = {10.1093/mnras/stx2656},
archivePrefix = {arXiv},
       eprint = {1703.02970},
 primaryClass = {astro-ph.GA},
       adsurl = {https://ui.adsabs.harvard.edu/abs/2018MNRAS.473.4077P}
}

@ARTICLE{Schaye_2015,
       author = {{Schaye}, Joop and {Crain}, Robert A. and {Bower}, Richard G. and {Furlong}, Michelle and {Schaller}, Matthieu and {Theuns}, Tom and {Dalla Vecchia}, Claudio and {Frenk}, Carlos S. and {McCarthy}, I.~G. and {Helly}, John C. and {Jenkins}, Adrian and {Rosas-Guevara}, Y.~M. and {White}, Simon D.~M. and {Baes}, Maarten and {Booth}, C.~M. and {Camps}, Peter and {Navarro}, Julio F. and {Qu}, Yan and {Rahmati}, Alireza and {Sawala}, Till and {Thomas}, Peter A. and {Trayford}, James},
        title = "{The EAGLE project: simulating the evolution and assembly of galaxies and their environments}",
      journal = {\mnras},
         year = 2015,
        month = jan,
       volume = {446},
       number = {1},
        pages = {521-554},
          doi = {10.1093/mnras/stu2058},
archivePrefix = {arXiv},
       eprint = {1407.7040},
 primaryClass = {astro-ph.GA},
       adsurl = {https://ui.adsabs.harvard.edu/abs/2015MNRAS.446..521S}
}

@ARTICLE{Dave_2019,
       author = {{Dav{\'e}}, Romeel and {Angl{\'e}s-Alc{\'a}zar}, Daniel and {Narayanan}, Desika and {Li}, Qi and {Rafieferantsoa}, Mika H. and {Appleby}, Sarah},
        title = "{SIMBA: Cosmological simulations with black hole growth and feedback}",
      journal = {\mnras},
         year = 2019,
        month = jun,
       volume = {486},
       number = {2},
        pages = {2827-2849},
          doi = {10.1093/mnras/stz937},
archivePrefix = {arXiv},
       eprint = {1901.10203},
 primaryClass = {astro-ph.GA},
       adsurl = {https://ui.adsabs.harvard.edu/abs/2019MNRAS.486.2827D}
}

@ARTICLE{Villaescusa-Navarro_2020,
       author = {{Villaescusa-Navarro}, Francisco and {Hahn}, ChangHoon and {Massara}, Elena and {Banerjee}, Arka and {Delgado}, Ana Maria and {Ramanah}, Doogesh Kodi and {Charnock}, Tom and {Giusarma}, Elena and {Li}, Yin and {Allys}, Erwan and {Brochard}, Antoine and {Uhlemann}, Cora and {Chiang}, Chi-Ting and {He}, Siyu and {Pisani}, Alice and {Obuljen}, Andrej and {Feng}, Yu and {Castorina}, Emanuele and {Contardo}, Gabriella and {Kreisch}, Christina D. and {Nicola}, Andrina and {Alsing}, Justin and {Scoccimarro}, Roman and {Verde}, Licia and {Viel}, Matteo and {Ho}, Shirley and {Mallat}, Stephane and {Wandelt}, Benjamin and {Spergel}, David N.},
        title = "{The Quijote Simulations}",
      journal = {\apjs},
         year = 2020,
        month = sep,
       volume = {250},
       number = {1},
          eid = {2},
        pages = {2},
          doi = {10.3847/1538-4365/ab9d82},
archivePrefix = {arXiv},
       eprint = {1909.05273},
 primaryClass = {astro-ph.CO},
       adsurl = {https://ui.adsabs.harvard.edu/abs/2020ApJS..250....2V}
}

@ARTICLE{Dubois_2014,
       author = {{Dubois}, Y. and {Pichon}, C. and {Welker}, C. and {Le Borgne}, D. and {Devriendt}, J. and {Laigle}, C. and {Codis}, S. and {Pogosyan}, D. and {Arnouts}, S. and {Benabed}, K. and {Bertin}, E. and {Blaizot}, J. and {Bouchet}, F. and {Cardoso}, J.-F. and {Colombi}, S. and {de Lapparent}, V. and {Desjacques}, V. and {Gavazzi}, R. and {Kassin}, S. and {Kimm}, T. and {McCracken}, H. and {Milliard}, B. and {Peirani}, S. and {Prunet}, S. and {Rouberol}, S. and {Silk}, J. and {Slyz}, A. and {Sousbie}, T. and {Teyssier}, R. and {Tresse}, L. and {Treyer}, M. and {Vibert}, D. and {Volonteri}, M.},
        title = "{Dancing in the dark: galactic properties trace spin swings along the cosmic web}",
      journal = {\mnras},
         year = 2014,
        month = oct,
       volume = {444},
       number = {2},
        pages = {1453-1468},
          doi = {10.1093/mnras/stu1227},
archivePrefix = {arXiv},
       eprint = {1402.1165},
 primaryClass = {astro-ph.CO},
       adsurl = {https://ui.adsabs.harvard.edu/abs/2014MNRAS.444.1453D}
}

@ARTICLE{Alsing_2019,
       author = {{Alsing}, Justin and {Charnock}, Tom and {Feeney}, Stephen and {Wandelt}, Benjamin},
        title = "{Fast likelihood-free cosmology with neural density estimators and active learning}",
      journal = {\mnras},
         year = 2019,
        month = sep,
       volume = {488},
       number = {3},
        pages = {4440-4458},
          doi = {10.1093/mnras/stz1960},
archivePrefix = {arXiv},
       eprint = {1903.00007},
 primaryClass = {astro-ph.CO},
       adsurl = {https://ui.adsabs.harvard.edu/abs/2019MNRAS.488.4440A}
}

@ARTICLE{Chandro-Gomez_2025,
       author = {{Chandro-G{\'o}mez}, {\'A}ngel and {Lagos}, Claudia del P. and {Power}, Chris and {Forouhar Moreno}, Victor J. and {Helly}, John C. and {Lacey}, Cedric G. and {McGibbon}, Robert J. and {Schaller}, Matthieu and {Schaye}, Joop},
        title = "{On the accuracy of dark matter halo merger trees and the consequences for semi-analytic models of galaxy formation}",
      journal = {\mnras},
         year = 2025,
        month = may,
       volume = {539},
       number = {2},
        pages = {776-807},
          doi = {10.1093/mnras/staf519},
archivePrefix = {arXiv},
       eprint = {2501.07677},
 primaryClass = {astro-ph.GA},
       adsurl = {https://ui.adsabs.harvard.edu/abs/2025MNRAS.539..776C}
}

@ARTICLE{Benson_2016,
       author = {{Benson}, Andrew J. and {Cannella}, Chris and {Cole}, Shaun},
        title = "{Achieving convergence in galaxy formation models by augmenting N-body merger trees}",
      journal = {Computational Astrophysics and Cosmology},
         year = 2016,
        month = aug,
       volume = {3},
       number = {1},
          eid = {3},
        pages = {3},
          doi = {10.1186/s40668-016-0016-3},
archivePrefix = {arXiv},
       eprint = {1604.02147},
 primaryClass = {astro-ph.GA},
       adsurl = {https://ui.adsabs.harvard.edu/abs/2016ComAC...3....3B}
}

@ARTICLE{Saoulis_2024,
       author = {{Saoulis}, Alex A. and {Piras}, Davide and {Jeffrey}, Niall and {Mancini}, Alessio Spurio and {Ferreira}, Ana M.~G. and {Joachimi}, Benjamin},
        title = "{Transfer learning for multifidelity simulation-based inference in cosmology}",
      journal = {\mnras},
         year = 2025,
        month = sep,
       volume = {542},
       number = {4},
        pages = {3231-3245},
          doi = {10.1093/mnras/staf1436},
archivePrefix = {arXiv},
       eprint = {2505.21215},
 primaryClass = {astro-ph.CO},
       adsurl = {https://ui.adsabs.harvard.edu/abs/2025MNRAS.542.3231S}
}

@ARTICLE{Massara_2025,
       author = {{Massara}, Elena and {Hahn}, ChangHoon and {Eickenberg}, Michael and {Ho}, Shirley and {Hou}, Jiamin and {Lemos}, Pablo and {Modi}, Chirag and {Moradinezhad Dizgah}, Azadeh and {Parker}, Liam and {R{\'e}galdo-Saint Blancard}, Bruno},
        title = "{Cosmological constraints using simulation-based inference of galaxy clustering with marked power spectra}",
      journal = {\prd},
         year = 2025,
        month = oct,
       volume = {112},
       number = {8},
          eid = {083507},
        pages = {083507},
          doi = {10.1103/9l4n-3pcx},
archivePrefix = {arXiv},
       eprint = {2404.04228},
 primaryClass = {astro-ph.CO},
       adsurl = {https://ui.adsabs.harvard.edu/abs/2025PhRvD.112h3507M}
}

@ARTICLE{White_1991,
       author = {{White}, Simon D.~M. and {Frenk}, Carlos S.},
        title = "{Galaxy Formation through Hierarchical Clustering}",
      journal = {\apj},
         year = 1991,
        month = sep,
       volume = {379},
        pages = {52},
          doi = {10.1086/170483},
       adsurl = {https://ui.adsabs.harvard.edu/abs/1991ApJ...379...52W}
}

@ARTICLE{Kauffmann_1993,
       author = {{Kauffmann}, G. and {White}, S.~D.~M. and {Guiderdoni}, B.},
        title = "{The formation and evolution of galaxies within merging dark matter haloes.}",
      journal = {\mnras},
         year = 1993,
        month = sep,
       volume = {264},
        pages = {201-218},
          doi = {10.1093/mnras/264.1.201},
       adsurl = {https://ui.adsabs.harvard.edu/abs/1993MNRAS.264..201K}
}

@ARTICLE{Cole_2000,
       author = {{Cole}, Shaun and {Lacey}, Cedric G. and {Baugh}, Carlton M. and {Frenk}, Carlos S.},
        title = "{Hierarchical galaxy formation}",
      journal = {\mnras},
         year = 2000,
        month = nov,
       volume = {319},
       number = {1},
        pages = {168-204},
          doi = {10.1046/j.1365-8711.2000.03879.x},
archivePrefix = {arXiv},
       eprint = {astro-ph/0007281},
 primaryClass = {astro-ph},
       adsurl = {https://ui.adsabs.harvard.edu/abs/2000MNRAS.319..168C}
}

@ARTICLE{Zennaro_2019,
       author = {{Zennaro}, Matteo and {Angulo}, Ra{\'u}l E. and {Aric{\`o}}, Giovanni and {Contreras}, Sergio and {Pellejero-Ib{\'a}{\~n}ez}, Marcos},
        title = "{How to add massive neutrinos to your {\ensuremath{\Lambda}}CDM simulation - extending cosmology rescaling algorithms}",
      journal = {\mnras},
         year = 2019,
        month = nov,
       volume = {489},
       number = {4},
        pages = {5938-5951},
          doi = {10.1093/mnras/stz2612},
archivePrefix = {arXiv},
       eprint = {1905.08696},
 primaryClass = {astro-ph.CO},
       adsurl = {https://ui.adsabs.harvard.edu/abs/2019MNRAS.489.5938Z}
}

@ARTICLE{Ni_2023,
       author = {{Ni}, Yueying and {Genel}, Shy and {Angl{\'e}s-Alc{\'a}zar}, Daniel and {Villaescusa-Navarro}, Francisco and {Jo}, Yongseok and {Bird}, Simeon and {Di Matteo}, Tiziana and {Croft}, Rupert and {Chen}, Nianyi and {de Santi}, Natal{\'\i} S.~M. and {Gebhardt}, Matthew and {Shao}, Helen and {Pandey}, Shivam and {Hernquist}, Lars and {Dave}, Romeel},
        title = "{The CAMELS Project: Expanding the Galaxy Formation Model Space with New ASTRID and 28-parameter TNG and SIMBA Suites}",
      journal = {\apj},
         year = 2023,
        month = dec,
       volume = {959},
       number = {2},
          eid = {136},
        pages = {136},
          doi = {10.3847/1538-4357/ad022a},
archivePrefix = {arXiv},
       eprint = {2304.02096},
 primaryClass = {astro-ph.CO},
       adsurl = {https://ui.adsabs.harvard.edu/abs/2023ApJ...959..136N}
}

@ARTICLE{Springel_2010,
       author = {{Springel}, Volker},
        title = "{E pur si muove: Galilean-invariant cosmological hydrodynamical simulations on a moving mesh}",
      journal = {\mnras},
         year = 2010,
        month = jan,
       volume = {401},
       number = {2},
        pages = {791-851},
          doi = {10.1111/j.1365-2966.2009.15715.x},
archivePrefix = {arXiv},
       eprint = {0901.4107},
 primaryClass = {astro-ph.CO},
       adsurl = {https://ui.adsabs.harvard.edu/abs/2010MNRAS.401..791S}
}

@ARTICLE{Schaye_2023,
       author = {{Schaye}, Joop and {Kugel}, Roi and {Schaller}, Matthieu and {Helly}, John C. and {Braspenning}, Joey and {Elbers}, Willem and {McCarthy}, Ian G. and {van Daalen}, Marcel P. and {Vandenbroucke}, Bert and {Frenk}, Carlos S. and {Kwan}, Juliana and {Salcido}, Jaime and {Bah{\'e}}, Yannick M. and {Borrow}, Josh and {Chaikin}, Evgenii and {Hahn}, Oliver and {Hu{\v{s}}ko}, Filip and {Jenkins}, Adrian and {Lacey}, Cedric G. and {Nobels}, Folkert S.~J.},
        title = "{The FLAMINGO project: cosmological hydrodynamical simulations for large-scale structure and galaxy cluster surveys}",
      journal = {\mnras},
         year = 2023,
        month = dec,
       volume = {526},
       number = {4},
        pages = {4978-5020},
          doi = {10.1093/mnras/stad2419},
archivePrefix = {arXiv},
       eprint = {2306.04024},
 primaryClass = {astro-ph.CO},
       adsurl = {https://ui.adsabs.harvard.edu/abs/2023MNRAS.526.4978S}
}

@ARTICLE{Pakmor_2023,
       author = {{Pakmor}, R{\"u}diger and {Springel}, Volker and {Coles}, Jonathan P. and {Guillet}, Thomas and {Pfrommer}, Christoph and {Bose}, Sownak and {Barrera}, Monica and {Delgado}, Ana Maria and {Ferlito}, Fulvio and {Frenk}, Carlos and {Hadzhiyska}, Boryana and {Hern{\'a}ndez-Aguayo}, C{\'e}sar and {Hernquist}, Lars and {Kannan}, Rahul and {White}, Simon D.~M.},
        title = "{The MillenniumTNG Project: the hydrodynamical full physics simulation and a first look at its galaxy clusters}",
      journal = {\mnras},
         year = 2023,
        month = sep,
       volume = {524},
       number = {2},
        pages = {2539-2555},
          doi = {10.1093/mnras/stac3620},
archivePrefix = {arXiv},
       eprint = {2210.10060},
 primaryClass = {astro-ph.CO},
       adsurl = {https://ui.adsabs.harvard.edu/abs/2023MNRAS.524.2539P}
}

@ARTICLE{Contreras_2023,
       author = {{Contreras}, Sergio and {Angulo}, Raul E. and {Springel}, Volker and {White}, Simon D.~M. and {Hadzhiyska}, Boryana and {Hernquist}, Lars and {Pakmor}, R{\"u}diger and {Kannan}, Rahul and {Hern{\'a}ndez-Aguayo}, C{\'e}sar and {Barrera}, Monica and {Ferlito}, Fulvio and {Delgado}, Ana Maria and {Bose}, Sownak and {Frenk}, Carlos},
        title = "{The MillenniumTNG Project: inferring cosmology from galaxy clustering with accelerated N-body scaling and subhalo abundance matching}",
      journal = {\mnras},
         year = 2023,
        month = sep,
       volume = {524},
       number = {2},
        pages = {2489-2506},
          doi = {10.1093/mnras/stac3699},
archivePrefix = {arXiv},
       eprint = {2210.10075},
 primaryClass = {astro-ph.GA},
       adsurl = {https://ui.adsabs.harvard.edu/abs/2023MNRAS.524.2489C}
}

@ARTICLE{Pandya_2020,
       author = {{Pandya}, Viraj and {Somerville}, Rachel S. and {Angl{\'e}s-Alc{\'a}zar}, Daniel and {Hayward}, Christopher C. and {Bryan}, Greg L. and {Fielding}, Drummond B. and {Forbes}, John C. and {Burkhart}, Blakesley and {Genel}, Shy and {Hernquist}, Lars and {Kim}, Chang-Goo and {Tonnesen}, Stephanie and {Starkenburg}, Tjitske},
        title = "{First Results from SMAUG: The Need for Preventative Stellar Feedback and Improved Baryon Cycling in Semianalytic Models of Galaxy Formation}",
      journal = {\apj},
         year = 2020,
        month = dec,
       volume = {905},
       number = {1},
          eid = {4},
        pages = {4},
          doi = {10.3847/1538-4357/abc3c1},
archivePrefix = {arXiv},
       eprint = {2006.16317},
 primaryClass = {astro-ph.GA},
       adsurl = {https://ui.adsabs.harvard.edu/abs/2020ApJ...905....4P}
}

@ARTICLE{Oren_2026,
       author = {{Oren}, Yossi and {Pandya}, Viraj and {Somerville}, Rachel S. and {Genel}, Shy and {Omoruyi}, Osase and {Sternberg}, Amiel},
        title = "{The Cosmic Baryon Cycle in IllustrisTNG: Flows of Mass, Energy, and Metals}",
      journal = {\apj},
         year = 2026,
        month = mar,
       volume = {999},
       number = {2},
          eid = {259},
        pages = {259},
          doi = {10.3847/1538-4357/ae41bc},
archivePrefix = {arXiv},
       eprint = {2510.23343},
 primaryClass = {astro-ph.GA},
       adsurl = {https://ui.adsabs.harvard.edu/abs/2026ApJ...999..259O}
}

@ARTICLE{Hadzhiyska_2021,
       author = {{Hadzhiyska}, Boryana and {Liu}, Sonya and {Somerville}, Rachel S. and {Gabrielpillai}, Austen and {Bose}, Sownak and {Eisenstein}, Daniel and {Hernquist}, Lars},
        title = "{Galaxy assembly bias and large-scale distribution: a comparison between IllustrisTNG and a semi-analytic model}",
      journal = {\mnras},
         year = 2021,
        month = nov,
       volume = {508},
       number = {1},
        pages = {698-718},
          doi = {10.1093/mnras/stab2564},
archivePrefix = {arXiv},
       eprint = {2108.00006},
 primaryClass = {astro-ph.CO},
       adsurl = {https://ui.adsabs.harvard.edu/abs/2021MNRAS.508..698H}
}

@ARTICLE{Pandya_2026,
       author = {{Pandya}, Viraj and {Bryan}, Greg L. and {Makinen}, T. Lucas and {Gabrielpillai}, Austen and {Carr}, Christopher and {Fielding}, Drummond B. and {Hernquist}, Lars and {Ho}, Matthew and {Iyer}, Kartheik and {Jespersen}, Christian Kragh and {Koudmani}, Sophie and {Laska}, Marta and {Lemos}, Pablo and {Lovell}, Christopher C. and {Perez}, Lucia A. and {Robinson}, Jr., William F. and {Somerville}, Rachel S. and {Starkenburg}, Tjitske K. and {Stiskalek}, Richard and {Terrazas}, Bryan and {Voit}, G. Mark},
        title = "{Introducing sapphire: Towards Hybrid Physics-Informed, Data-Driven Modeling of Galaxy Formation}",
      journal = {arXiv e-prints},
         year = 2026,
        month = apr,
          eid = {arXiv:2604.06318},
        pages = {arXiv:2604.06318},
          doi = {10.48550/arXiv.2604.06318},
archivePrefix = {arXiv},
       eprint = {2604.06318},
 primaryClass = {astro-ph.GA},
       adsurl = {https://ui.adsabs.harvard.edu/abs/2026arXiv260406318P}
}

@ARTICLE{Bett_2007,
       author = {{Bett}, Philip and {Eke}, Vincent and {Frenk}, Carlos S. and {Jenkins}, Adrian and {Helly}, John and {Navarro}, Julio},
        title = "{The spin and shape of dark matter haloes in the Millennium simulation of a {\ensuremath{\Lambda}} cold dark matter universe}",
      journal = {\mnras},
         year = 2007,
        month = mar,
       volume = {376},
       number = {1},
        pages = {215-232},
          doi = {10.1111/j.1365-2966.2007.11432.x},
archivePrefix = {arXiv},
       eprint = {astro-ph/0608607},
 primaryClass = {astro-ph},
       adsurl = {https://ui.adsabs.harvard.edu/abs/2007MNRAS.376..215B}
}

@ARTICLE{Vitvitska_2002,
       author = {{Vitvitska}, Maya and {Klypin}, Anatoly A. and {Kravtsov}, Andrey V. and {Wechsler}, Risa H. and {Primack}, Joel R. and {Bullock}, James S.},
        title = "{The Origin of Angular Momentum in Dark Matter Halos}",
      journal = {\apj},
         year = 2002,
        month = dec,
       volume = {581},
       number = {2},
        pages = {799-809},
          doi = {10.1086/344361},
archivePrefix = {arXiv},
       eprint = {astro-ph/0105349},
 primaryClass = {astro-ph},
       adsurl = {https://ui.adsabs.harvard.edu/abs/2002ApJ...581..799V}
}

@ARTICLE{DOnghia_2007,
       author = {{D'Onghia}, Elena and {Navarro}, Julio F.},
        title = "{Do mergers spin-up dark matter haloes?}",
      journal = {\mnras},
         year = 2007,
        month = sep,
       volume = {380},
       number = {1},
        pages = {L58-L62},
          doi = {10.1111/j.1745-3933.2007.00348.x},
archivePrefix = {arXiv},
       eprint = {astro-ph/0703195},
 primaryClass = {astro-ph},
       adsurl = {https://ui.adsabs.harvard.edu/abs/2007MNRAS.380L..58D}
}

@ARTICLE{Jespersen_2022,
       author = {{Jespersen}, Christian Kragh and {Cranmer}, Miles and {Melchior}, Peter and {Ho}, Shirley and {Somerville}, Rachel S. and {Gabrielpillai}, Austen},
        title = "{Mangrove: Learning Galaxy Properties from Merger Trees}",
      journal = {\apj},
         year = 2022,
        month = dec,
       volume = {941},
       number = {1},
          eid = {7},
        pages = {7},
          doi = {10.3847/1538-4357/ac9b18},
archivePrefix = {arXiv},
       eprint = {2210.13473},
 primaryClass = {astro-ph.GA},
       adsurl = {https://ui.adsabs.harvard.edu/abs/2022ApJ...941....7J}
}

@ARTICLE{Croton_2006,
       author = {{Croton}, Darren J. and {Springel}, Volker and {White}, Simon D.~M. and {De Lucia}, G. and {Frenk}, C.~S. and {Gao}, L. and {Jenkins}, A. and {Kauffmann}, G. and {Navarro}, J.~F. and {Yoshida}, N.},
        title = "{The many lives of active galactic nuclei: cooling flows, black holes and the luminosities and colours of galaxies}",
      journal = {\mnras},
         year = 2006,
        month = jan,
       volume = {365},
       number = {1},
        pages = {11-28},
          doi = {10.1111/j.1365-2966.2005.09675.x},
archivePrefix = {arXiv},
       eprint = {astro-ph/0508046},
 primaryClass = {astro-ph},
       adsurl = {https://ui.adsabs.harvard.edu/abs/2006MNRAS.365...11C}
}

@ARTICLE{Bower_2006,
       author = {{Bower}, R.~G. and {Benson}, A.~J. and {Malbon}, R. and {Helly}, J.~C. and {Frenk}, C.~S. and {Baugh}, C.~M. and {Cole}, S. and {Lacey}, C.~G.},
        title = "{Breaking the hierarchy of galaxy formation}",
      journal = {\mnras},
         year = 2006,
        month = aug,
       volume = {370},
       number = {2},
        pages = {645-655},
          doi = {10.1111/j.1365-2966.2006.10519.x},
archivePrefix = {arXiv},
       eprint = {astro-ph/0511338},
 primaryClass = {astro-ph},
       adsurl = {https://ui.adsabs.harvard.edu/abs/2006MNRAS.370..645B}
}

@ARTICLE{DeLucia_2007,
       author = {{De Lucia}, Gabriella and {Blaizot}, J{\'e}r{\'e}my},
        title = "{The hierarchical formation of the brightest cluster galaxies}",
      journal = {\mnras},
         year = 2007,
        month = feb,
       volume = {375},
       number = {1},
        pages = {2-14},
          doi = {10.1111/j.1365-2966.2006.11287.x},
archivePrefix = {arXiv},
       eprint = {astro-ph/0606519},
 primaryClass = {astro-ph},
       adsurl = {https://ui.adsabs.harvard.edu/abs/2007MNRAS.375....2D}
}

@ARTICLE{Guo_2011,
       author = {{Guo}, Qi and {White}, Simon and {Boylan-Kolchin}, Michael and {De Lucia}, Gabriella and {Kauffmann}, Guinevere and {Lemson}, Gerard and {Li}, Cheng and {Springel}, Volker and {Weinmann}, Simone},
        title = "{From dwarf spheroidals to cD galaxies: simulating the galaxy population in a {\ensuremath{\Lambda}}CDM cosmology}",
      journal = {\mnras},
         year = 2011,
        month = may,
       volume = {413},
       number = {1},
        pages = {101-131},
          doi = {10.1111/j.1365-2966.2010.18114.x},
archivePrefix = {arXiv},
       eprint = {1006.0106},
 primaryClass = {astro-ph.CO},
       adsurl = {https://ui.adsabs.harvard.edu/abs/2011MNRAS.413..101G}
}

@ARTICLE{Benson_2012,
       author = {{Benson}, Andrew J.},
        title = "{GALACTICUS: A semi-analytic model of galaxy formation}",
      journal = {\na},
         year = 2012,
        month = feb,
       volume = {17},
       number = {2},
        pages = {175-197},
          doi = {10.1016/j.newast.2011.07.004},
archivePrefix = {arXiv},
       eprint = {1008.1786},
 primaryClass = {astro-ph.CO},
       adsurl = {https://ui.adsabs.harvard.edu/abs/2012NewA...17..175B}
}

@ARTICLE{Henriques_2015,
       author = {{Henriques}, Bruno M.~B. and {White}, Simon D.~M. and {Thomas}, Peter A. and {Angulo}, Raul and {Guo}, Qi and {Lemson}, Gerard and {Springel}, Volker and {Overzier}, Roderik},
        title = "{Galaxy formation in the Planck cosmology - I. Matching the observed evolution of star formation rates, colours and stellar masses}",
      journal = {\mnras},
         year = 2015,
        month = aug,
       volume = {451},
       number = {3},
        pages = {2663-2680},
          doi = {10.1093/mnras/stv705},
archivePrefix = {arXiv},
       eprint = {1410.0365},
 primaryClass = {astro-ph.GA},
       adsurl = {https://ui.adsabs.harvard.edu/abs/2015MNRAS.451.2663H}
}

@ARTICLE{Lacey_2016,
       author = {{Lacey}, Cedric G. and {Baugh}, Carlton M. and {Frenk}, Carlos S. and {Benson}, Andrew J. and {Bower}, Richard G. and {Cole}, Shaun and {Gonzalez-Perez}, Violeta and {Helly}, John C. and {Lagos}, Claudia D.~P. and {Mitchell}, Peter D.},
        title = "{A unified multiwavelength model of galaxy formation}",
      journal = {\mnras},
         year = 2016,
        month = nov,
       volume = {462},
       number = {4},
        pages = {3854-3911},
          doi = {10.1093/mnras/stw1888},
archivePrefix = {arXiv},
       eprint = {1509.08473},
 primaryClass = {astro-ph.GA},
       adsurl = {https://ui.adsabs.harvard.edu/abs/2016MNRAS.462.3854L}
}

@ARTICLE{Croton_2016,
       author = {{Croton}, Darren J. and {Stevens}, Adam R.~H. and {Tonini}, Chiara and {Garel}, Thibault and {Bernyk}, Maksym and {Bibiano}, Antonio and {Hodkinson}, Luke and {Mutch}, Simon J. and {Poole}, Gregory B. and {Shattow}, Genevieve M.},
        title = "{Semi-Analytic Galaxy Evolution (SAGE): Model Calibration and Basic Results}",
      journal = {\apjs},
         year = 2016,
        month = feb,
       volume = {222},
       number = {2},
          eid = {22},
        pages = {22},
          doi = {10.3847/0067-0049/222/2/22},
archivePrefix = {arXiv},
       eprint = {1601.04709},
 primaryClass = {astro-ph.GA},
       adsurl = {https://ui.adsabs.harvard.edu/abs/2016ApJS..222...22C}
}

@ARTICLE{Hirschmann_2016,
       author = {{Hirschmann}, Michaela and {De Lucia}, Gabriella and {Fontanot}, Fabio},
        title = "{Galaxy assembly, stellar feedback and metal enrichment: the view from the GAEA model}",
      journal = {\mnras},
         year = 2016,
        month = sep,
       volume = {461},
       number = {2},
        pages = {1760-1785},
          doi = {10.1093/mnras/stw1318},
archivePrefix = {arXiv},
       eprint = {1512.04531},
 primaryClass = {astro-ph.GA},
       adsurl = {https://ui.adsabs.harvard.edu/abs/2016MNRAS.461.1760H}
}

@ARTICLE{Lagos_2018,
       author = {{Lagos}, Claudia del P. and {Tobar}, Rodrigo J. and {Robotham}, Aaron S.~G. and {Obreschkow}, Danail and {Mitchell}, Peter D. and {Power}, Chris and {Elahi}, Pascal J.},
        title = "{Shark: introducing an open source, free, and flexible semi-analytic model of galaxy formation}",
      journal = {\mnras},
         year = 2018,
        month = dec,
       volume = {481},
       number = {3},
        pages = {3573-3603},
          doi = {10.1093/mnras/sty2440},
archivePrefix = {arXiv},
       eprint = {1807.11180},
 primaryClass = {astro-ph.GA},
       adsurl = {https://ui.adsabs.harvard.edu/abs/2018MNRAS.481.3573L}
}

@ARTICLE{Zentner_2005,
       author = {{Zentner}, Andrew R. and {Berlind}, Andreas A. and {Bullock}, James S. and {Kravtsov}, Andrey V. and {Wechsler}, Risa H.},
        title = "{The Physics of Galaxy Clustering. I. A Model for Subhalo Populations}",
      journal = {\apj},
         year = 2005,
        month = may,
       volume = {624},
       number = {2},
        pages = {505-525},
          doi = {10.1086/428898},
archivePrefix = {arXiv},
       eprint = {astro-ph/0411586},
 primaryClass = {astro-ph},
       adsurl = {https://ui.adsabs.harvard.edu/abs/2005ApJ...624..505Z}
}

@ARTICLE{Boylan_Kolchin_2008,
       author = {{Boylan-Kolchin}, Michael and {Ma}, Chung-Pei and {Quataert}, Eliot},
        title = "{Dynamical friction and galaxy merging time-scales}",
      journal = {\mnras},
         year = 2008,
        month = jan,
       volume = {383},
       number = {1},
        pages = {93-101},
          doi = {10.1111/j.1365-2966.2007.12530.x},
archivePrefix = {arXiv},
       eprint = {0707.2960},
 primaryClass = {astro-ph},
       adsurl = {https://ui.adsabs.harvard.edu/abs/2008MNRAS.383...93B}
}

@ARTICLE{Trenti_2010,
       author = {{Trenti}, Michele and {Smith}, Britton D. and {Hallman}, Eric J. and {Skillman}, Samuel W. and {Shull}, J. Michael},
        title = "{How Well do Cosmological Simulations Reproduce Individual Halo Properties?}",
      journal = {\apj},
         year = 2010,
        month = mar,
       volume = {711},
       number = {2},
        pages = {1198-1207},
          doi = {10.1088/0004-637X/711/2/1198},
archivePrefix = {arXiv},
       eprint = {1001.5037},
 primaryClass = {astro-ph.CO},
       adsurl = {https://ui.adsabs.harvard.edu/abs/2010ApJ...711.1198T}
}
\bibliographystyle{aasjournal}

\appendix

\section{Convergence with 79 base simulations}\label{app:medium_base}

\begin{figure*}
    \centering
    \textbf{Increasing training set size through rescaling from 79 base simulations}
    \vspace{-0.1em}

    \begin{subfigure}{0.48\textwidth}
        \centering
        \includegraphics[width=\textwidth]{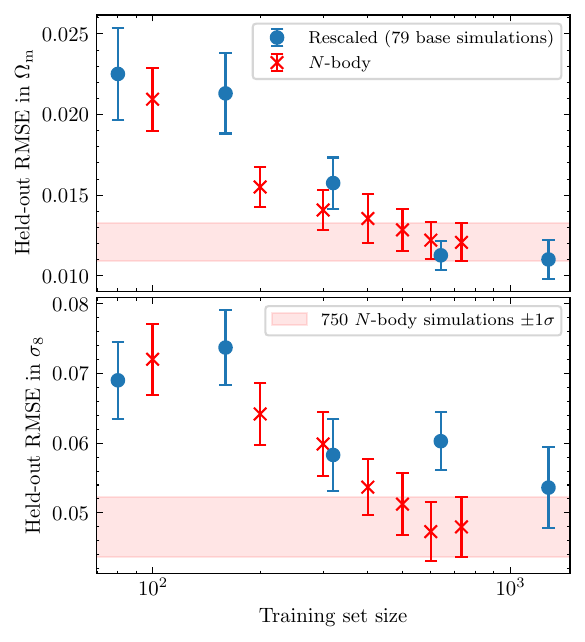}
        \caption{Models \textbf{trained on the SMF.}}
        \label{fig:SMF_vary_rescaled_medium}
    \end{subfigure}
    \hfill
    \begin{subfigure}{0.48\textwidth}
        \centering
        \includegraphics[width=\textwidth]{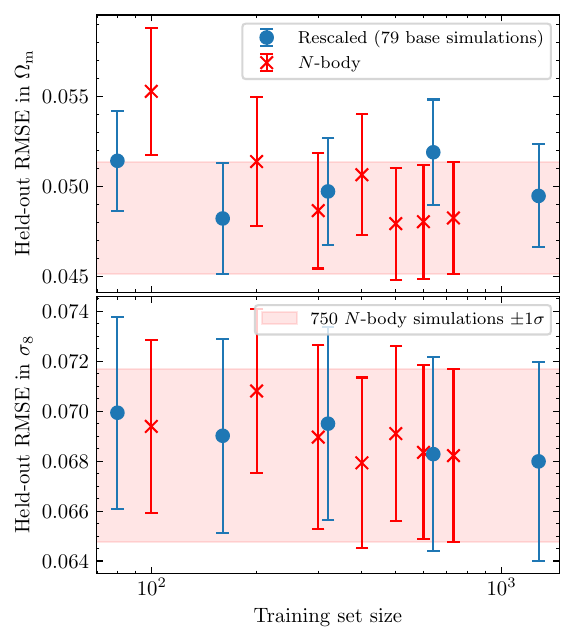}
        \caption{Models \textbf{trained on the galaxy 2PCF.}}
        \label{fig:2PCF_vary_rescaled_medium}
    \end{subfigure}
    \caption{
        RMSE in $\Om$ and $\sigma_8$ as a function of training-set size.
        Blue points: models trained on rescaled simulations generated from a fixed pool of 79 base $N$-body simulations (64 from \CS\ plus 15 at $\sigma_8 = 1$), with the number of rescaled realisations varying along the $x$-axis.
        Red points: models trained directly on $N$-body simulations without rescaling.
        The RMSE is evaluated on \num{150} held-out SAM galaxy populations derived from direct $N$-body simulations.
        Error bars show $1\sigma$ bootstrap uncertainties; the red shaded band shows the RMSE $\pm 1\sigma$ when training on all \num{750} $N$-body simulations.
        }
    \label{fig:vary_rescaled_medium}
\end{figure*}

\autoref{fig:vary_rescaled_medium} shows the RMSE in $\Om$ and $\sigma_8$ as a function of training-set size for models trained on rescaled simulations generated from 79 base $N$-body simulations (64 from \CS\ plus 15 at $\sigma_8 = 1$), with the number of rescaled realisations varying from 100 to \num{1600}.
Compared to the 25-base scenario (\autoref{fig:vary_base_small}), the SMF-trained models show marginal improvements, and in both cases the RMSE is now comparable to that obtained with \num{750} base simulations, though we have not tested whether further increasing the number of rescaled realisations to \num{2400} or \num{3200} would yield the additional gains observed in the \num{750}-base case (\autoref{fig:vary_rescaled}).
For the 2PCF, the improvement over the 25-base scenario is more pronounced: with 79 base simulations, the rescaled models no longer exhibit the systematically worse performance in $\sigma_8$ that was observed with fewer base simulations, and instead achieve RMSE values comparable to models trained directly on $N$-body simulations.

\end{document}